\documentclass{aa}

\usepackage{graphicx,textcomp}
\usepackage[varg]{txfonts}
\usepackage[colorlinks=true, linkcolor=blue, citecolor=blue]{hyperref}
\usepackage{multirow}
\usepackage{natbib}
\bibpunct{(}{)}{;}{a}{}{,} % to follow the A&A style
\usepackage{threeparttable}

\usepackage{natbib,twoopt}
\bibpunct{(}{)}{;}{a}{}{,}             %% natbib format for A&A and ApJ
\hypersetup{
  colorlinks,
  citecolor=[rgb]{0.06, 0.2, 0.65},
  linkcolor=[rgb]{0.8, 0.2, 1.0},
  urlcolor=[rgb]{0.06, 0.2, 0.65},
}
\makeatletter
  \newcommandtwoopt{\citeads}[3][][]{\href{http://adsabs.harvard.edu/abs/#3}%
    {\def\hyper@linkstart##1##2{}%
     \let\hyper@linkend\@empty\citealp[#1][#2]{#3}}}
  \newcommandtwoopt{\citepads}[3][][]{\href{http://adsabs.harvard.edu/abs/#3}%
    {\def\hyper@linkstart##1##2{}%
     \let\hyper@linkend\@empty\citep[#1][#2]{#3}}}
  \newcommandtwoopt{\citetads}[3][][]{\href{http://adsabs.harvard.edu/abs/#3}%
    {\def\hyper@linkstart##1##2{}%
     \let\hyper@linkend\@empty\citet[#1][#2]{#3}}}
  \newcommandtwoopt{\citeyearads}[3][][]%
    {\href{http://adsabs.harvard.edu/abs/#3}
    {\def\hyper@linkstart##1##2{}%
     \let\hyper@linkend\@empty\citeyear[#1][#2]{#3}}}
\makeatother

\newcommand{\kms}{km\,s$^{-1}$\xspace}
\newcommand{\tengs}[1]{$\times$ 10$^{#1}$ g\,s$^{-1}$}
\newcommand{\mdot}{$\mathrm{\dot{m}}$\xspace}
\newcommand{\emin}[2]{{#1}$\times10^{#2}$}
\newcommand{\chisquare}{$\mathrm{\chi^2}$\xspace}
\newcommand{\mc}{\multicolumn}
\newcommand{\change}[1]{\color{black}#1\color{black}\xspace} 
\begin{document}

   \title{Probing atmospheric escape through metastable He\,I triplet lines in 15 exoplanets observed with \mbox{SPIRou}}

   \author{A. Masson\inst{1}
          \and S. Vinatier\inst{1}
          \and B. Bézard\inst{1}
          \and M. López-Puertas\inst{2}
          \and M. Lampón\inst{2}
          \and F. Debras\inst{3}
          \and A. Carmona\inst{4}
          \and B. Klein\inst{5}
          \and E. Artigau\inst{6}
          \and W. Dethier\inst{4}
          \and S. Pelletier\inst{6}
          \and T. Hood\inst{3}
          \and R. Allart\inst{6}
          \and V. Bourrier\inst{7}
          \and C. Cadieux\inst{6}
          \and B. Charnay\inst{1}
          \and N. B. Cowan\inst{8}
          \and N. J. Cook\inst{6}
          \and X. Delfosse\inst{4}
          \and J.-F. Donati\inst{3}
          \and P.-G. Gu\inst{9}
          \and G. H\'ebrard\inst{10,11}
          \and E. Martioli\inst{10,12}
          \and C. Moutou\inst{3}
          \and O. Venot\inst{13}
          \and A. Wyttenbach\inst{4,14}
          }
   \institute{LESIA, Observatoire de Paris, Université PSL, CNRS, Sorbonne Université, Université Paris Cité, 5 place Jules Janssen, 92195 Meudon, France\\       % 1
              \email{adrien.masson@obspm.fr}
              \and Instituto de Astrofísica de Andalucía (IAA-CSIC), Glorieta de la Astronomía s/n, 18008 Granada, Spain                                          % 2
              \and IRAP, Université de Toulouse, CNRS, UPS, Toulouse, France                                                                                      % 3
              \and Univ. Grenoble Alpes, CNRS, IPAG, 38000 Grenoble, France                                                                                       % 4
              \and Department of Physics, University of Oxford, Oxford OX1 3RH, UK                                                                                % 5
              \and D\'epartement de Physique, Institut Trottier de Recherche sur les Exoplan\`etes, Universit\'e de Montr\'eal, Montr\'eal, Qu\'ebec, H3T 1J4, Canada                                                                                                                                              % 6
              \and Observatoire Astronomique de l’Université de Genève, Chemin Pegasi 51b, CH-1290 Versoix, Switzerland                                           % 7
              \and Department of Physics and Department of Earth \& Planetary Sciences, McGill University, 3550 rue University, Montr\'eal, QC H3A 2A7, Canada    % 8
              \and Institute of Astronomy and Astrophysics, Academia Sinica, Taipei 10617, Taiwan                                                                % 9
              \and Institut d'astrophysique de Paris, UMR7095 CNRS, Universit\'e Pierre \& Marie Curie, 98bis boulevard Arago, 75014 Paris, France                % 10
              \and Observatoire de Haute-Provence, CNRS, Universit\'e d'Aix-Marseille, 04870 Saint-Michel-l'Observatoire, France                                  % 11
              \and Laborat\'{o}rio Nacional de Astrof\'{i}sica, Rua Estados Unidos 154, 37504-364, Itajub\'{a} - MG, Brazil                                       % 12
              \and Université Paris Cité and Univ Paris Est Creteil, CNRS, LISA, F-75013 Paris, France                                                            % 13
              \and Leiden Observatory, Leiden University, Postbus 9513, 2300 RA Leiden, The Netherlands                                                           % 14
              }

   \date{Received Month dd, yyyy; accepted Month dd, yyyy}

  \abstract
  {For several years, the metastable helium triplet line has been successfully used as a tracer to probe atmospheric escape in transiting exoplanets. This absorption in the near-infrared (1083.3\,nm) can be observed from the ground using high-resolution spectroscopy, providing new constraints on the mass-loss rate and the temperature characterizing the upper atmosphere of close-in exoplanets.
  % aims heading (mandatory)
  The aim of this work is to search for the He triplet signature in 15 transiting exoplanets ---ranging from super-Earths to ultrahot Jupiters---  observed with \mbox{SPIRou}, a high-resolution ($R\sim$70\,000) near-infrared spectropolarimeter at the CFHT, in order to bring new constraints or to improve existing ones regarding atmospheric escape through a homogeneous study.
  % methods heading (mandatory)
  We developed a full data processing and analysis pipeline to correct for the residual telluric and stellar contributions. We then used two different 1D models based on the Parker-wind equations and nonlocal thermodynamic equilibrium (NLTE) radiative transfer to interpret the observational results.
  % results heading (mandatory)
  We confirm published He triplet detections for HAT-P-11\,b, HD\,189733\,b, and WASP-69\,b. We tentatively detect the signature of escaping He in HD\,209458\,b, GJ\,3470\,b, and WASP-76\,b. We report new constraints on the mass-loss rate and temperature for our three detections and set upper limits for the tentative and nondetections. We notably report improved constraints on the mass-loss rate and temperature of the escaping gas for TOI-1807\,b, and report a nondetection for the debated atmospheric escape in GJ\,1214\,b. We also conducted the first search for the He signature in GJ\,486\,b since its discovery and report a nondetection of the He triplet. Finally, we studied the impact of important model assumptions on our retrieved parameters, notably the limitations of 1D models and the influence of the H/He ratio on the derived constraints.}


  \maketitle
  %______________________________________________________________
  \section{Introduction}
  Exoplanets orbiting very close to their host star can be subject to atmospheric escape, which entails a collection of thermal and nonthermal processes in which the stellar irradiation, by injecting energy into the planet's atmosphere, causes some of the upper atmosphere constituents to overcome the planet's gravity \citep{Gronoff2020}. Erosion of a significant fraction of the envelope mass through atmospheric escape occurs generally on timescales of $\sim$\,100\,Myr to 1\,Gyr \citep{Owen2019,King2021,Attia2021}, and has a large impact on planetary evolution and affects the composition and structure of planetary atmospheres \citep[\change{e.g.,}][]{Watson1981,Yelle2004,Garcia2007,Shaikhislamov2014,Kubyshkina2018,Gu2023,Malsky2023}. It is thought that atmospheric evaporation plays a major role in the formation of super-Earths from sub-Neptune mass planets \citep{Owen2017}. This, combined with high-eccentricity migration, could explain some of the features of the observed population of exoplanets (\change{e.g.,} in mass--period or radius--period distributions), such as the close-in sub-Jovian planet desert \citep{Owen2013,Lecavelier2007,Beauge2013,Lundkvist2016,Mazeh2016,Fulton2017,Owen2018,Bourrier2018c}.

  Hydrogen, being the lightest and most abundant element in primordial atmospheres, was the first tracer used to constrain atmospheric escape. Successful measurements were obtained with Ly-$\alpha$ observations in the UV (at 0.1215\,$\mu$m), starting with the first detection of escaping hydrogen by \citet{Vidal2003} in HD\,209458\,b using HST/STIS. This was followed by several other detections, for example in HD\,189733\,b \citep{Lecavelier2010,Lecavelier2012,Bourrier2013_mar}, GJ\,436\,b \citep{Kulow2014,Ehrenreich2015,LaVie2017, DosSantos2019}, and GJ\,3470\,b \citep{Bourrier2018_3470}. However, Ly-$\alpha$ measurements are contaminated by emission from the Earth's exosphere (geocoronal emission) and are restricted to space-based observations. Furthermore, this emission line is absorbed by the interstellar medium (ISM). As a consequence, only the line's wings (\change{i.e.,} corresponding to radial velocities beyond $\pm$30\,km\,s$^{-1}$ in the exoplanet rest frame; \citealp{Lecavelier2010,DosSantos2019}) of nearby (\change{i.e.,} up to $\sim$50 pc) stars can be characterized. Atmospheric escape can also be probed by the H$\alpha$ line in the visible and metallic lines in the UV (\change{e.g.,} O I and C II in HD\,209458\,b and HD\,189733\,b, \citealp{Vidal2004,BenJaffel2013}), and a few other lines in the near-ultraviolet (NUV; \change{e.g.,} Fe II, Mg II, \citealp{Fossati2010,Haswell2012,Sing2019,Cubillos2020}).

  For some years now, the metastable He triplet lines (1083.3\,nm in vacuum) have been a fruitful near-infrared (NIR) tracer for atmospheric escape studies. This tracer is indeed weakly affected by interstellar absorption \citep{Indriolo2009} and lies in a spectral region with relatively little telluric absorption. It can thus be observed from the ground using high-resolution spectroscopy, and generally occurs in a region of the stellar continuum with a larger flux with respect to UV observations, thus enhancing the observed signal-to-noise ratio (\change{S/N}). Originally based on the simulations of \citet{SeagerSasselov2000}, early attempts to detect these absorption lines using ground-based low-resolution spectrographs were unsuccessful \citep{Moutou2003}, until \citet{Spake2018} obtained the first detection of escaping He in an exoplanet with HST.

  Since \citet{Oklopcic2018} introduced the formalism of He escape modeling, this tracer has been widely used for atmospheric escape detection and characterization. The detection of escaping He needs sufficient stellar X-ray and extreme ultraviolet (EUV) radiation to populate the metastable triplet state, which requires active or moderately active stars \citep{Oklopcic2019}. The metastable triplet population also depends on the NUV (1000-2600\,\AA) stellar flux, which depopulates the triplet through ionization \citep[\change{e.g.,}][]{Lampon2020,Shaikhislamov2021,Khodachenko2021}. Observation of escaping metastable He should thus be favored for planets with extended atmospheres orbiting stars with strong extreme-ultraviolet (XUV) and low NUV emission, which is typically the case for active K-stars.

  Detections have been reported both from space ---using the Wide Field Camera 3 (WFC3) on board HST (observations of WASP-107 b and HAT-P-11\,b by \citealp{Spake2018} and \citealp{Mansfield2018})--- and from the ground using NIR high-resolution spectroscopy (with CARMENES, GIANO, Keck/NIRSPEC, \mbox{SPIRou}) in numerous targets, such as HAT-P-11\,b \citep{Allart2018}, WASP-69\,b \citep{Nortmann2018}, HD\,209458\,b \citep{Alonso2019}, HD 1897333 b \citep{Salz2018}, and WASP-107 b \citep{Allart_2019,Kirk2020}. When analyzed by simulating hydrodynamic escape and nonlocal thermodynamic equilibrium (NLTE) radiative transfer, these observations can provide constraints on the mass-loss rate (\mdot) and temperature of the escaping gas. These parameters are key to understanding the underlying atmospheric escape processes and their role in planetary evolution.

  However, the constraints reported for most of these targets are based on a small number of observations, and would therefore benefit from additional measurements to strengthen the reported detections. Furthermore, the variety of instruments, reduction processes, and model assumptions used in He escape studies poses a challenge to the comparison of results. Recent efforts have led to homogenous surveys of atmospheric escape, where several targets were analyzed using a given methodology and/or a given instrument. \citet{Kasper2020} searched for helium signatures in three sub-Neptune-sized planets using Keck/NIRSPEC and did not report helium absorption for any of the targets. \citet{Vissapragada2022} targeted seven gas-giant planets orbiting K-type host stars with Palomar/WIRC and detected helium absorption for WASP-69\,b, HAT-P-18\,b, and HAT-P-26\,b, and tentatively detected signals for WASP-52\,b and NGTS-5\,b. \citet{Zhang2023} detected helium absorption in four young short-period mini-Neptunes (TOI\,560\,b, TOI\,1430\,b, TOI\,1683\,b, and TOI\,2076\,b) with Keck/NIRSPEC. \citet{Allart_2023} searched for atmospheric escape in 11 exoplanets observed with \mbox{SPIRou} (Spectro-Polarimètre InfraROUge; \citealp{Donati2020}), confirming published detections for HAT-P-11\,b, HD\,189733\,b, and WASP-69\,b. \citet{Guilluy2023} targeted nine planets along the edges of the sub-Jovian planet desert using the GIANO-B spectrograph mounted on the Telescopio Nazionale Galileo telescope and reported no signatures of He absorption in any of the targets. Both studies employed the 1D \textit{p-winds} model to derive constraints on the atmospheric escape properties ---mainly its temperature and mass-loss rate--- assuming a solar H/He ratio  in both studies and limiting the vertical extension of the model to the Roche lobe. These studies pave the way toward a statistical understanding of atmospheric escape in close-in exoplanets, calling for more observations and studies through global surveys. We provide here a homogeneous study of atmospheric escape of the atmosphere of 15 exoplanets (including different kinds of hot Jupiters, Saturns, and Neptunes or super-Earths) through the observation of metastable helium with \mbox{SPIRou}. For comparison purposes with previous studies, we used two 1D spherically symmetric models (the p-winds code from \citealp{DosSantos2022} and an adapted \texttt{Python} version of the model used in \citealp{Lampon2023}), comprising hydrodynamic and NLTE radiative transfer processes, to retrieve constraints on the mass-loss rate and temperature of the escaping atmosphere. We also studied the impacts of our main model parameters and common assumptions in 1D helium escape studies, notably the influence of fixing the H/He ratio to a solar value and limiting the model's radial extension to the Roche lobe, and the impact of center-to-limb variations and the Rossiter-Mclaughlin effect on the He signature \citep{Dethier2023}, for which we applied the correction from \citet{Guilluy2023}.
  
  This paper is organized as follows: Section~\ref{observation} presents the observations, while Section~\ref{reduction} describes our data reduction and processing pipeline. Section~\ref{Modeling} presents the models we used for the retrieval analysis, and Section~\ref{results} shows the results obtained for each of the studied targets. We discuss the influence of our main model parameters in Section~\ref{discussion} and summarize our work with perspectives for future atmospheric escape observations in Section~\ref{summary}.

  %______________________________________________________________
  \section{Observations}

  \label{observation}
    All data used in this work were acquired with \mbox{SPIRou}, a near-infrared cryogenic spectropolarimeter with a continuous spectral range between 0.95 and 2.50 $\mu$m. The instrument is mounted on the 3.6\,m Canada France Hawaii Telescope (CFHT), in Maunakea. It covers the \textit{Y}, \textit{J}, \textit{H}, and \textit{K} bands with a high resolving power of 70~000, each pixel corresponding to a $\sim$2.28\,\kms velocity bin \citep{Donati2020}. Data are distributed over 50 spectral orders, indexed from \#31 in the red to \#79 in the blue, each containing 4088 pixels. In this study, we only used the echelle order \#71, spanning the spectral range from 1063.9 to 1097.6\,nm, as it is centered around the helium triplet wavelengths.

\begin{figure}[h]
        \centering
        \includegraphics[width=\hsize]{"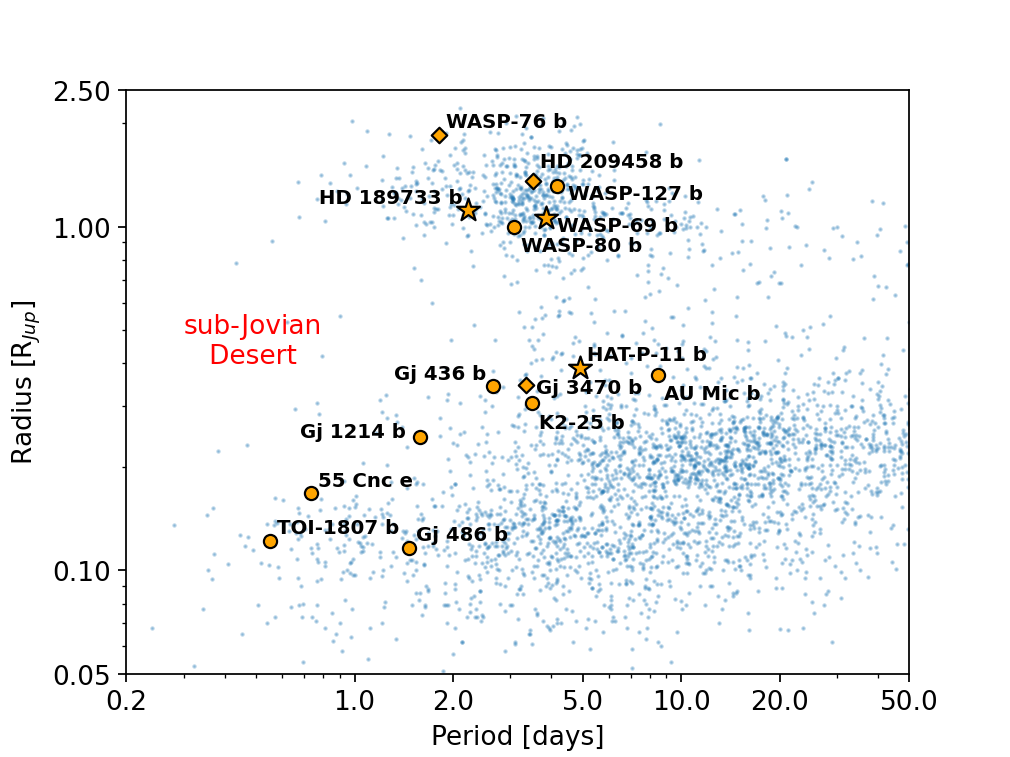"}
         \caption{Period--radius diagram. The blue dots are exoplanets from the \url{exoplanet.eu} catalog. The 15 exoplanets studied here are represented with orange markers (stars for helium detections, diamonds for tentative detections, and dots for nondetections), and the position of the sub-Jovian planet desert is highlighted in red.}
         \label{period_radius}
      \end{figure}

    We focused our analysis on 15 transiting short-period exoplanets observed with \mbox{SPIRou} through two large programs: the \mbox{SPIRou} Legacy Survey (SLS, PI: J.-F. Donati) and ATMOSPHERIX \citep[PI: F. Debras, see][]{Debras_2023,Klein_2023}, completed with the reanalysis of a few publicly available data (PI: A. Boucher, E. Deibert, G. Hebrard, M. Radica). Our target list contains one ultrahot Jupiter (WASP-76b), three hot Jupiters (HD\,189733\,b, HD\,209458\,b, and the inflated hot Jupiter WASP-127\,b), one warm Jupiter (WASP-80\,b), one hot Saturn (WASP-69\,b), five warm Neptunes (AU\,Mic\,b, GJ\,436\,b, GJ\,3470\,b, HAT-P-11\,b, K2-25\,b), two mini-Neptunes (GJ\,1214\,b, and TOI-1807\,b, which may also be a super-Earth as no atmospheric signature has been detected yet; see \citealp{Gaidos2023}), and two super-Earths (55\,Cnc\,e and GJ\,486\,b). These targets are displayed in a period--radius diagram in Fig.~\ref{period_radius}. The list of the observed transits for each target is given in Table~\ref{transit_ppties_tab}, along with each transit's UT date, midpoint (in BJD$_\mathrm{TDB}$), duration, exposure time, and the number of exposures. A complete list of the planetary and stellar parameters for each target is given in Table~\ref{table_stellar_ppties} and Table~\ref{table_planetary_ppties}.

    \begin{table*}[]
      \begin{threeparttable}
       \caption[]{Transit characteristics of the studied targets}
       \label{transit_ppties_tab}
       \begin{tabular}{llllllll}
        \hline \hline
        Target                        & Program ID      & PI                      & UT date         & Midpoint                   & Transit duration        & Texp            & Nexp            \\
                                      &                 &                         & [yyyy-mm-dd]    & [BJD$_\mathrm{TDB}$]       & [hours]                 & [sec]           &                 \\
        \hline
        55\,Cnc\,e                    & 20BF09          & F. Debras               & 2020-12-30      & 2459213.9232               & 1.59                    & 44.575          & 176             \\ \smallskip
        AU\,Mic\,b$^{\dagger}$        & 19AP42          & J.-F. Donati            & 2019-06-17      & 2458651.9845               & 1.66                    & 122.582         & 116             \\ \smallskip
        GJ\,1214\,b                   & 19AP40          & J.-F. Donati            & 2019-04-18      & 2458591.97315              & 0.86                    & 601.767         & 11              \\ \smallskip
        GJ\,1214\,b$^{\dagger}$       & 20AP40          & J.-F. Donati            & 2020-05-14      & 2458983.91348              & 0.86                    & 250.736         & 30              \\ \smallskip
        GJ\,1214\,b                   & 23BF16          & F. Debras               & 2023-09-21      & 2460208.72703              & 0.86                    & 406.75          & 15              \\ \smallskip
        GJ\,3470\,b$^{\dagger}$       & 19AP40          & J.-F. Donati            & 2019-02-18      & 2458532.9062               & 1.39                    & 434.61          & 32              \\ \smallskip
        GJ\,3470\,b$^{\dagger}$       & 21BP40          & J.-F. Donati            & 2021-12-15      & 2459563.9318               & 1.39                    & 300.884         & 43              \\ \smallskip
        GJ\,3470\,b                   & 23BF16          & F. Debras               & 2023-12-29      & 2460308.0053               & 1.39                    & 551.62          & 20              \\ \smallskip
        GJ\,436\,b                    & 19AQ57          & QSO Team                & 2019-02-25      & 2458540.1018               & 0.72                    & 122.582         & 36              \\ \smallskip
        GJ\,486\,b                    & 21AD97          & F. Debras               & 2021-03-24      & 2459297.9391               & 0.97                    & 217.305         & 30              \\ \smallskip
        GJ\,486\,b                    & 21BC34          & E. Deibert              & 2022-01-23      & 2459603.0999               & 0.97                    & 183.873         & 42              \\ \smallskip
        HAT-P-11\,b                   & 21AF18          & F. Debras               & 2021-06-30      & 2459395.93748              & 2.41                    & 323.171         & 44              \\ \smallskip
        HAT-P-11\,b$^{\dagger}$       & 21BC09          & M. Radica               & 2021-08-13      & 2459439.9277               & 2.41                    & 284.168         & 46              \\ \smallskip
        HAT-P-11\,b$^{\dagger}$       & 21BF19          & F. Debras               & 2021-08-18      & 2459444.8155               & 2.41                    & 323.171         & 44              \\ \smallskip
        HAT-P-11\,b                   & 22AF11          & F. Debras               & 2022-06-12      & 2459742.97142              & 2.41                    & 323.171         & 37              \\ \smallskip
        HD\,189733\,b$^{\dagger}$     & 18BD50          & CFHT/\mbox{SPIRou} team & 2018-09-22      & 2458383.80116              & 1.97                    & 250.736         & 36              \\ \smallskip
        HD\,189733\,b$^{\dagger}$     & 19AP40          & J.-F. Donati            & 2019-06-15      & 2458650.03015              & 1.97                    & 250.736         & 50              \\ \smallskip
        HD\,189733\,b$^{\dagger}$     & 20AC01          & E. Deibert              & 2020-07-03      & 2459033.84362              & 1.97                    & 94.723          & 56              \\ \smallskip
        HD\,189733\,b$^{\dagger}$     & 20AC01          & E. Deibert              & 2020-07-05      & 2459036.06219              & 1.97                    & 94.723          & 52              \\ \smallskip
        HD\,189733\,b$^{\dagger}$     & 20AC01          & E. Deibert              & 2020-07-25      & 2459056.02937              & 1.97                    & 94.723          & 124             \\ \smallskip
        HD\,189733\,b$^{\dagger}$     & 21BC16          & E. Deibert              & 2021-08-24      & 2459450.93571              & 1.97                    & 94.723          & 107             \\ \smallskip
        HD\,209458\,b                 & 19AF16          & G. Hebrard              & 2019-06-18      & 2458653.04                 & 3.09                    & 902.651         & 18              \\ \smallskip
        K2-25\,b                      & 19BP40          & J.-F. Donati            & 2019-10-06      & 2458763.047                & 1.29                    & 300.884         & 20              \\ \smallskip
        TOI-1807\,b                   & 22AF11          & F. Debras               & 2022-04-15      & 2459684.947                & 1.00                    & 362.175         & 20              \\ \smallskip
        TOI-1807\,b                   & 22AF11          & F. Debras               & 2022-06-09      & 2459739.8842               & 1.00                    & 362.175         & 20              \\ \smallskip
        WASP-127\,b$^{\dagger}$       & 20AP42          & J.-F. Donati            & 2020-03-11      & 2458919.9682               & 4.54                    & 300.884         & 50              \\ \smallskip
        WASP-127\,b$^{\dagger}$       & 21AC02          & A. Boucher              & 2021-03-22      & 2459295.994                & 4.54                    & 501.473         & 28              \\ \smallskip
        WASP-127\,b$^{\dagger}$       & 21AC02          & A. Boucher              & 2021-05-03      & 2459337.7747               & 4.54                    & 501.473         & 28              \\ \smallskip
        WASP-69\,b$^{\dagger}$        & 19BP40          & J.-F. Donati            & 2019-10-13      & 2458769.8511               & 2.23                    & 122.582         & 93              \\ \smallskip
        WASP-76\,b                    & 20BF09          & F. Debras               & 2020-10-31      & 2459153.8849               & 3.75                    & 507.045         & 27              \\ \smallskip
        WASP-76\,b                    & 21BF19          & F. Debras               & 2021-10-28      & 2459515.861                & 3.75                    & 774.497         & 26              \\ \smallskip
        WASP-80\,b$^{\dagger}$        & 19BP40          & J.-F. Donati            & 2019-10-07      & 2458763.77156              & 1.27                    & 183.873         & 74              \\
        \hline
       \end{tabular}

       \begin{tablenotes}[flushleft]
          \footnotesize{\item{\textbf{Notes.}  The planetary and stellar parameters are respectively given in Tables \ref{table_planetary_ppties} and \ref{table_stellar_ppties}. Transit midpoints were computed using the periods and midpoint reference values from the ExoClock\tablefootmark{*} database \citep{Kokori2022_a,Kokori2022_b,Kokori2023}. "Texp" is the exposure time, in seconds, of a single exposure. "Nexp" is the total number of exposures collected during the transit. Data from transits marked with a dagger ($\dagger$) have previously been analyzed by \citet{Allart_2023} using an older (v0.7.179) version of \mbox{APERO}, the \mbox{SPIRou} Data Reduction Software.}
          \\
          \tablefootmark{*}{\url{https://www.exoclock.space/}}
          }
       \end{tablenotes}

     \end{threeparttable}
    \end{table*}

    % \onecolumn
    % For changing the vertical spacing, see the beginning of this document.
    \begin{table*}
      \begin{threeparttable}[b]
      \caption{\label{table_stellar_ppties} Host star properties used in this study}

      \centering
      \begin{tabular}{|l|*{12}{c|}}
        \hline
        Host star                              & \mc{3}{c|}{ 55 Cnc                                   } & \mc{3}{c|}{ AU Mic                          }         & \mc{3}{c|}{ GJ 436                          }            & \mc{3}{c|}{ GJ 486                          }              \\
        \hline
        M$_\mathrm{*}$ \hfill [M$_{\odot}$]    & \mc{3}{c|}{$ 0.905 \pm 0.015         $ \hfill (B18a) } & \mc{3}{c|}{$ 0.50 \pm 0.03  $ \hfill (Pl20)         } & \mc{3}{c|}{$ 0.445 \pm 0.018 $ \hfill (M22) }    & \mc{3}{c|}{$ 0.323 \pm 0.015 $ \hfill (T21)        }              \\
        R$_\mathrm{*}$ \hfill [R$_{\odot}$]    & \mc{3}{c|}{$ 0.943 \pm 0.010         $ \hfill (B18a) } & \mc{3}{c|}{$ 0.75 \pm 0.03  $ \hfill (W19)          } & \mc{3}{c|}{$ 0.425 \pm 0.006 $ \hfill (M22) }    & \mc{3}{c|}{$ 0.328 \pm 0.011 $ \hfill (T21)        }              \\
        Ks \hfill [m/s]                        & \mc{3}{c|}{$ 6.02^{+0.24}_{-0.23}    $ \hfill (B18a) } & \mc{3}{c|}{$ 5.8 \pm 2.5  $ \hfill (Z22)            } & \mc{3}{c|}{$ 17.38 \pm 0.17  $ \hfill (T18) }    & \mc{3}{c|}{$ 3.370^{+0.078}_{-0.080} $ \hfill (T21) }   \\
        $\mathrm{v_{sys}}$ \hfill [km/s]       & \mc{3}{c|}{$ 27.19 \pm 0.12          $ \hfill (GDR3) } & \mc{3}{c|}{$ -6.90 \pm 0.37  $ \hfill (GDR3)        } & \mc{3}{c|}{$ 8.87\pm0.16$ \hfill (GDR3)              }    & \mc{3}{c|}{$ 19.20 \pm 0.17$ \hfill (GDR3)         }              \\
        $\mathrm{v.sin(i)}$ \hfill [km/s]      & \mc{3}{c|}{$ 1.7 \pm 0.5             $ \hfill (B16)  } & \mc{3}{c|}{$ 7.8\pm0.3 $ \hfill (K21)      } & \mc{3}{c|}{$ (293.5^{+43.7}_{-52.2}).10^{-3}$ \hfill (B22)}    & \mc{3}{c|}{$ \leq 0.128$ \hfill (T21)              }              \\
        c1                                     & \mc{3}{c|}{$ 0.544 \pm 0.008         $ \hfill (B18a) } & \mc{3}{c|}{$ 0.17^{+0.22}_{-0.12} $ \hfill (Pl20)   } & \mc{3}{c|}{$ 0.151 $ \hfill (EXO)          }   & \mc{3}{c|}{$ 0.169 $ \hfill (EXO)                  }              \\
        c2                                     & \mc{3}{c|}{$ 0.186 \pm 0.004         $ \hfill (B18a) } & \mc{3}{c|}{$ 0.15^{+0.27}_{-0.21} $ \hfill (Pl20)   } & \mc{3}{c|}{$ 0.431 $ \hfill (EXO)          }   & \mc{3}{c|}{$  0.424 $ \hfill (EXO)                 }              \\
        Teff \hfill [K]                        & \mc{3}{c|}{$ 5172 \pm 18             $ \hfill (B18a) } & \mc{3}{c|}{$ 3700 \pm 100  $ \hfill (P9)            } & \mc{3}{c|}{$ 3505 \pm 51 $ \hfill (S19)    }   & \mc{3}{c|}{$ 3340 \pm 54  $ \hfill (T21)           }              \\
        log(g) \hfill [cgs]                    & \mc{3}{c|}{$ 4.43 \pm 0.02           $ \hfill (B18a) } & \mc{3}{c|}{$4.39\pm0.03  $ \hfill (Z22)    } & \mc{3}{c|}{$ 4.91 \pm 0.07 $ \hfill (S19)  }   & \mc{3}{c|}{$ 4.92 \pm 0.03 $ \hfill (T21)          }              \\
        {[Fe/H]}                               & \mc{3}{c|}{$ 0.35 \pm 0.1            $ \hfill (B18a) } & \mc{3}{c|}{$ -0.12  $ \hfill (G14)                  } & \mc{3}{c|}{$ -0.04 \pm 0.16 $ \hfill (S19) }   & \mc{3}{c|}{$ 0.070 \pm 0.160 $ \hfill (T21)        }              \\
        \hline  \hline
        Host star                              & \mc{3}{c|}{ GJ 1214                                  } & \mc{3}{c|}{ GJ 3470                         } & \mc{3}{c|}{ HAT-P-11                        }    & \mc{3}{c|}{ HD 189733                       }              \\
        \hline
        M$_\mathrm{*}$ \hfill [M$_{\odot}$]    & \mc{3}{c|}{$ 0.178 \pm 0.010         $ \hfill (C21) } & \mc{3}{c|}{$ 0.476 \pm 0.019  $ \hfill (Pa20)       } & \mc{3}{c|}{$ 0.802 \pm 0.028 $ \hfill (A18)        }    & \mc{3}{c|}{$ 0.846 \pm 0.049 $ \hfill (B15)        }              \\
        R$_\mathrm{*}$ \hfill [R$_{\odot}$]    & \mc{3}{c|}{$ 0.215 \pm 0.008         $ \hfill (C21) } & \mc{3}{c|}{$ 0.474 \pm 0.014  $ \hfill (Pa20)       } & \mc{3}{c|}{$ 0.683 \pm 0.009 $ \hfill (D11)        }    & \mc{3}{c|}{$ 0.805 \pm 0.016 $ \hfill (B15)        }              \\
        Ks \hfill [m/s]                        & \mc{3}{c|}{$ 14.36 \pm 0.53          $ \hfill (C21) } & \mc{3}{c|}{$ 8.9 \pm 1.1  $ \hfill (D13)            } & \mc{3}{c|}{$ 12.01 \pm 1.38 $ \hfill (A18)         }    & \mc{3}{c|}{$ 208 \pm 6 $ \hfill (A19)              }              \\
        $\mathrm{v_{sys}}$ \hfill [km/s]       & \mc{3}{c|}{$ 20.91\pm0.65            $ \hfill (GDR3)} & \mc{3}{c|}{$ 25.95\pm0.25  $ \hfill (GDR3)          } & \mc{3}{c|}{$ -63.46 \pm0.13$ \hfill (GDR3)         }    & \mc{3}{c|}{$ -2.53 \pm 0.12$ \hfill (GDR3)         }              \\
        $\mathrm{v.sin(i)}$ \hfill [km/s]      & \mc{3}{c|}{$ \sim 0.084              $ \hfill (M18) } & \mc{3}{c|}{$ \sim 2  $ \hfill (Pa20)                } & \mc{3}{c|}{$ \sim 1.5$ \hfill (Y18)                }    & \mc{3}{c|}{$ 3.5\pm1.0 $ \hfill (B17)              }              \\
        c1                                     & \mc{3}{c|}{$ -0.0210 \pm 0.0052      $ \hfill (C21) } & \mc{3}{c|}{$ 0.033 \pm 0.015 $ \hfill (D13)         } & \mc{3}{c|}{$ 0.267$ \hfill (A18)                   }    & \mc{3}{c|}{$ 0.2248$ \hfill (A23)                  }              \\
        c2                                     & \mc{3}{c|}{$ 0.1852 \pm 0.0050       $ \hfill (C21) } & \mc{3}{c|}{$ 0.181 \pm 0.010  $ \hfill (D13)        } & \mc{3}{c|}{$ 0.265 $ \hfill (A18)                  }    & \mc{3}{c|}{$ 0.2795 $ \hfill (A23)                 }              \\
        Teff \hfill [K]                        & \mc{3}{c|}{$ 3250 \pm 100            $ \hfill (C21) } & \mc{3}{c|}{$ 3725 \pm 54  $ \hfill (Pa20)           } & \mc{3}{c|}{$ 4780 \pm 50 $ \hfill (B10)            }    & \mc{3}{c|}{$ 4875 \pm 43 $ \hfill (B15)            }              \\
        log(g) \hfill [cgs]                    & \mc{3}{c|}{$ 5.026 \pm 0.040         $ \hfill (C21) } & \mc{3}{c|}{$ 4.65 \pm 0.06  $ \hfill (Pa20)         } & \mc{3}{c|}{$ 4.59 \pm 0.03 $ \hfill (B10)          }    & \mc{3}{c|}{$ 4.56 \pm 0.03 $ \hfill (B15)          }              \\
        {[Fe/H]}                               & \mc{3}{c|}{$ 0.29 \pm 0.12           $ \hfill (C21) } & \mc{3}{c|}{$ 0.420 \pm 0.019  $ \hfill (Pa20)       } & \mc{3}{c|}{$ 0.31 \pm 0.05 $ \hfill (B10)          }    & \mc{3}{c|}{$ -0.03 \pm 0.08 $ \hfill (B15)         }              \\
        \hline  \hline
        Host star                              & \mc{3}{c|}{ HD 209458                               } & \mc{3}{c|}{ K2-25                           } & \mc{3}{c|}{ TOI-1807                        }    & \mc{3}{c|}{ WASP-69                         }              \\
        \hline
        M$_\mathrm{*}$ \hfill [M$_{\odot}$]    & \mc{3}{c|}{$ 1.119 \pm 0.033         $ \hfill (B17) } & \mc{3}{c|}{$ 0.2634 \pm 0.0077     $ \hfill (T20)   } & \mc{3}{c|}{$ 0.76\pm0.03  $ \hfill (N22)           }    & \mc{3}{c|}{$ 0.826\pm0.029     $ \hfill (A14)      }              \\
        R$_\mathrm{*}$ \hfill [R$_{\odot}$]    & \mc{3}{c|}{$ 1.155^{+0.014}_{-0.016} $ \hfill (B17) } & \mc{3}{c|}{$ 0.2932 \pm 0.0093     $ \hfill (T20)   } & \mc{3}{c|}{$ 0.690\pm0.036   $ \hfill (N22)        }    & \mc{3}{c|}{$ 0.813\pm0.028     $ \hfill (A14)      }              \\
        Ks \hfill [m/s]                        & \mc{3}{c|}{$ 84.27^{+0.69}_{-0.70}   $ \hfill (B17) } & \mc{3}{c|}{$ 27.9^{+6.5}_{-6.0}    $ \hfill (St20)  } & \mc{3}{c|}{$ 2.39^{+0.45}_{-0.46} $ \hfill (N22)   }    & \mc{3}{c|}{$ 38.1\pm2.4  $ \hfill (A14)            }              \\
        $\mathrm{v_{sys}}$ \hfill [km/s]       & \mc{3}{c|}{$ -14.78\pm0.16           $ \hfill (GDR3)} & \mc{3}{c|}{$ 42.92\pm4.11 $ \hfill (GDR3)           } & \mc{3}{c|}{$ -7.14 \pm 0.15$ \hfill (GDR3)         }    & \mc{3}{c|}{$ -9.83\pm0.13 $ \hfill (GDR3)          }              \\
        $\mathrm{v.sin(i)}$ \hfill [km/s]      & \mc{3}{c|}{$ 4.49\pm0.50             $ \hfill (B17) } & \mc{3}{c|}{$ 8.8\pm0.6 $ \hfill (St20)              } & \mc{3}{c|}{$ 4.2\pm0.5  $ \hfill (N22)             }    & \mc{3}{c|}{$ 2.2\pm0.4  $ \hfill (A14)             }              \\
        c1                                     & \mc{3}{c|}{$ 0.212                   $ \hfill (EXO) } & \mc{3}{c|}{$ 0.179 $ \hfill (EXO)                   } & \mc{3}{c|}{$ 0.46\pm0.05$ \hfill (N22)             }    & \mc{3}{c|}{$ 0.394$ \hfill (EXO)                   }              \\
        c2                                     & \mc{3}{c|}{$ 0.290                   $ \hfill (EXO) } & \mc{3}{c|}{$ 0.420 $ \hfill (EXO)                   } & \mc{3}{c|}{$ 0.17\pm0.05  $ \hfill (N22)           }    & \mc{3}{c|}{$ 0.197    $ \hfill (EXO)               }              \\
        Teff \hfill [K]                        & \mc{3}{c|}{$ 6065 \pm 50             $ \hfill (B17) } & \mc{3}{c|}{$ 3207 \pm 58     $ \hfill (T20)         } & \mc{3}{c|}{$ 4730\pm75  $ \hfill (N22)             }    & \mc{3}{c|}{$ 4715\pm50     $ \hfill (A14)          }              \\
        log(g) \hfill [cgs]                    & \mc{3}{c|}{$ 4.45 \pm 0.02           $ \hfill (S17) } & \mc{3}{c|}{$ 4.944 \pm 0.031      $ \hfill (St20)   } & \mc{3}{c|}{$ 4.55\pm±0.05  $ \hfill (N22)          }    & \mc{3}{c|}{$ 4.535\pm0.023    $ \hfill (A14)       }              \\
        {[Fe/H]}                               & \mc{3}{c|}{$ 0.00 \pm 0.05           $ \hfill (B17) } & \mc{3}{c|}{$ 0.15 \pm 0.03     $ \hfill (St20)      } & \mc{3}{c|}{$ -0.04\pm0.02  $ \hfill (N22)          }    & \mc{3}{c|}{$ 0.144\pm0.077    $ \hfill (A14)       }              \\
        \hline  \hline
        Host star                              & \mc{4}{c|}{ WASP-76                                 } & \mc{4}{c|}{  WASP-80                        } & \mc{4}{c|}{  WASP-127                       }    \\
        \hline
        M$_\mathrm{*}$ \hfill [M$_{\odot}$]    & \mc{4}{c|}{$ 1.458\pm0.021           $ \hfill (E20) } & \mc{4}{c|}{$ 0.577^{+0.051}_{-0.054} $ \hfill (T15) } & \mc{4}{c|}{$ 1.08\pm0.03        $ \hfill (L17)     }    \\
        R$_\mathrm{*}$ \hfill [R$_{\odot}$]    & \mc{4}{c|}{$ 1.756\pm0.071           $ \hfill (E20) } & \mc{4}{c|}{$ 0.586^{+0.017}_{-0.018} $ \hfill (T15) } & \mc{4}{c|}{$ 1.39\pm0.03       $ \hfill (L17)      }    \\
        Ks \hfill [m/s]                        & \mc{4}{c|}{$ 116.02^{+1.29}_{-1.35}  $ \hfill (E20) } & \mc{4}{c|}{$ 109.0^{+3.1}_{-4.4}   $ \hfill (T15)   } & \mc{4}{c|}{$ 22^{+3}_{-2}     $ \hfill (L17)       }    \\
        $\mathrm{v_{sys}}$ \hfill [km/s]       & \mc{4}{c|}{$ -0.79\pm0.25            $ \hfill (GDR3)} & \mc{4}{c|}{$ 9.91 \pm 0.31$ \hfill (GDR3)          } & \mc{4}{c|}{$ -8.97 \pm 0.16 $ \hfill (GDR3)        }    \\
        $\mathrm{v.sin(i)}$ \hfill [km/s]      & \mc{4}{c|}{$ 1.48\pm0.28             $ \hfill (E20) } & \mc{4}{c|}{$  1.27^{+0.14}_{-0.17}  $ \hfill (T15) } & \mc{4}{c|}{$ 0.3\pm0.2  $ \hfill (L17)             }    \\
        c1                                     & \mc{4}{c|}{$ 0.190                   $ \hfill (EXO) } & \mc{4}{c|}{$ 0.298$ \hfill (EXO)                   } & \mc{4}{c|}{$ 0.256       $ \hfill (EXO)            }    \\
        c2                                     & \mc{4}{c|}{$  0.305                  $ \hfill (EXO) } & \mc{4}{c|}{$  0.282 $ \hfill (EXO)                 } & \mc{4}{c|}{$  0.266       $ \hfill (EXO)           }    \\
        Teff \hfill [K]                        & \mc{4}{c|}{$ 6329\pm65               $ \hfill (E20) } & \mc{4}{c|}{$ 4143^{+92}_{-94}  $ \hfill (T15)      } & \mc{4}{c|}{$ 5620\pm85   $ \hfill (L17)            }    \\
        log(g) \hfill [cgs]                    & \mc{4}{c|}{$ 4.196\pm0.106           $ \hfill (E20) } & \mc{4}{c|}{$ 4.663^{+0.015}_{-0.016}$ \hfill (T15) } & \mc{4}{c|}{$ 4.18\pm0.01        $ \hfill (L17)     }    \\
        {[Fe/H]}                               & \mc{4}{c|}{$ 0.366\pm0.053           $ \hfill (E20) } & \mc{4}{c|}{$ -0.13^{+0.15}_{-0.17}$ \hfill (T15)   } & \mc{4}{c|}{$ -0.18\pm0.06       $ \hfill (L17)     }    \\
        \hline

       % We insert a "ghost" line here to specify the size of the individual columns to ensure that all multi-columns above will fit properly (otherwise we get )
       \mc{1}{p{8mm}}{} &\mc{1}{p{8mm}}{} &\mc{1}{p{8mm}}{} &\mc{1}{p{8mm}}{} &\mc{1}{p{8mm}}{} &\mc{1}{p{8mm}}{} &\mc{1}{p{8mm}}{} &\mc{1}{p{8mm}}{} &\mc{1}{p{8mm}}{} &\mc{1}{p{8mm}}{} &\mc{1}{p{8mm}}{} &\mc{1}{p{8mm}}{} &\mc{1}{p{8mm}}{}

      \end{tabular}

      \begin{tablenotes}[flushleft]
         \footnotesize{\item{\textbf{Notes.} "M$_\mathrm{*}$" is the stellar mass, "R$_\mathrm{*}$" the stellar radius, "Ks" the semi-amplitude of the motion-reflex radial-velocity induced by the planet, "$\mathrm{v_{sys}}$" the systemic radial velocity, and "$\mathrm{v.sin(i)}$" the stellar rotational velocity. "c1, c2" are the quadratic limb darkening coefficients, "Teff" is the effective temperature, "log(g)" the decimal logarithm of surface gravity (directly computed from M$_\mathrm{*}$ and R$_\mathrm{*}$), and {[Fe/H]} the metallicity with respect to solar value.}} % URL for EXOFAST: (\url{https://astroutils.astronomy.osu.edu/exofast/limbdark.shtml}. We used channel SDSS z' as it is the closest to the He triplet lines wavelength)}
      \end{tablenotes}

      \tablebib{A14: \citet{Anderson2014}; A18: \citet{Allart2018}; A19: \citet{Addison2019}; A23: \citet{Allart_2023}; B10: \citet{Bakos2010}; B15: \citet{Boyajian2015}; B16: \citet{Brewer2016}; B17: \citet{Bonomo2017}; B18a: \citet{Bourrier2018a}; B18c: \citet{Bourrier2018c}; B22: \citet{Bourrier2022}; C21: \citet{Cloutier2021}; D11: \citet{Deming2011}; D13: \citet{Demory2013}; E20: \citet{Ehrenreich2020}; (EXO): EXOFAST\footnote{\url{https://astroutils.astronomy.osu.edu/exofast/limbdark.shtml}}, code from \citet{Eastman2013}, we used channel SDSS z' as it is the closest to the He triplet lines wavelength; GDR3: Gaia DR3 \citep{GaiaDR3}; G14: \citet{Gaidos2014}; K21: \citet{Klein2021}; L17: \citet{Lam2017}; M18: \citet{Mallonn2018}, M22: \citet{Maxted2022}; N22: \citet{Nardiello2022}; P9: \citet{Plavchan2009}; Pl20: \citet{Plavchan2020}; Pa20: \citet{Palle2020}; S17: \citet{Stassun2017}; S19: \citet{Schweitzer2019}; St20: \citet{Stefansson2020}; T8: \citet{Torres2008}; T15: \citet{Triaud2015}; T18: \citet{Trifonov2018}; T20: \citet{Thao2020}; T21: \citet{Trifonov2021}; W19: \citet{White2019}; Y18: \citet{Yee2018}; Z22: \citet{Zicher2022}}

    \end{threeparttable}
    \end{table*}

    \begin{table*}[]
      \begin{threeparttable}
      \caption[]{\label{table_planetary_ppties} Planet properties used in this study}

      \centering
      \begin{tabular}{|l|c|c|c|c|c|c|c|}
        \hline
        Target                             & 55\,Cnc\,e                                           & AU\,Mic\,b                                           & GJ\,436\,b                                           \\
        \hline
        Mp \hfill [M$_\mathrm{Jup}$]       &$ 0.0251^{+0.0010}_ {-0.0010} $ \hfill (B18a)       &$ 0.037 \pm 0.015 $ \hfill (Z22)           &$ 0.068\pm0.002  $ \hfill (M22)                     \\
        Rp \hfill [R$_\mathrm{Jup}$]       &$ 0.1673 \pm {0.0026} $ \hfill (B18a)               &$ 0.371 \pm 0.016 $ \hfill (M21)                    &$ 0.343 \pm {0.008} $ \hfill (M22)                 \\
        P$_\mathrm{orb}$ \hfill [days]     &$ 0.73654625 \pm $(\emin{1.5}{-7}) \hfill (K22)     &$ 8.463 \pm $(\emin{1}{-3})  \hfill (M21)            &$ 2.643897621 \pm $(\emin{9.6}{-8}) \hfill (K22)   \\
        a \hfill [AU]                      &$ 0.01544 \pm 0.00005 $ \hfill (B18a)               &$ 0.0645 \pm 0.0013 $ \hfill (M21)                  &$ 0.028 \pm 0.001 $ \hfill (T18)                   \\
        i \hfill [°]                       &$ 83.59^{+0.47}_{-0.44} $ \hfill (B18a)             &$ 89.5 \pm 0.3 $ \hfill (M21)                       &$ 86.84 \pm 0.11 $ \hfill (M22)                    \\
        e                                  &$ 0.05 \pm 0.03 $ \hfill (B18a)                     &$ 0.041^{+0.047}_{-0.026} $ \hfill (Z22)            &$ 0.152^{+0.009}_ {-0.008} $ \hfill (T18)          \\
        w \hfill [°]                       &$ 86.0^{+30.7}_{-33.4} $ \hfill (B18a)              &$ 179^{+128}_{-125} $ \hfill (Z22)                  &$ 325.8^{+5.4}_{-5.7} $ \hfill (T18)               \\
        $\lambda$ \hfill [°]               &$ 72.4^{+12.7}_{-11.5} $ \hfill (BH14)              &$ -4.7^{+6.8}_{-6.4} $ \hfill (H20)                 &$ 103.2^{+12.8}_{-11.5} $ \hfill (B22)             \\
        \hline  \hline
        Target                              & GJ\,486\,b                                           & GJ\,1214\,b                                         & GJ\,3470\,b                                          \\
        \hline
        Mp \hfill [M$_\mathrm{Jup}$]       &$ (8.87^{+0.35}_{-0.38}).10^{-3} $ \hfill (T21)     &$ 0.0257 \pm 0.0014 $ \hfill (C21)                  &$ 0.03958^{+0.00412}_{-0.00403} $ \hfill (K19)      \\
        Rp \hfill [R$_\mathrm{Jup}$]       &$ (0.1164^{+0.0056}_{-0.0060}).10^{-1} $ \hfill (T21)  &$ 0.2446^{+0.0045}_{-0.0047} $ \hfill (C21)         &$ 0.346 \pm 0.029 $ \hfill (K19)                    \\
        P$_\mathrm{orb}$ \hfill [days]     &$ 1.467119  \pm $(\emin{3.1}{-5})   \hfill (T21)    &$ 1.580404571 \pm $(\emin{4.2}{-8})   \hfill (K22)  &$ 3.3366524 \pm $(\emin{1.4}{-7})  \hfill (K22)     \\
        a \hfill [AU]                      &$ 0.01734^{+0.00026}_{-0.00027} $ \hfill (T21)      &$ 0.01490 \pm 0.00026 $ \hfill (C21)                &$ 0.03557^{+0.00096}_{-0.00100} $ \hfill (D13)      \\
        i \hfill [°]                       &$ 88.4^{+1.1}_{-1.4} $ \hfill (T21)                 &$ 88.7 \pm 0.1 $ \hfill (C21)                       &$ 88.3^{+0.5}_{-0.4} $ \hfill (D13)                 \\
        e                                  &$ 0 $ \hfill (fixed)                                &$ <0.063 $ \hfill (C21)                             &$ 0.114^{+0.052}_{-0.051} $ \hfill (K19)            \\
        w \hfill [°]                       &$ 0 $ \hfill (fixed)                                &$ 0 $ \hfill (fixed)                                &$ -82.5^{+5.7}_{-2.3} $ \hfill (K19)                \\
        $\lambda$ \hfill [°]               &$ 0 $ \hfill (fixed)                                &$ 0 $ \hfill (fixed)                                &$ 0  $ \hfill (fixed)                               \\
        \hline  \hline
        Target                             & HAT-P-11\,b                                         & HD\,189733\,b                                        & HD\,209458\,b                                        \\
        \hline
        Mp \hfill [M$_\mathrm{Jup}$]       &$ 0.08728 \pm 0.00979 $ \hfill (A18)                &$ 1.166^{+0.052}_{-0.049} $ \hfill (A19)            &$ 0.685^{+0.015}_{-0.014} $ \hfill (T8)             \\
        Rp \hfill [R$_\mathrm{Jup}$]       &$ 0.389 \pm 0.006 $ \hfill (A18)                    &$ 1.119 \pm 0.038 $ \hfill (A19)                    &$ 1.359^{+0.016}_{-0.019} $ \hfill (T8)             \\
        P$_\mathrm{orb}$ \hfill [days]     &$ 4.88780201 \pm $(\emin{1.7}{-7}) \hfill (K22)     &$ 2.218574944 \pm $(\emin{3}{-8}) (K22)             &$ 3.52474955 \pm $(\emin{3.2}{-7}) \hfill (K22)    \\
        a \hfill [AU]                      &$ 0.05254^{+0.00064}_{-0.00066} $ \hfill (Y18)      &$ 0.03106^{+0.00051}_{-0.00049} $ \hfill (A19)      &$ 0.04707^{+0.00046}_{-0.00047} $ \hfill (T8)       \\
        i \hfill [°]                       &$ 89.00 $ \hfill (H17)                              &$ 85.690^{+0.095}_{-0.097} $ \hfill (A19)           &$ 86.71 \pm 0.05 $ \hfill (T8)                      \\
        e                                  &$ 0.26528 $ \hfill (H17)                            &$ <0.0039 $ \hfill (B17)                            &$ <0.0081 $ \hfill (B17)                            \\
        w \hfill [°]                       &$ -162.157 $ \hfill (H17)                           &$ 90 $ \hfill (C16)                                 &$ 0 \pm 63 $ \hfill (R21)                           \\
        $\lambda$ \hfill [°]               &$ 100 $ \hfill (Y18)                                &$ -0.4 \pm 0.2 $ \hfill (C16)                       &$ 0.1 \pm 2.4 $ \hfill (F9)                         \\
        \hline \hline
        Target                             & K2-25\,b                                            & TOI-1807\,b                                         & WASP-69\,b                                          \\
        \hline
        Mp \hfill [M$_\mathrm{Jup}$]       &$ 0.0771^{+0.0179}_{-0.0164} $ \hfill (St20)        &$ 0.00809\pm0.00157    $ \hfill (N22)               &$ 0.260\pm0.017         $ \hfill (A14)              \\
        Rp \hfill [R$_\mathrm{Jup}$]       &$ 0.306 \pm 0.011 $ \hfill (St20)                   &$ 0.122\pm0.008   $ \hfill (N22)                    &$ 1.057\pm0.047         $ \hfill (A14)              \\
        P$_\mathrm{orb}$ \hfill [days]     &$ 3.48456248 \pm $(\emin{6.4}{-7})   \hfill (K22)   &$ 0.549372\pm$(\emin{7}{-6}  )\hfill (H21)          &$ 3.86813888\pm$(\emin{9.1}{-7}) (K22)              \\
        a \hfill [AU]                      &$ 0.0287 \pm 0.0012   $ \hfill (St20)               &$ 0.0120\pm0.0003    $ \hfill (N22)                 &$ 0.04525\pm0.00053 $ \hfill (A14)                  \\
        i \hfill [°]                       &$ 88.3^{+1.2}_{-0.7} $ \hfill (M16)                 &$ 82.0\pm2.0    $ \hfill (N22)                      &$ 86.71\pm0.20 $ \hfill (A14)                       \\
        e                                  &$ 0.27^{+0.16}_{-0.21} $ \hfill (M16)               &$ 0      $ \hfill (fixed)                           &$ 0      $ \hfill (fixed)                           \\
        w \hfill [°]                       &$ 62^{+44}_{-39} $ \hfill (M16)                     &$ 0   $ \hfill (fixed)                              &$ 90        $ \hfill (fixed)                        \\
        $\lambda$ \hfill [°]               &$ 3 \pm 16  $ \hfill (St20)                         &$ 0   $ \hfill (fixed)                              &$ 0.4^{+2.0}_{-1.9} $ \hfill (C17)                  \\
        \hline \hline
        Target                             & WASP-76\,b                                                       & WASP-80\,b                                          & WASP-127\,b                                         \\
        \hline
        Mp \hfill [M$_\mathrm{Jup}$]       &$ 0.894^{+0.014}_{-0.013}         $   \hfill (E20)     &$ 0.538^{+0.035}_{-0.036} $ \hfill (T15)            &$0.1647^{+0.0214}_{-0.0172} $ \hfill (S20)          \\
        Rp \hfill [R$_\mathrm{Jup}$]       &$ 1.854^{+0.077}_{-0.076}         $   \hfill (E20)     &$ 0.9990^{+0.0300}_{-0.0310} $ \hfill (T15)         &$1.311^{+0.025}_{-0.029}$ \hfill (S20)              \\
        P$_\mathrm{orb}$ \hfill [days]     &$ 1.8098806\pm$(\emin{7}{-7})         \hfill (K22)     &$ 3.06785251\pm $(\emin{1.8}{-7})  \hfill (K22)     &$4.17806513\pm$(\emin{5.7}{-7})  \hfill (S20)       \\
        a \hfill [AU]                      &$ 0.0330\pm0.0002                 $   \hfill (E20)     &$ 0.0344^{+0.0010}_{-0.0011} $ \hfill (T15)         &$0.04840^{+0.00136}_{-0.00095}$ \hfill (S20)        \\
        i \hfill [°]                       &$ 89.623^{+0.005}_{-0.034}        $   \hfill (E20)     &$ 89.02^{+0.11}_{-0.10} $ \hfill (T15)              &$87.84^{+0.36}_{-0.33}$ \hfill (S20)                \\
        e                                  &$ 0     $ \hfill (fixed)                                         &$ 0.0020^{+0.0100}_{-0.0020} $ \hfill (T15)         &$0 $   \hfill (fixed)                               \\
        w \hfill [°]                       &$ 0            $  \hfill (fixed)                                 &$ 94^{+120}_{-21}$ \hfill (T15)                     &$0 $ \hfill (fixed)                                 \\
        $\lambda$ \hfill [°]               &$ 61.28^{+7.61}_{-5.06}           $ \hfill (E20)       &$ -14^{+15}_{-14} $ \hfill (T15)                    &$0 $ \hfill (fixed)                                 \\
        \hline
      \end{tabular}

      \begin{tablenotes}[flushleft]
        \footnotesize{\item{\textbf{Notes.} "Mp" is the planetary mass, "Rp" the planetary radius, "P$_\mathrm{orb}$" the orbital period, "a" the semi-major axis, "i" the orbital inclination, "e" the eccentricity, "w" the longitude of periapse of the planet orbit, and "$\lambda$" the spin-orbit angle.}}
      \end{tablenotes}

      \tablebib{A14: \citet{Anderson2014}; A18: \citet{Allart2018}; A19: \citet{Addison2019}; B12: \citet{VonBraun2012}; B17: \citet{Bonomo2017}; B18a: \citet{Bourrier2018a}; B18c: \citet{Bourrier2018c}; B22: \citet{Bourrier2022}; BH14: \citet{Bourrier&Hebrard2014}; C16: \citet{Cegla2016}; C17: \citet{Casasayas2017}; C21: \citet{Cloutier2021}; D13: \citet{Demory2013}; E20: \citet{Ehrenreich2020}; F9: \citet{Fabrycky_2009}; H17: \citet{Huber2017}; H20: \citet{Hirano2020}; H21: \citet{Hedges2021}; K19: \citet{Kosiarek2019}; K22: \citet{Kokori2022_a}; M16: \citet{Mann2016}; M21: \citet{Martioli2021}; M22: \citet{Maxted2022}; N22: \citet{Nardiello2022}; R21: \citet{Rosenthal2021}; S20: \citet{Seidel2020}; St20: \citep{Stefansson2020}; Sz21: \citet{Szabo2021}; T8: \citet{Torres2008}; T15: \citet{Triaud2015}; T18: \citet{Trifonov2018}; T21: \citet{Trifonov2021}; Y18: \citet{Yee2018}; Z22: \citet{Zicher2022} }

    \end{threeparttable}
    \end{table*}

    We stress the importance of using midpoints defined in the Barycentric Julian Date (BJD) reference frame and using the Barycentric Dynamical Time (TDB) as the time standard instead of the UTC standard. The BJD reference frame corrects for the delay induced by the Earth's movement in its orbit and the finite speed of light. Because the UTC standard is based on atomic clock measurements from the Earth, UTC's rate changes over time when observed from the BJD reference frame due to relativity. The latter effect is corrected by the TDB time standard, making the BJD$_\mathrm{TDB}$ (\change{i.e.,} Barycentric Julian Date in the Barycentric Dynamical Time standard) time definition suitable for observing interstellar events \citep{Eastman2010}. BJD$_\mathrm{UTC}$ events should never be compared with BJD$_\mathrm{TDB}$ events, as they may differ by a few minutes. This difference may for example introduce fake Doppler shifts of up to a few km/s in transiting exoplanetary signatures.

    The observation conditions of each transit are summarized in terms of \change{S/N} and airmass in Fig~\ref{observation_condition}, with the \change{S/N} being directly measured at $\mathrm{1.65\, \mu m}$ by the instrument from the amount of collected photons. Due to scheduling and visibility windows constraints, some transit observations are incomplete while for some targets only one transit was available at the time of this study.

    We note that, among the 32 transits constituting our sample, 17 transits (identified with a dagger symbol in Table~\ref{transit_ppties_tab}) are in common with the study of \citet{Allart_2023}. Although these transits have already been analyzed in the frame of a global He escape study by these authors, we decided to include them in our analysis for several reasons. First, we did not use the same version of the data: \citet{Allart_2023} used data processed with the version 0.7.179 of \mbox{APERO} (A PipelinE to Reduce Observations, \citealp{Cook2022}), the \mbox{SPIRou} data reduction software (DRS). In this version, correction for the OH emission lines from the Earth's atmosphere was not optimized for strong OH lines like the ones around the helium triplet, and \citet{Allart_2023} developed a manual procedure to avoid contamination of the planetary metastable He signature by these lines. We used data from version 0.7.288 of \mbox{APERO}, which handled the correction of the OH emission line in the majority of our targets (see Section~\ref{OH_correction}). Second, we complemented these data with additional transits from the ATMOSPHERIX large program in the case of GJ\,1214\,b, GJ\,3470\,b, and HAT-P-11\,b. This allowed us to compare individual transits and look for potential temporal variations of the signal for a given target, while also strengthening the constraints on the derived parameters by increasing the number of transits stacked together for a given target. Third, our main model's assumptions differ from \citet{Allart_2023} in terms of stellar flux, H/He ratio, and vertical extension of the hydrodynamical outflow, thus providing a different estimation of the upper atmosphere's properties for the transits in common. We justify our choices regarding these assumptions in Section~\ref{Modeling}. For comparison purposes, we also explored the influence of these parameters and common assumptions made about them in the literature, and discuss the results in Section~\ref{discussion}.

  %______________________________________________________________
  \section{Data cleaning}
    \label{reduction}
      \subsection{Data reduction with APERO} % \mbox prevent the "APERO" name being cut in line break
      As the use of \mbox{APERO} in the case of atmospheric characterization with \mbox{SPIRou} has already been described in \citet{Boucher2023} and \citet{Allart_2023}, we briefly summarize the main points here. APERO provides data calibrated in wavelength and corrected from most detector effects such as bad pixels, instrumental background, and nonlinearities \citep{Artigau2018}. The software also handles geometrical corrections in the image plane due to the location and extraction of the orders, and the correction of the flat, hot pixels, and cosmic rays. The wavelength calibration is done using a Fabry-Pérot étalon \citep{Cersullo2017} and a hollow-cathode UNe lamp \citep{Hobson2021}.

         We used \mbox{APERO}'s output E2DS data files\footnote{Once publicly available, these files will be distributed on \url{https://www.cadc-ccda.hia-iha.nrc-cnrc.gc.ca/} }, which refer to 2D optimally extracted spectra. Each E2DS file corresponds to a given observed transit phase and contains the 49 extracted orders, each one being sampled over 4088 wavelength channels. The E2DS are produced for the two science fibers A and B (each corresponding to one of each orthogonal state of the selected polarization) and for the combined flux AB. We only used the latter in the present analysis. \mbox{APERO} also provides telluric-corrected versions of the data \citep{Cook2022}, which first consists of fitting a telluric transmission model from the TAPAS code (Transmissions of the AtmosPhere for AStromomical data; \citealp{Bertaux2014}) to the science targets and remove it from the observed spectra, leaving percent-level residuals. The same process is applied to a large set of rapidly rotating hot stars in order to derive a library of the residuals, from which a model is computed with three degrees of freedom for each pixel (optical depth for water, optical depth for the other atmospheric components, and an additive constant) and is used to correct each observation. The resulting \mbox{APERO} outputs are two versions of each observation: the so-called "t" files, corrected from the telluric lines using the above procedure, and the "e" files corresponding to the uncorrected spectra. One can also access the final telluric model applied by \mbox{APERO} to the data as one of the components of the telluric-corrected E2DS spectra.

        \subsection{OH correction}
      \label{OH_correction}
      The emission of two OH doublets from the Earth's atmosphere, centered at 1083.2 and 1083.4\,nm \citep{Oliva2015}, can overlap with the He signature depending on the barycentric Earth radial velocity (BERV), the systemic velocity, and the Doppler shift of the planetary signature itself. This emission is corrected by \mbox{APERO} as of version 0.7.288. However, we note the presence of OH emission residuals left by \mbox{APERO}'s correction in the case of HAT-P-11\,b and GJ\,436\,b transits.

      We hence directly worked with the telluric corrected "t" files for all transits, except for the four transits of HAT-P-11\,b and the single transit of GJ\,436\,b, for which we used the "e" files (\change{i.e.,} not corrected from tellurics) and manually corrected for the OH emission lines. For both "t" and "e" files, data are first corrected from the instrumental blaze function and normalized by their median value along the spectral axis. After this, a first sigma-clipping is applied to each pixel along the time axis to remove outliers and bad pixels. This is done by removing any value farther than 4\,$\sigma$ from a second-order polynomial fit of each pixel intensity evolution over time, iterating the procedure on the remaining values until no more values are rejected.

      \begin{table}
         \caption[]{Positions of the individual OH emission lines, as described in \citet{Oliva2015}.}
            \label{OH_pos}
              \centering
              \begin{tabular}{lcr}
               \hline
               % \noalign{\smallskip}
               Line identification &  Designation        & Position [nm] \\
               % \noalign{\smallskip}
               \hline
               % \noalign{\smallskip}
               (5-2)Q$_{1}$(0.5)e  &  Q$_{1}$e & 1083.4338   \\
               (5-2)Q$_{1}$(1.5)f  &  Q$_{1}$f & 1083.4241   \\
               \hline
               (5-2)Q$_{2}$(0.5)e  &  Q$_{2}$e & 1083.2412   \\
               (5-2)Q$_{2}$(1.5)f  &  Q$_{2}$f & 1083.2103   \\
               % \noalign{\smallskip}
               \hline
             \end{tabular}
    \end{table}

    The following procedure was only applied to the "e" files, corresponding to the transits of HAT-P-11\,b and GJ\,436\,b. The OH emissions come from two so-called $\Lambda$-doublets, resulting from spin-orbit coupling and belonging to the Q-branch. In order to correct the telluric OH emission, we applied a similar method to \citet{Oliva2015}, \citet{Czesla2022}, and \citet{Allart_2023}. Following the same nomenclature as in \citet{Czesla2022}, each doublet is designated using the corresponding spin quantum number as a Q-subscript, the individual lines of the doublet being identified using the letters e and f to indicate the corresponding lower state parity. The positions of the individual lines as given by \citet{Oliva2015} are listed in Table~\ref{OH_pos}. We modeled the OH emission with four Gaussians, one per individual OH line, with FWHM fixed at the instrumental resolution and positions fixed at their theoretical values from Table~\ref{OH_pos}. As the e and f lines of the same doublet have equal amplitudes, the two strongest OH lines from the $Q_{1}$ doublet (near 1083.4nm) were simultaneously fitted with the two weakest OH lines from the $Q_{2}$ doublet (near 1083.2nm), fixing the amplitude ratio between the amplitudes of $Q_{2}$ and $Q_{1}$ lines to the value of 0.0482 computed by \citet{Czesla2022} and thus only letting the amplitude of the $Q_{1}e$ line as a free parameter.

    \begin{figure}
      \centering
      \includegraphics[width=\hsize]{"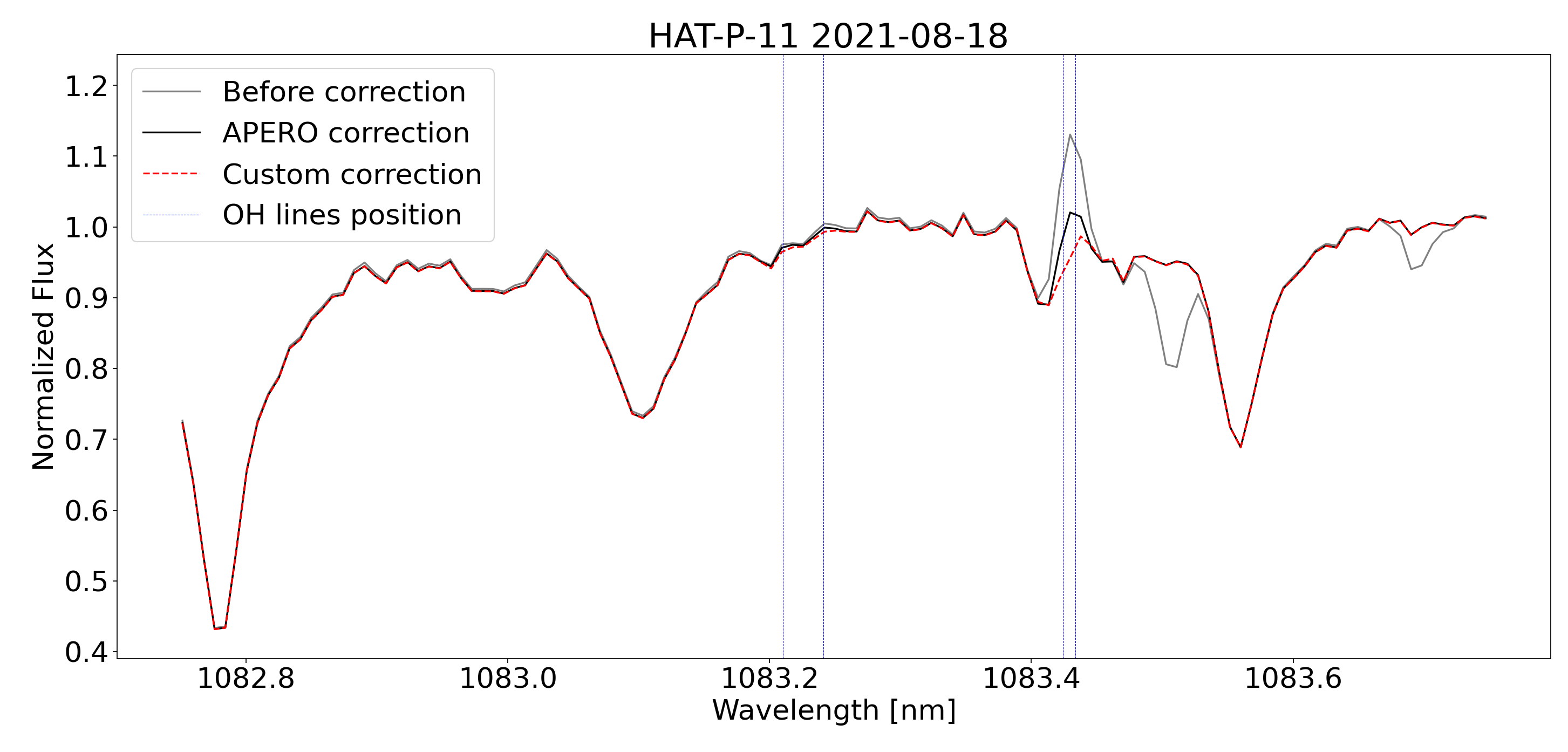"}
       \caption{Data from the normalized E2DS "e" files before (solid gray) and after ("t" files, solid black) correction of the telluric lines by \mbox{APERO}. Our custom correction of the OH emission lines is shown in dotted red, with the position of the OH lines indicated with vertical blue lines. We plot here an arbitrary observation in the Earth rest frame as an example. Absorption lines from stellar Si and He can respectively be seen near 1083.1 and 1082.8\,nm, while a telluric absorption line is visible near 1083.5\,nm in the spectrum before correction.}
       \label{OH_correction_example}
    \end{figure}

    Depending on the BERV and systemic velocity of the observed target, the helium triplet signature from the host star can affect the shape of the $Q_{1}$ doublet. To reproduce this stellar He absorption, we thus added a component to our model that contains a constant continuum (left as a free parameter) and an absorption feature also modeled with a Gaussian, with a position fixed at its expected value in the Earth rest frame and with free amplitude and FWHM for each observation. The resulting model is thus the sum of a constant continuum, a Gaussian absorption for the stellar He, and four Gaussian emissions for the OH lines. This model is fitted to each observation of the target, and the four OH Gaussian emission components are removed from the data. After this first correction from telluric OH, we apply the telluric correction model (the so-called "recon file") computed by \mbox{APERO} (except for its OH emission component that we have manually corrected) to remove the other remaining telluric lines. An illustration of this correction is shown in Fig.~\ref{OH_correction_example}, and the before/after telluric correction for each transit can be found in Appendix~\ref{telluric_correction_grid}.

    \subsection{Division by a master spectrum}
    \label{master_spectrum_reduction}

    For each transit, we first computed a master spectrum by averaging the out-of-transit observations together. The spectra were kept in the Earth rest frame instead of being shifted in the stellar rest frame, as it would require interpolating the observed spectra and, thus, adding an interpolation error. Ideally, this master spectrum should represent the mean value of the stellar lines and telluric residuals. Dividing the data by this "master-out" spectrum should thus correct for the mean component of the stellar lines and telluric residuals, removing the majority of their contribution while not affecting the planetary signature which is not present in the out-of-transit spectra. However, this method requires enough out-of-transit observations both before and after the transit event to ensure that the master-out spectrum correctly reflects the mean value of the stellar lines, which was not the case for about one-third of the studied transits (see Fig.~\ref{observation_condition}).

          To better correct the stellar lines, we thus included in-transit observations to the master spectrum computation. However, including the in-transit observations at the planetary He wavelengths would make its He signal contribute to the master spectrum, and thus affect the recovered planetary signal. We thus computed the master spectrum by averaging together both off- and in-transit spectra for wavelengths farther than 0.1\,nm from the theoretical position of the planetary metastable He triplet lines, while only averaging out-of-transit observations for wavelengths in a spectral bin of $\pm0.1$\,nm around the theoretical position of the planetary He triplet lines. This 0.2\,nm bin total size was chosen to be larger than the expected span of the triplet lines.

    \subsection{Impact of center to limb variations and the Rossiter-McLaughlin effect}

    A common hypothesis in helium triplet studies is to neglect transit-induced deformations of the stellar lines, such as CLVs and the RME (see \citealp{Chiavassa2019}), as their impact has been estimated to be negligible in some of the targets studied here (e.g \citealp{Alonso2019} for HD\,209458\,b, \citealp{Salz2018} and \citealp{Guilluy2020} for HD\,189733\,b, \citealp{Nortmann2018} for WASP-69\,b, \citealp{Allart2018} for HAT-P-11\,b). However, these effects depend on the strength of the stellar He triplet lines, which itself may strongly vary among several observations of the same target due to stellar activity \citep[see][]{Guilluy2020,Guilluy2023}. The impact of these planetary-induced stellar lines' deformation and the need for taking them into account in both molecular and helium studies were recently outlined by \citet{Dethier2023} and \citet{Guilluy2023}. We thus decided to include the correction of CLV and RME in our master spectrum computation procedure following a method similar to that of \citet{Brogi2016}, \citet{Guilluy2023}, and \citet{Dethier2023}.

          We simulated the stellar surface intensity map in a 2D grid of pixels using PYSME\footnote{\url{https://github.com/AWehrhahn/SME/}}, the \texttt{Python} implementation of Spectroscopy Made Easy (SME) from \citet{SME1996}. PYSME computes wavelength-dependent continuum and stellar lines opacities in a self-consistent way depending on the limb-darkening angle. Each pixel thus contains the model stellar spectrum with limb-darkening and the CLV effects on the stellar lines shape directly computed by PYSME, and shifted by the star's rotational velocity (projected on the line of sight), all depending on the longitude and latitude of the pixel on the stellar surface.

    \begin{figure}[h]
      \centering
      \includegraphics[width=\hsize]{"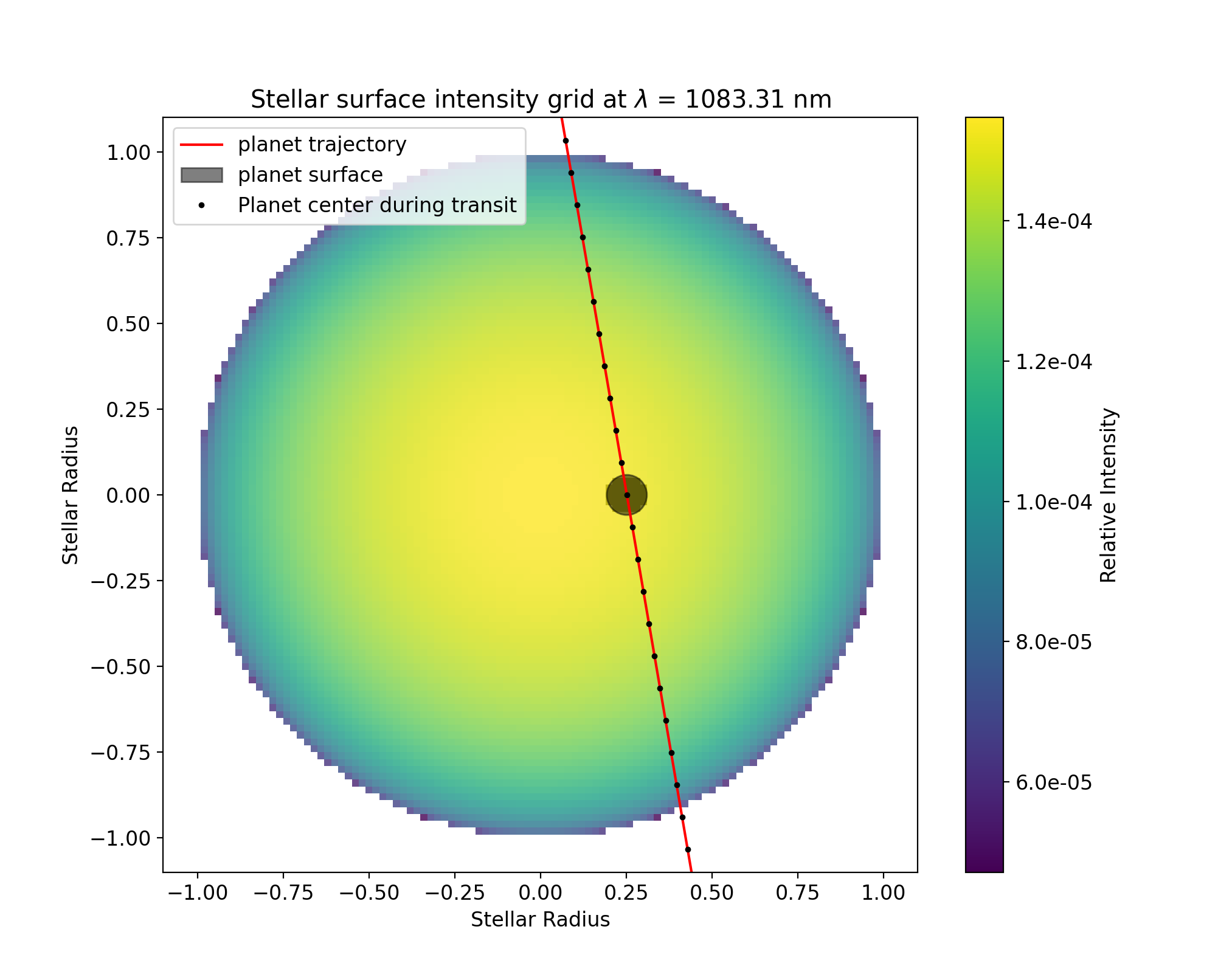"}
       \caption{Stellar surface intensity map at 1083.31nm computed with PYSME in the case of HAT-P-11\,b. The intensity map is relative to the integrated intensity over the stellar map at 1083.31nm. The grid contains 101x101 cells, and the planet's trajectory is shown with a red solid line with black dots indicating the planet's center at each observed phase. The planet solid disk is shown at a given phase, with the underlying stellar cells being excluded from the master spectrum computation.}
       \label{stellar_grid}
    \end{figure}

    We note, however, that PYSME is limited to the photospheric stellar lines. To include the metastable helium triplet lines from the stellar chromosphere in our correction, we followed a procedure similar to the one used in  \citet{Guilluy2023}. We set up a model\footnote{Original code from W. Dethier (priv. communication)} that, taking a given metastable helium column density and chromospheric temperature as inputs, computes the line profiles of the stellar helium triplet based on the resulting optical depth. This helium signature is then incorporated in the local stellar spectra contained in the pixels of the stellar grid. This is done by shifting the helium signature by the star's rotational velocity and then multiplying it by the local stellar spectrum to account for the limb-darkening. We then average the spectra of all pixels in the grid to obtain, for a given He column density and chromospheric temperature, an out-of-transit model of the disk-averaged stellar spectrum that includes the RME and CLV effect for the photospheric stellar lines, and the RME and limb darkening effects for the chromospheric helium. We repeat the whole process using a $\chi^2$ method to fit this modeled disk-averaged stellar spectrum to the mean out-of-transit observed spectrum, letting the stellar He column density and chromospheric temperature as free parameters.

          Using the best-fit values for the stellar He column density and chromospheric temperature, we then compute the in-transit stellar spectrum at a given phase by averaging the local stellar spectra of the grid's pixels not occulted by the planetary solid disk (\change{i.e.,} at one planetary radius from the planet’s center, see Fig.~\ref{stellar_grid}). To estimate the planet-induced deformation of the stellar line profiles at a given phase\footnote{The planet trajectory in front of the stellar disk was computed using the Kepler ellipse model from PyAstronomy (\citet{pya}, repository: \url{https://github.com/sczesla/PyAstronomy}).}, we divide the corresponding in-transit model of the disk-averaged stellar spectrum by the out-of-transit disk-averaged model. In doing so, we obtain a deformation model of the stellar lines' profiles at a given transit phase. For a given transit phase, we then include the RME and the CLV effect in the master spectrum (computed in Section~\ref{master_spectrum_reduction} as the average of the observed out-of-transit spectra) by multiplying it by the corresponding deformation model. We note that we were not able to estimate the influence of more complex phenomena such as convection or center-to-limb variation on the stellar He line shape as it would require 3D modeling of the stellar surface and the inclusion of a realistic chromosphere model, which is beyond the scope of this paper.

    \subsection{Continuum flattening with a moving average}
    \label{average_transits_together}

    After division by the master spectrum, we flattened the continuum by dividing it using a low-pass-filtered version of the data computed with a moving average. This technique is commonly used to remove the modal noise in \mbox{SPIRou} data \citep[\change{e.g.,}][]{Pelletier2021,Boucher2021,Boucher2023,Klein_2023}, and we chose a kernel width of 119 pixels to ensure that this step would not affect the He line shape (whose FWHM is typically of 10 pixels). We then shifted the data in the planetary rest frame by correcting for the BERV, the systemic velocity, and the planetary orbital motion projected along the line of sight ($V_{P}$). The total Doppler shift velocity $V_{tot}$, applied to the data to shift them from the Earth rest-frame to the exoplanetary one, is thus defined as: $V_{tot} = V_{BERV} - V_{sys} - V_{P}$, with $V_{BERV}$ and $V_{sys}$ the BERV and systemic velocity respectively. To ensure the data were properly coadded in the planetary rest frame, we estimated the influence of the uncertainties on the orbital eccentricity and found that the variations in the planet's radial velocity are typically one hundred m/s. This is far below the resolution of a single SPIRou pixel ($\sim$2.3\,km/s), and we are thus confident that our data are correctly coadded in the planetary rest frame even considering the uncertainties on the orbital parameters. The mean spectrum of a given transit was computed by averaging all in-transit observations (\change{i.e.,} between contact point T1, when the planetary solid disk first crosses the edge of the host star, and contact point T4, when the planetary solid disk fully exits the host star's disk), weighting each observation by its \change{S/N}². If several transits were available for a given target, we averaged the spectra together, weighting each spectrum by the inverse square of its estimated error (see Appendix~\ref{error_bar_estimation} for more details), to get the final spectrum for each target. The time series of the individual transit's data after reduction are shown in Fig.~\ref{2D_grid_reduced_data}, and the final averaged spectra for each target\footnote{Will be made available in electronic form at the CDS via anonymous ftp to cdsarc.u-strasbg.fr (130.79.128.5) or via \url{http://cdsweb.u-strasbg.fr/cgi-bin/qcat?J/A+A/}} are displayed in Fig.~\ref{1d_reduced_data}.

  %______________________________________________________________
  \section{Modeling the He I triplet lines}

    \label{Modeling}
    \subsection{Methodology}

    For comparison purposes, we used two different models to reproduce the He triplet: one computed using the p-winds code \citep{DosSantos2022}, while the other one is an adapted \texttt{Python} version of the model used in \citet{Lampon2023}. Both are 1D spherical NLTE radiative transfer models coupled with a hydrodynamic escape model. NLTE processes are modeled from \citet{Oklopcic2018} formulations and the equations from \citet{Lampon2020} describing the physical processes driving the population of metastable He. Modeling the He I triplet lines involves a model for the expansion of the upper atmosphere (the "hydrodynamical module"), and a model computing the associated radiative transfer taking into account the geometry of the transit (the "radiative transfer module").

    We used a least-squares approach to constrain the atmospheric escape of a given target from the fit of its metastable He triplet lines: we fixed the H/He ratio and generated a grid of models of the He triplet lines in a 2D parameter space. The two parameters are the mass-loss rate \mdot (in g.s$^{-1}$) and the temperature T of the upper atmosphere. We then computed the $\chi^2$ value between the data and each model in the (\mdot,T) parameter space, thus obtaining the $\chi^2$ maps presented in Section~\ref{results}, from which we derived the best (\mdot,T) model, the error bars, and the corresponding atmospheric structure.

    \subsection{Hydrodynamical module}
    Given the planet parameters, the stellar XUV flux, and the H/He ratio as inputs, the hydrodynamical modules of both models use the isothermal Parker wind approximation \citep{Parker1958} along with the hypothesis of a constant speed of sound to solve for the abundance profiles of hydrogen (neutral and ionized) and helium (neutral, metastable and ionized), and for the velocity profile of the escaping atmosphere. Each hydrodynamical simulation corresponds to a given mass-loss rate \mdot and temperature T. As explained in \cite{Lampon2020}, this temperature approximates the maximum of the temperature profile obtained by \cite{Salz2016} when solving for the energy equation in a computationally more expensive way (\change{i.e.,} without the constant speed of sound approximation), and corresponds to the location of the atmospheric photo-ionization region where most of the stellar radiation is absorbed.

          \begin{figure*}[]
            \centering
            \includegraphics[width=\hsize]{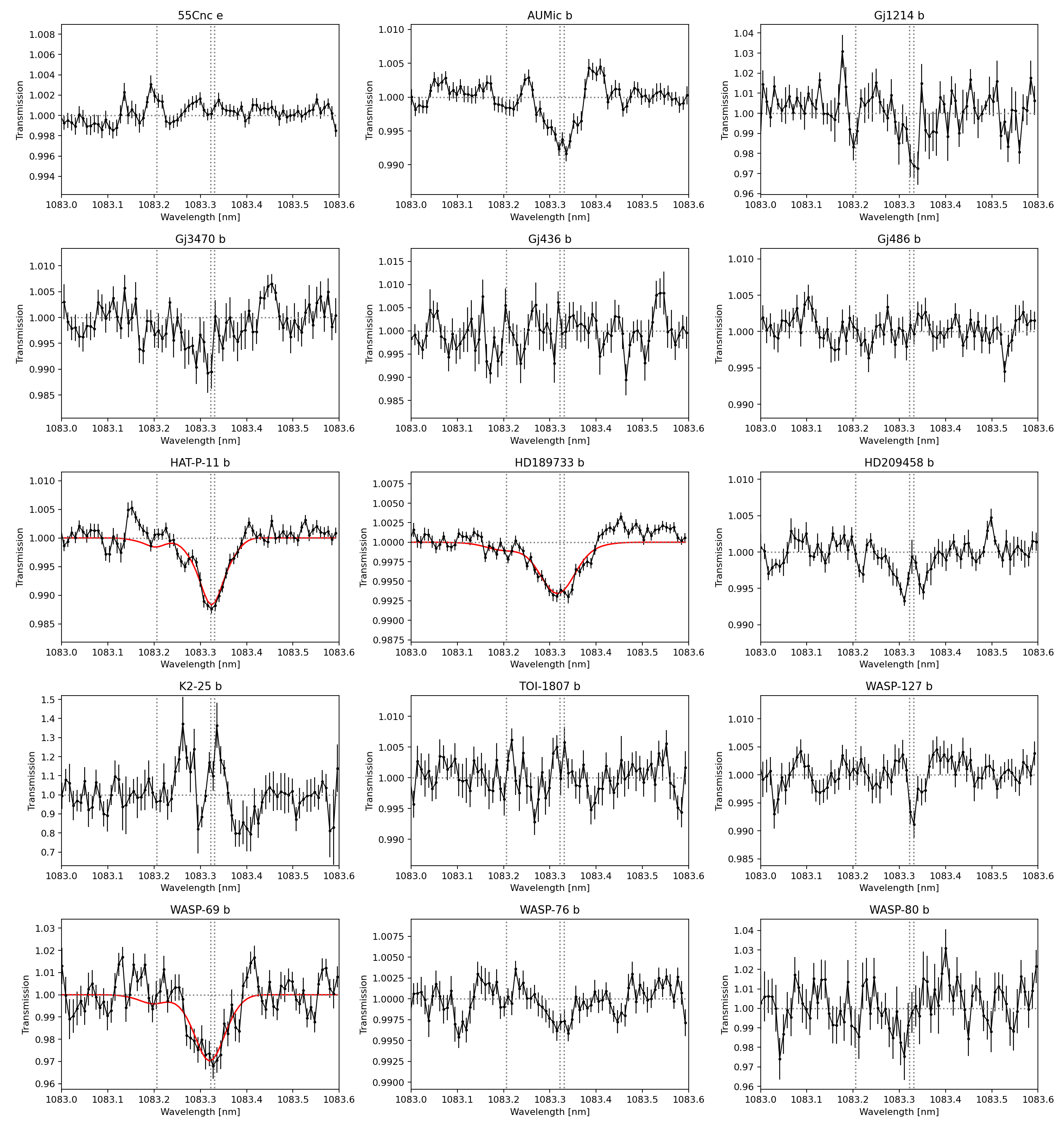}
            \caption{Final averaged spectra for each target. If several transits were available for a given target, the final spectrum was obtained by averaging the spectra together, weighting each spectrum by the inverse square of its estimated error. The vertical dotted gray lines show the theoretical position of the He triplet lines, and the horizontal dotted gray line represents the null model (no absorption). The best fit (corresponding to our main model M1, see Section~\ref{models_list}) is shown in red for the three detections. Wavelengths are given in vacuum. Error bars are given for a 95\,\% confidence interval. The calculation of the error bars is detailed in Appendix~\ref{error_bar_estimation}.}
            \label{1d_reduced_data}
          \end{figure*}

    \subsubsection{Stellar flux}
    The hydrodynamical modules of both models require the stellar flux (up to 2600\,{\AA}) received by the planet, as it drives both the population of the metastable triplet and the structure of the upper atmosphere \citep{Oklopcic2018}. With the absence of tight-enough observational constraints on the stellar EUV flux \citep{France2022}, semi-empirical models are required to estimate the EUV part of the stellar SED. \citet{Allart_2023} and \citet{Guilluy2023} for example directly used the equations from \citet{Linsky2014} and \citet{Sanz2011} to estimate the stellar X-EUV flux (0$-$1170\,{\AA}). 
    
    We estimated the stellar flux using the MUSCLES database\footnote{\url{https://cos.colorado.edu/~kevinf/muscles.html}} \citep{France2016,Loyd2018}, similarly to \citet{Kasper2020} and \citet{Vissapragada2020}, in order to include the stellar \change{spectral energy distribution (SED)} up to 2600\,{\AA}. The stellar spectra from the MUSCLES database indeed cover wavelengths from 5\,{\AA} to 5.5\,$\mu$m, combining measurements from Chandra/XMM-Newton/Swift and APEC models \citep{Smith2001} for the X-rays (<100\,{\AA}) part, semi-empirical models from \citet{Linsky2014} and \citet{Duvvuri2021} for the EUV part, and HST COS and STIS measurements along with a model of the Ly$\alpha$ line for the far-UV (FUV, 912$-$1700\,{\AA}) and NUV (1700$-$3200\,{\AA}) parts \citep{Youngblood2016}. For each of our science targets, we used the stellar SED corresponding to the closest stellar type, log(g) and effective temperature, and rescaling it at the target's distance from its host star. Two exceptions are for HAT-P-11, for which we used a spectrum based on the computations of Sanz-Forcada (priv. comm.), and for HD 189733 for which we used the stellar SED computed by \citet{Bourrier2020}. The MUSCLES stellar references are given for each target, along with other target-specific model parameters, in Table~\ref{table_model_parameter_targets}.

    \subsubsection{H/He ratio}
    In the absence of precise observational constraints on the H/He ratio in the upper atmosphere of close-in exoplanets, a common assumption in previous studies of He escape has been to set this ratio at a solar (H/He = 90/10) value (\change{e.g.,} \citealp{Vissapragada2022,Gaidos2023,Allart_2023,Guilluy2023}). However, the H/He ratio is known to be highly degenerated with the temperature and the mass-loss rate when it comes to characterizing the escape of metastable He \citep{Lampon2021_mar}, and the value to which this ratio is fixed hence has an important influence on the retrieved mass-loss rate and temperature of the outflow (see for example its influence on line shape and intensity in Fig.~\ref{param_influence_on_line_shape}). Additional information on the upper atmosphere is thus required to lift the degeneracy, for example by deriving the neutral hydrogen abundance profile from H$\alpha$ or Ly-$\alpha$ measurements. This has been done in previous studies for HD\,209458\,b \citep{Lampon2020}, HD\,189733\,b and GJ\,3470\,b \citep{Lampon2021_mar}, and for HAT-P-32 b \citep{Lampon2023}. In the latter study, the authors also used the relationship between the heating efficiency and the H/He ratio to constrain its value in GJ\,1214\,b. In this series of papers, and despite the diversity of bulk parameters in the studied targets, the authors observed a trend toward larger H/He values with the majority of published measurements pointing to a ratio higher or equal to 98/2 rather than the widely assumed 90/10 ratio. 
    
    We thus adopted the H/He values constrained by \citet{Lampon2020,Lampon2021_mar,Lampon2023} for the targets for which it had been derived (HD\,209458\,b, HD\,189733\,b, GJ\,3470\,b and GJ\,1214\,b), and assumed a H/He ratio of 98/2 for the other targets. The corresponding ratio for each target, along with other model parameters, are listed in Table~\ref{table_model_parameter_targets}. For comparison purposes with previous studies assuming a solar 90/10 value, we also performed our analysis using the solar ratio and discussed the resulting differences in the retrieved parameters in Section~\ref{discussion}.

    \subsection{Radiative transfer module}
    Using the H and He abundance profiles (simulated from the hydrodynamical module) as inputs, the radiative transfer module computes the He triplet lines profile assuming an altitude grid defined by the user. In both codes, the radiative transfer module takes into account the transit geometry (see Section~\ref{transit_geometry_Section}), a global Doppler shift of the whole signature defined by the user, the instrumental spectral resolution, and the broadening of the He triplet due to the projected outflow's velocity along the line of sight (see Fig. 7 in \citealp{Lampon2020}).

      \subsubsection{Transit geometry}
      \label{transit_geometry_Section}
\begin{figure}
        \includegraphics[width=\hsize]{"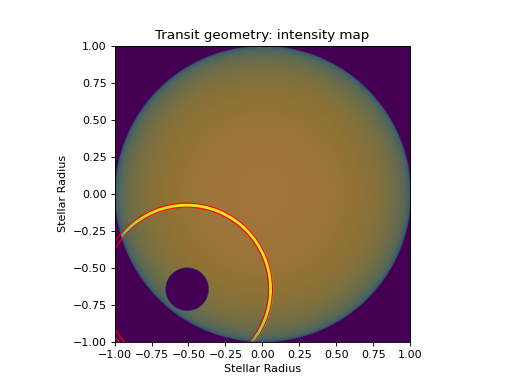"}
        \caption{Example of an intensity grid computed for HD\,189733\,b and used in the transit geometry module. The stellar intensity is computed over a grid, the number of cells in the grid being set such that the thinnest planetary atmospheric layer covers at least two pixels. The map corresponds to a given transit phase, the cells covered by the planetary solid disk at 1\,Rp being masked out. One of the altitude layers is represented here, limited by the two red rings. The contribution of each altitude layer is computed using the intensity map to take into account the intersection (highlighted pixels) of each altitude layer with the stellar intensity surface.}
        \label{transit_geometry_model}
      \end{figure}

      Taking into account the geometry of the transit in the simulated He triplet absorption is one of the main steps in the radiative transfer module for both codes. The following calculations are performed in the planetary rest frame. The method is similar to the one used for the correction of RME and the CLV effect (see Section~\ref{master_spectrum_reduction}) and consists in computing the transit geometry map at different phases in a grid of pixels mapping the stellar surface intensity. For a given transit phase, the stellar cells occulted by the planetary solid disk (defined by convention at 1\,Rp) are masked-out \citep{DosSantos2022}. The metastable helium absorption is then integrated over a grid of altitude layers, taking into account the intersection of each altitude layer with the stellar intensity map, and the corresponding abundance of metastable helium along the line of sight from the vertical profiles computed by the hydrodynamical module (see Fig.~\ref{transit_geometry_model}). p-winds requires a limb-darkening law to compute the stellar intensity map: we used a quadratic law with the limb-darkening coefficients listed in Table~\ref{table_stellar_ppties}. %This method also takes into account limb darkening and can be used to estimate the contribution of stellar spots by directly adding some on the stellar surface.

      As of its version 1.3.2, in use in this study, p-winds' transit geometry module does not take into account the eccentricity and spin-orbit angle of the planet's orbit. While these hypotheses are valid for the majority of short-period gas giants with circularized and coplanar orbits, one of the targets studied here, HAT-P-11\,b, has an eccentricity of 0.265 and a spin-orbit angle of $\sim$100° (see Tab. \ref{table_planetary_ppties}). For both models, we thus adapted the transit geometry module to account for the eccentricity and spin-orbit angle of each target. Doing so, the difference in the modeled helium absorption over the transit is at the 0.01\,\% level for the most extreme case of HAT-P-11\,b, which is one order of magnitude below the noise level and thus does not significantly impact the retrieved parameters of the escape.

      \subsubsection{Effective transit phase}
      The above method consists in sampling the transit geometry by computing the model for a given number of transit phases and then averaging them to get the final model. When performing our tests, we found that changing the number of sampling phases could affect the depth of the final averaged He lines, with relative variations up to 20\,\% in the line depth\footnote{The comparison was done between an averaged model using ten sampling phases (value by default in p-winds) and another one using 100 sampling phases. The phases were linearly spaced between T1 and T4 in both cases.}. Therefore, we estimate that at least a hundred phases must be sampled and averaged to properly model the He line. Yet, doing so would significantly increase the computation time of our $\chi^2$ maps as for each model in our (\mdot, T) parameter space, a hundred radiative transfer calculations would have to be done. % or in Section{results} to avoid spoiling the figure ?

      To significantly reduce the computation time, we applied the same method as in \citet{Lampon2023}: we first computed an averaged model of the He triplet, for a fixed set of (\mdot, T), and using a hundred phases to sample the transit geometry. We then used a least-squares method to find an effective transit phase that reproduced the 100-phase averaged model. We then explored the full (\mdot, T) parameter space using the effective transit phase instead of the 100 phases to account for the transit geometry. By doing so, we reduced the computation time of a $\chi^2$ map down to a few minutes.

      This method assumes that the effective transit phase, computed for an arbitrary set of (\mdot, T), is the same for any other model in the (\mdot, T) parameter space. We tested the validity of this hypothesis by computing a $\chi^2$ map for which each (\mdot, T) model was the average of a hundred transit phases, and computing the same $\chi^2$ map using the effective transit phase method. We found a relative difference at the 1\,\% level in the best-fit values of the two parameters in both methods, which is within our 1-$\sigma$ level of confidence. We therefore used the effective transit phase approach to quickly explore the full (\mdot, T) parameter space and estimate the region of minimum $\chi^2$ in a first step. In a second step, we refined the position of the best fit by performing another $\chi^2$ minimization, this time computing the transit geometry with 100 averaged phases in a small (\mdot, T) region centered on the position of the minimum $\chi^2$ inferred using the effective transit phase method.

      \subsubsection{Altitude grid}
      We used a log-space altitude grid to ensure sampling of the abundance profiles with a high enough resolution in the lower altitude layers, where the maximum helium absorption occurs. A common hypothesis when using 1D models for helium characterization is to set an upper limit for the atmospheric extension far beyond the Roche lobe of the planet in order to take into account all of the helium absorption in front of the stellar disk \citep{Oklopcic2018, Lampon2020, DosSantos2022, Kirk2022, Lampon2023}.

      As reminded in the recent study of \cite{Allart_2023}, 1D Parker-wind models are not fully valid beyond this limit where the effect of stellar wind requires more complex 3D modeling for a more detailed description of the escape, which lies beyond the scope of this study. To account for this limitation, \citet{Guilluy2023} and \citet{Allart_2023} set the radius at the top of the simulated atmosphere to the Roche lobe. Yet, 3D hydrodynamic models show that spherical symmetry can still be found at distances beyond the Roche lobe \citep{Shaikhislamov2018,Shaikhislamov2020,Wang2021,Rumenskikh2022}. Therefore, although complete 3D models are necessary for a detailed spatial distribution, 1D models still provide reasonably accurate constraints on the main parameters of the upper atmosphere such as the mass-loss rate and the temperature (\change{e.g.,} \citealp{Owen2020,MacLeod2022}). In our targets sample, we estimated that the absorption occurring beyond the Roche lobe contributes up to $\sim$50\,\% to the total signature in the most extreme cases (see Fig.~\ref{weight_density_grid}), which may thus affect the retrieved parameters of the escape in a non-negligible way. Thus, even though our 1D model is not fully valid beyond the Roche lobe, we decided to include the corresponding layers in our main analysis. We also tested the influence of limiting the extension of the model to the Roche lobe on our results, and discuss this case in Section~\ref{discussion}.

  %______________________________________________________________
  \section{Results}

    \label{results}
    Our final spectra, obtained after applying the reduction steps detailed in Section~\ref{reduction}, are presented in Fig.~\ref{1d_reduced_data}. We started by fitting a simple Gaussian model to each of the reduced spectra to estimate the depth, equivalent width (EW), and Doppler shift of the He signature for each detection and derive upper limits for non- and tentative detections. We set two conditions to claim for a detection: i) an \change{S/N} of 5 or above and ii) the absence of features with similar amplitudes in the surrounding continuum. This is the case for HAT-P-11\,b, HD\,189733\,b, and WASP-69\,b, for which an excess of He absorption is detected during transit due to the planetary escaping atmosphere absorptions. Signatures meeting only the first condition (\change{S/N} threshold of 5) are considered as tentative detections: this is the case for HD\,209458\,b, GJ\,3470\,b, and WASP-76\,b, for which the presence of residuals due to uncorrected stellar signal or correlated noise with amplitudes similar to the potential He signals prevents us from claiming detections. We provide uncertainties on the signature's depth and EW corresponding to a 95\,\% confidence interval for the detections. For tentative and nondetections, upper limits correspond to a 3-$\sigma$ confidence interval and were computed assuming that a signature should have an FWHM covering at least 10 SPIRou pixels (width of $\sim$ 0.08\,nm), which is the typical FWHM of the He signature in our detections. Our method for the errors and upper limit estimations is further detailed in Appendix~\ref{error_estimation_depth_EW}. Results corresponding to our main model (with H/He $\geq$ 98/2) and for a solar H/He value are given in Table~\ref{table_results} for comparison and are further discussed in Section~\ref{discussion}.

    The transit light curves were computed by integrating the signal on a bin of size equal to the FWHM of the Gaussian fit of the He triplet in case of detection, which provided the best compromise between transit depth and \change{S/N}. In case of non- and tentative detections, the bin size was set to 0.1\,nm. The bin was centered at the measured He triplet position using the center of the Gaussian fit in case of detections, and at the theoretical position of the triplet otherwise. Transit light curves (TLC) of each transit are shown in Fig.~\ref{TLC_grid}. Excess absorption in the helium TLC is detected during transit in the case of HAT-P-11\,b, HD\,189733\,b, and WASP-69\,b, with no visible asymmetries.
    An excess of absorption is also detected during the transit of HD\,209458\,b, comforting us in the attribution of this signature to a planetary origin. These signatures are also visible in the time series of the reduced data in Fig.~\ref{2D_grid_reduced_data}. For AU\,Mic\,b, we note an excess of absorption occurring during the first half of the transit only. The fact that this excess of absorption is not present during the entire transit duration and the presence of uncorrected stellar residuals in AU\,Mic\,b time series (Fig.~\ref{2D_grid_reduced_data}) suggest that this signature may be of stellar origin. This hypothesis is supported by the young age of this system, whose host star is known to be very active \citep{Martioli2020,Klein2021,Klein2022}.

\begin{figure*}[]
      \centering
      \includegraphics[width=0.85\hsize]{"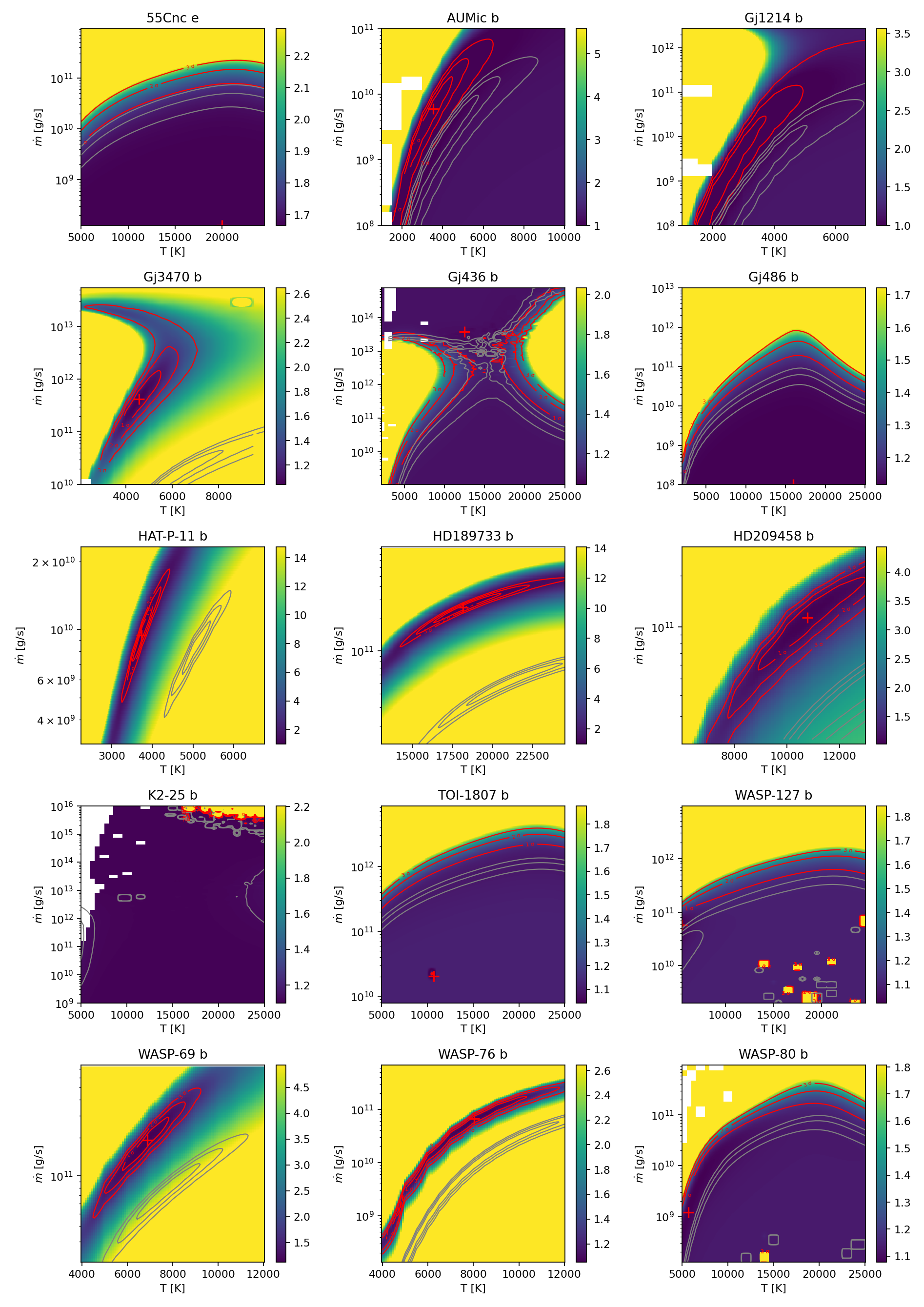"}
      \caption{Reduced \chisquare maps obtained for each target with the model parameters from Table~\ref{table_model_parameter_targets}. The vertical axis shows the mass-loss rate (\mdot) in log-scale, and the horizontal axis shows the temperature in linear scale. Higher \chisquare values are shown in yellow and lower values are in blue. Parts of the grids contain NaN values (white areas) where the model failed to converge to a physical solution. The minimum \chisquare value position is shown with a red cross, encircled by 1, 2, and 3$\sigma$ contours in red and corresponding to our main model with a high H/He ratio ($\geq$ 98/2, corresponding to model M1 in Section~\ref{models_list}). The confidence intervals overplotted in gray show the best fits obtained with the same model but using a solar H/He ratio (90/10, corresponding to model M4 in Section~\ref{models_list}) for comparison. Our estimation of the error that we used for the reduced $\chi^2$ computation is detailed in Appendix~\ref{error_estimation_mlr_T}. We note the apparent correlation between \mdot and T in the case of detections and tentative detections, which is further detailed in Section~\ref{mlr_T_correlation}.}
      \label{chi2_map_grid}
    \end{figure*}

    For each target, we performed a $\chi^2$ minimization using both models described in Section~\ref{Modeling} to explore the (\mdot,T) parameter space. Our resulting reduced $\chi^2$ maps obtained with p-winds are shown in Fig.~\ref{chi2_map_grid}. The main model parameters used for the computation of each map are given in Table~\ref{table_model_parameter_targets}, and we summarize the corresponding best-fit and upper-limits we obtained in Table~\ref{table_results}. The influence of these parameters on our results, in particular the H/He ratio and the extension of the modeled atmosphere, are further discussed in Section~\ref{discussion}.

    We note the presence of residuals in our final reduced spectra, which may be of stellar origin or could be due to the instrument's modal noise \citep{Oliva2019}. These systematics can reach amplitudes higher than the estimated error bars, for which the computation was based on the standard deviation of each individual spectral bin along the time axis (see Appendix~\ref{error_bar_estimation}). To better account for these systematics in our reduced $\chi^2$ calculation, we reestimated the error using the standard deviation along the spectral axis. Our error bars and \change{S/N} calculation method is detailed in Appendix~\ref{error_estimation_mlr_T}. To compute our final $\chi^2$ maps (Fig.~\ref{chi2_map_grid}), we used these reestimated error bars in place of the previous ones shown in Fig.~\ref{1d_reduced_data} (estimated from the standard deviation of the transit observation data set).

    \begin{table*}[]
      \begin{threeparttable}
       \caption[]{Model parameters for the studied targets}
       \label{table_model_parameter_targets}
       \begin{tabular}{llccccr}
        \hline \hline
        Target            & H/He         & Extension [Rp] & R$_{lobe}$ [Rp] & V$_{shift}$ [km/s] & Detection   & Stellar SED              \\
        \hline
        55\,Cnc\,e        & 98/2         & 110            & 2.8             & 0                  & N           & HD 97658                 \\ \smallskip
        AU\,Mic\,b        & 98/2         & 40             & 8.5             & 0                  & N           & GJ 832                   \\ \smallskip
        GJ\,1214\,b       & 98/2*        & 18             & 3.2             & 0                  & N           & GJ 1214                  \\ \smallskip
        GJ\,3470\,b       & 99.8/0.2*    & 26             & 4.1             & \change{$-4\pm2    $}& T           & GJ 832                   \\ \smallskip
        GJ\,436\,b        & 98/2         & 25             & 4.5             & 0                  & N           & GJ 436                   \\ \smallskip
        GJ\,486\,b        & 98/2         & 55             & 4.5             & 0                  & N           & GJ 436                   \\ \smallskip
        HAT-P-11\,b       & 98/2         & 35             & 6.5             & \change{$-0.8\pm1.8$}& Y           & HAT-P-11$^{\dagger}$     \\ \smallskip
        HD\,189733\,b     & 99/1*        & 14             & 3.0             & \change{$-2.5\pm1.4$}& Y           & HD 189733$^{\ddagger}$   \\ \smallskip
        HD\,209458\,b     & 98/2*        & 18             & 3.0             & \change{$-2\pm2    $}& T           & HD 97658                 \\ \smallskip
        K2-25\,b          & 98/2         & 19             & 6.3             & 0                  & N           & GJ 1214                  \\ \smallskip
        TOI-1807\,b       & 98/2         & 81             & 1.1             & 0                  & N           & HD 85512                 \\ \smallskip
        WASP-127\,b       & 98/2         & 20             & 2.1             & 0                  & N           & HD 97658                 \\ \smallskip
        WASP-69\,b        & 98/2         & 15             & 2.9             & \change{$-1.8\pm1.6$}& Y           & HD 85512                 \\ \smallskip
        WASP-76\,b        & 98/2         & 19             & 1.5             & \change{$-1\pm2    $}& T           & HD 97658                 \\ \smallskip
        WASP-80\,b        & 98/2         & 12             & 3.4             & 0                  & N           & HD 85512                 \\
        \hline

       \tablefootmark{*}{\citet{Lampon2020,Lampon2021_mar,Lampon2023}}     \\
       \tablefootmark{$\dagger$}{Sanz-Forcada (priv. comm.)} \\
       \tablefootmark{$\ddagger$}{\citet{Bourrier2020}}

       \end{tabular}

       \begin{tablenotes}[flushleft]
         \footnotesize{\item{\textbf{Notes.} "H/He" is the hydrogen/helium ratio used for the computation of the abundance profiles: values with a "*" were constrained by \citet{Lampon2020,Lampon2021_mar,Lampon2023}, while the other ones were fixed to 98/2. "Extension" is the vertical extension, in planetary radius (Rp), of the atmosphere on which the abundance profiles were computed. We set it as the host star's diameter to include all possibly absorbing layers at any transit phase. "R$_{lobe}$" is the Roche lobe of the planet, computed using PyAstronomy's \citep{pya} implementation of Eggleton's formula \citep{Eggleton1983}. "V$_{shift}$" is a constant Doppler shift applied to the modeled He triplet lines: it was \change{fixed at a null value for nondetections, and was} estimated using a simple Gaussian fit of the He absorption triplet in the data \change{for detections and tentative detections} by comparing the theoretical position of the strongest He triplet lines with the center of the best fitting Gaussian function. \change{The error on V$_{shift}$ are given for a 95\% confidence interval.} "Detection" indicates whether planetary helium is detected (Y), tentatively detected (T) or nondetected (N). "Stellar SED" is the name of the star from which the spectral energy distribution (SED) was used, taken from the MUSCLES database unless otherwise stated.}}
       \end{tablenotes}

     \end{threeparttable}
    \end{table*}

    \subsection{Influence of the model parameters}
    \label{models_list}
    For comparison purposes with previous studies, we studied how our model parameters could influence the retrieved mass-loss rate and temperature. We tested the impact of the H/He ratio, the atmospheric extension, and the difference between our wrapper of the p-winds code (which takes into account the orbital eccentricity and spin-orbit angle in the transit geometry) and our adapted \texttt{Python} version of the model used by \citet{Lampon2023}. We focused on the differences obtained in terms of \mdot and T for each of the three detections (HAT-P-11\,b, HD\,189733\,b, and WASP-69\,b), with 5 different cases:
    \begin{itemize}
      \item{M1: Our main model, corresponding to the parameters presented in Table~\ref{table_model_parameter_targets}. This model corresponds to our wrapper of p-winds, with H/He $\geq 98/2$ and extending the atmospheric extension over the full stellar disk.}
      \item{M2: Same as M1 but using our adapted \texttt{Python} version of the model used in \cite{Lampon2023}. This model and these assumptions are similar to the ones used in \cite{Lampon2023}: the model is integrated over the full stellar disk and the H/He ratio is set to be higher than or equal to 98/2.}
      \item{M3: Same as M1 but limiting the atmospheric extension to the ionopause stand-off point altitude level, which is the altitude at which the stellar wind pressure balances the planetary escaping wind pressure. This model and these assumptions are similar to the "SW" model used in \cite{Lampon2023}.}
      \item{M4: Same as M1 but fixing the H/He ratio to its solar value (90/10).}
      \item{M5: Same as M4 but limiting the atmospheric extension to the Roche lobe. This model and these assumptions are similar to the ones used by \cite{Allart_2023} and \citet{Guilluy2023}.}
    \end{itemize}

    \subsection{Results for each target}

    \subsubsection{55\,Cnc\,e   }
    55\,Cnc\,e is a hot super-Earth orbiting a bright K0 main-sequence star \citep{McArthur2004,Bourrier2018a} with a period of 0.74 d. With a radius of $1.88\pm0.03$ $R_{\oplus}$, this target is an interesting case of transiting sub-Neptune-size planets whose atmospheric nature is still debated. Some observations suggest the absence of a hydrogen and helium-dominated atmosphere, such as the nondetection of hydrogen Ly-$\alpha$ absorption \citep{Ehrenreich2012} and the absence of metastable helium signature \citep{Zhang2021}. The latter observed two transits of 55\,Cnc\,e in 2019 using Keck/NIRSPEC and reported no absorption greater than 250 ppm for the helium triplet. Using a 1D Parker wind model, \citet{Zhang2021} reported an upper limit at $\sim$10$^9$ g.s$^{-1}$ for the mass-loss rate with a plausible range of temperature of 5000-6000 K.

    % \remove{From their phase curve measurement, \citet{Demory2016} reported a 41$\pm$12 degrees offset of the hot spot. This suggests a heavy molecular-weight atmosphere (also supported by the density measurements from \citealp{Bourrier2018}) or no atmosphere at all (\change{e.g.,} with a magma surface). The reanalysis of \citet{Mercier2022} indicates a smaller hotspot offset of $-12^{+21}_{-18}$ degrees, the nondetections of HCN, NH$_3$ and C$_2$H$_2$ by \citet{Deibert2021}, along the nondetections of metals like Na, Ca and Fe \citep{Ridden-Harper2016,Keles2022,Rasmussen2023}, favored the hypothesis of a very thin, if any, secondary atmosphere around this planet. \citet{Tsiaras2016} reported the detection of a spectral feature consistent with HCN absorption in a hydrogen-rich atmosphere using HST/WFC3.} % Remove this part? But is also relevant as proof of no atmosphere at all...

    We analyzed data from a single transit of 55\,Cnc\,e observed with \mbox{SPIRou} on \change{30 Dec 2020}. The time series has no baseline before transit and is missing data until the end of ingress. We do not report strong OH contamination (Fig.~\ref{telluric_correction_grid}). The time series in the planetary rest frame is featureless (Fig.~\ref{2D_grid_reduced_data}), which translates into no visible absorption in the transit light curve (Fig.~\ref{TLC_grid}) or in the transmission spectrum (Fig.~\ref{1d_reduced_data}). We report no detection of helium absorption with a 3-$\sigma$ upper limit of 0.1\,\% in excess absorption and 1 m\AA~in equivalent width (hereafter EW), which corresponds to upper limits of 2.2\,\tengs{11} and 0.8\,\tengs{11} in mass-loss rate for a H/He ratio of 98/2 (model M1) and 90/10 (model M4) respectively. We were not able to constrain the temperature. Our results are compatible with the constraints derived by \citet{Zhang2021} and support the absence of a thick, H/He-rich, atmosphere around this planet.

         \subsubsection{AU\,Mic\,b   }
         AU\, Mic\,b is a young \citep[22$\pm$3 Myr,][]{Mamajek2014} Neptune-sized planet closely orbiting a young M dwarf in 8.5\,d \citep{Martioli2021} that is still hosting a debris disk \citep{Kalas2004,Boccaletti2018}. Alignment of the planetary orbit \citep{Hirano2020,Martioli2020,Palle2020_aumic,Addison2021} is thought to be the result of formation and migration in the disc. AU\,Mic\,b is thus a key target for the study of atmospheric escape in young planets, and its influence on the super-Earth and mini-Neptune population distribution. Yet, intense activity from its host star (\change{e.g.,} large active regions, frequent flaring events, see \citealp{Plavchan2020,Martioli2021,Klein2021,Klein2022,Donati2023}) makes it difficult to characterize atmospheric escape in this target given the influence of the stellar flux on the atmospheric escape process.

         \citet{Rockcliffe2023} analyzed two transits of AU\,Mic\,b in July 2020 and October 2021 using HST/STIS, searching for Ly-$\alpha$ absorption to investigate atmospheric escape. They report a nondetection during the first visit and a neutral hydrogen detection ahead of the planet during the second visit with ~30\,\% excess absorption, corresponding to a mass-loss rate in the range \mbox{(2-4)\,\tengs{11}}. The difference between both visits was interpreted as stellar contamination occulting the planetary signature during visit 1, or a consequence of the stellar flux variability with an increase of flux during visit 2 similar to the Ly-$\alpha$ absorption variations observed by \citet{Lecavalier2012} in HD\,189733\,b. \citet{Hirano2020} used NIRSPEC on Keck II and reported nondetection of the helium triplet, placing an upper limit for the EW of 3.7 m\AA~and corresponding to a mass-loss rate lower than (0.3-0.9)\,\tengs{11}.

   We reanalyzed the single transit of AU\,Mic\,b observed by \mbox{SPIRou} on \change{17 Jun 2019} by \citet{Allart_2023}, applying our own reduction pipeline and modeling process. The time series starts just after the end of ingress and covers the rest of the transit plus a baseline after transit. Due to important stellar activity from the young host star, removal of the stellar signature was challenging and some features attributable to time-varying stellar residuals are still visible in the time series (Fig.~\ref{2D_grid_reduced_data}). In particular, we report an absorption feature occurring at the expected position of the He triplet lines during the first half of the transit only (Fig.~\ref{TLC_grid}). This feature was also observed by \citet{Allart_2023}, who derived upper limits of 0.26\,\% in excess of absorption, and a mass-loss rate smaller than 1.51\,\tengs{11}. Averaging all in-transit spectra together, we obtained a $0.5\pm0.2$\,\% absorption with a 95\% confidence interval. If atmospheric escape would be the whole responsible for this 8-sigma signature, our best fit (using our main model M1) would correspond to a mass-loss rate of $7^{+28}_{-6}$\,\tengs{9} and a temperature of $3500^{1700}_{-1900}$ K. However, we suspect this feature could be mainly, if not fully, of stellar origin, in which case we are unable to constrain the atmospheric escape. More observations of this target are thus required to investigate the stellar activity and probe for potential variations in the atmospheric escape of this target.

    \subsubsection{GJ\,1214\,b  }
    GJ\,1214\,b is a warm, sub-Neptune-sized, exoplanet orbiting an M dwarf with a period of 1.58\,d \citep{Charbonneau2009,Kokori2022_a}. Several studies attempted to detect atmospheric escape in this target through helium triplet absorption and reported no detection \citep{Crossfield2019,Petit2020,Kasper2020}. \citet{Orell2022} analyzed a single transit using CARMENES and reported a 4.6-$\sigma$ tentative detection of the helium triplet, corresponding to an excess of absorption of $2.10^{+0.45}_{-0.50}$\,\%, a mass-loss rate of (1.5-18)\,\tengs{10} and an outflow temperature in the range of 2900-4400 K. \citet{Lampon2023} reanalyzed this signature with new constraints on the H/He ratio and estimated a mass-loss rate of (1.3$\pm$1.1)\,\tengs{11} and a temperature of 3750$\pm$750 K.

        To reconcile with previous nondetections, \citet{Orell2022} showed that the measurements used by \citet{Kasper2020} and \citet{Petit2020} could have been contaminated by telluric contributions, mainly H$_2$O absorption and OH emission, overlapping with the expected helium triplet position at the corresponding observation date. A following study by \citet{Spake2022} used Keck/NIRSPEC in \change{May 2019} to observe another transit of GJ\,1214\,b. The observation time was chosen to minimize telluric line contaminations, however, the authors did not observe any planetary helium signature. They report an upper limit of 1.22\,\% on the excess absorption, in disagreement with the detection of \citet{Orell2022}.

        We analyzed three transits of GJ\,1214\,b observed with \mbox{SPIRou} on \change{18 Apr 2019}, \change{14 May 2020}, and \change{21 Sep 2023}. The \change{14 May 2020} transit was analyzed by \citet{Allart_2023} who reported an excess of absorption of $1.59\pm0.97$\,\% at the He triplet position. Yet, due to the presence of systematics, they set an upper limit at 2.92\,\%. The time series of our reduced data (Fig.~\ref{2D_grid_reduced_data}) both cover the full transit with observations before and after transit, but with high \change{S/N} variations (Fig.~\ref{observation_condition}) during the first visit. Although our data would have been sensitive to the reported tentative detection of \citet{Orell2022}, we report no visible excess of absorption during any of the three visits (Fig.~\ref{2D_grid_reduced_data} and Fig.~\ref{TLC_grid}), nor in the final combined spectrum (Fig.~\ref{1d_reduced_data}). We set 3-$\sigma$ upper limits of 3.7\,\% in excess of absorption and 16 m\AA~in EW, corresponding to a mass-loss rate lower than 3\tengs{12} for a H/He ratio of 98/2 (model M1). Our upper limit in excess of absorption is not constraining enough to support either the tentative detection of \citet{Orell2022} or the nondetections of \citet{Kasper2020} and \citet{Spake2022}.

        \subsubsection{GJ\,3470\,b  }
        GJ\,3470\,b is a warm sub-Neptune closely orbiting an M-dwarf with an orbital period of 3.3\,d \citep{Bonfils2012}. \citet{Bourrier2018_3470} observed three transits in the Ly-$\alpha$ line using HST and detected absorption signatures in all three epochs, deriving a mass-loss rate of about (1.5-8.5)\,\tengs{10} which would have blown out up to 35\,\% of the planet current mass over its $\sim$2\,Gyr lifetime. \citet{Nina2020} and \citet{Palle2020} both reported the detection of helium triplet absorption in this exoplanet. The first study analyzed three transits observed with the Habitable Zone Planet Finder spectrograph and measured a broad absorption with an EW $\sim 0.012\pm0.002$~\AA~and a blueshifted wing extending to -36\,\kms. The second one used CARMENES observations of three transits and measured an excess of absorption of $1.5\pm0.3$\,\%. They obtained a mass-loss rate in the range (3-10)\,\tengs{10} and a temperature range of 6000-9000 K, in agreement with the estimations of the hydrogen escaping rate by \citet{Bourrier2018_3470}. \citet{Palle2020} data were reanalyzed by \citet{Lampon2021_mar} and \citet{Lampon2023}, revising the H/He ratio value and obtaining much colder temperature in the range of $3400\pm350$ K and a similar mass-loss rate of ($1.3\pm0.6$)\,\tengs{11}. \citet{Allart_2023} analyzed two transits observed with \mbox{SPIRou} on \change{18 Feb 2019} and \change{15 Dec 2021} and reported a slight excess of absorption with insufficient \change{S/N} to claim for detection. They report upper limits of 0.64\,\% in excess absorption, corresponding to a mass-loss rate lower than 1.41\,\tengs{11}.

        We reanalyzed the two transits of GJ\,3470\,b observed with \mbox{SPIRou} on \change{18 Feb 2019} and \change{15 Dec 2021}, to which we added a new transit observed with \mbox{SPIRou} on \change{29 Dec 2023}. The three transits were fully covered with baselines before and after transit. Similarly to \citet{Allart_2023}, a slight excess of absorption, hard to spot in the time series (Fig.~\ref{2D_grid_reduced_data}), is observed in the transit light curves (Fig.~\ref{TLC_grid}). This results in a broad 5.6-$\sigma$ signature in the final spectrum (Fig.~\ref{1d_reduced_data}), reaching $0.6\pm0.4$\,\% excess of absorption (with a 95\% confidence interval) and corresponding to a range of 1\,\tengs{11} - 4\,\tengs{12} in mass-loss rate and 3200-6300 K in temperature. Our tentative detection is compatible with the parameters derived by \citet{Lampon2023}. Yet, given the presence of residuals with amplitudes at the $\sim$1\,\% level, we cautiously set a 3-$\sigma$ upper limit of 1.1\,\% in excess of absorption and 19 m\AA~in EW, leading to upper limits of 2.6\,\tengs{13} in mass-loss rate for a H/He ratio of 99.8/0.2, and a temperature lower than 7000 K.

    \subsubsection{GJ\,436\,b   }
    GJ\,436\,b is a Neptune-sized planet orbiting a quiet M dwarf star with an orbital period of 2.6\,d \citep{Butler2004}. This warm Neptune is known for hosting a large, comet-like hydrogen exosphere transiting for more than 5h and absorbing up to 56\,\% of the flux in the blue wing of the stellar Ly-$\alpha$ line \citep{Kulow2014, Ehrenreich2015, LaVie2017, DosSantos2019}. The corresponding mass-loss rate has been estimated to span from $\sim 10^{6}$ g s$^{-1}$ to $\sim 10^{10}$ g $s^{-1}$ \citep{Kulow2014, Ehrenreich2015, Bourrier2015, Bourrier2016, Villarreal2021}. \citet{DosSantos2019} reported the atmospheric loss to be stable over a timescale of a few years and did not detect metallic ions in absorption. The latter suggests that the escape is not hydrodynamic, or that metals are ineffectively dragged in the exosphere by the atmospheric mixing to produce a detectable escape rate.

    \citet{Oklopcic2018} predicted a large absorption depth in the He I line for GJ\,436\,b with an EW up to $\sim$105~m\AA~and excess of absorption reaching $\sim$8\,\%. This prediction was supported by the planet's low gravitational potential and favorable radiation flux from the host star. Yet, observation of GJ\,436\,b using CARMENES \citep{Nortmann2018} did not detect He I absorption, yielding an upper limit of 0.41\,\% for excess absorption depth. The nondetection of absorption in the metastable He lines despite the well-pronounced absorption measured in Ly-$\alpha$ and the predictions from \citet{Oklopcic2018} could be explained by a H/He ratio three times larger than the solar value \citep{Rumenskikh2023}.

    We analyzed the \change{25 Feb 2019} transit of GJ\,436\,b observed with \mbox{SPIRou}. The observations cover the full transit with baselines before and after transit (Fig.~\ref{observation_condition}). We do not detect an excess absorption during transit for the helium triplet when looking at the time series, the transit light curve, or the final spectrum (Fig.~\ref{2D_grid_reduced_data}, Fig.~\ref{TLC_grid} and Fig.~\ref{1d_reduced_data} respectively). We report 3-$\sigma$ upper limits of 0.3\,\% in excess of absorption and 2.6\,m\AA~in EW but were not able to put constraints on the mass-loss rate at the 3-$\sigma$ level. Our results, like those from \citet{Nortmann2018}, disagree with the 1D model predictions of \citet{Oklopcic2018} and are surprising given the large Ly-$\alpha$ absorption depth reported by \citet{Kulow2014, Ehrenreich2015, LaVie2017, DosSantos2019}. Similarly to \citet{Nortmann2018}, only one transit of GJ\,436\,b was available for this study, limiting the reproducibility of this result and calling for more observations of this target to understand the absence of helium signature despite its strong hydrogen escape.

    \subsubsection{GJ\,486\,b   }
    GJ\,486\,b is a 1.3 R$_\oplus$ and 2.8 M$_\oplus$ exoplanet orbiting in 1.5\,d around an M star \citep{Trifonov2021}. Previous studies reported different possible atmospheric compositions for this planet, spanning from bare rock with a tenuous atmosphere to a H/He-dominated atmosphere with solar or enhanced metallicity depending on formation and evolution scenarios, the most probable scenario being an H$_2$O-dominated atmosphere \citep{Caballero2022}. Currently, the only attempt to characterize GJ\,486\,b's atmosphere using high-resolution spectroscopy was made by \citet{Ridden2023} who analyzed three transits gathered with the spectrographs IRD, IGRINS, and \mbox{SPIRou}. They searched for several molecular absorption lines, notably H$_2$O, and did not report any detection at the 3-$\sigma$ level. From their results, it seems unlikely that GJ\,486\,b possesses a cloud-free H/He-dominated atmosphere.

    We conducted here the first search for He signature in this target since its discovery. We analyzed two transits of GJ\,486\,b observed with \mbox{SPIRou} on \change{24 Mar 2021} and \change{23 Jan 2022}. Both observations cover the full transit, with baselines before and after transit in both cases (Fig.~\ref{observation_condition}). Time-series and transit light curves are featureless for both transits (Fig.~\ref{2D_grid_reduced_data} and Fig.~\ref{TLC_grid}), and the final spectrum shows no sign of He I absorption (Fig.~\ref{1d_reduced_data}). We set 3-$\sigma$ upper limits of 0.2\,\% in excess of absorption and 1.4\,m\AA~in EW, corresponding to a mass-loss rate lower than 8.3\,\tengs{11} for a H/He ratio of 98/2 (model M1) and 0.9\,\tengs{11} for a solar H/He ratio (Model M4, see Fig.~\ref{chi2_map_grid}). Our results are in agreement with \citet{Ridden2023} and indicate that GJ\,486\,b is unlikely to have a primary atmosphere undergoing strong escape.

    \subsubsection{HAT-P-11\,b }
    HAT-P-11\,b is a low-mass (0.08 M$_{Jup}$) hot-Neptune ($\sim$0.4\,R$_{Jup}$) orbiting a bright K star in 4.89\,d \citep{Bakos2010}. Its $\sim$0.26 eccentricity \citep{Huber2017} allows for a clear separation between the planetary signal and absorption lines from the host star during transit when using high-resolution spectroscopy \citep{Allart2018}. This planet is thus an ideal case for atmospheric escape studies. \citet{BenJaffel2022} reported strong, phase-extended transit absorption of neutral hydrogen with excess absorption up to $32\pm4$\,\% and a cometary-like tail. \citet{Allart2018} and \citet{Mansfield2018} both detected the presence of metastable helium using respectively CARMENES and HST.

    \citet{Mansfield2018} reported an excess of absorption of $162\pm36$ ppm using HST (and thus observing unresolved lines), corresponding to a mass-loss rate in the range $10^{9} - 10^{11}$ g s$^{-1}$. \citet{Allart2018} measured an excess absorption of 1.08 $\pm$ 0.05\,\% with CARMENES. Using the 3D code EVE \citep{Bourrier2013_sep}, they set an upper limit of 3\,\tengs{5} in helium mass-loss rate, reporting a net blue-shift compatible with high-altitude winds flowing at $\sim$3 km s$^{-1}$ from day to night-side. \citet{DosSantos2022} reanalyzed \citet{Allart2018} data using the 1D code p-winds \citep{DosSantos2022} and estimated a mass-loss rate of 2.5\,\tengs{10} and an outflow temperature of 7200 K. \citet{Allart_2023} analyzed two transits of HAT-P-11\,b obtained with \mbox{SPIRou} on \change{13 Aug 2021} and \change{18 Aug 2021}. They measured an excess absorption of $0.76\pm0.07$\,\%, corresponding to a mass-loss rate of $(0.67)^{+0.27}_{-0.24}$\,\tengs{11}, a temperature of $8100 - 8900$ K and a $-5.6\pm0.8$ km s$^{-1}$ blueshift similar to the one found by \citet{Allart2018}.

    \begin{figure}[h]
      \centering
      \includegraphics[width=\hsize]{"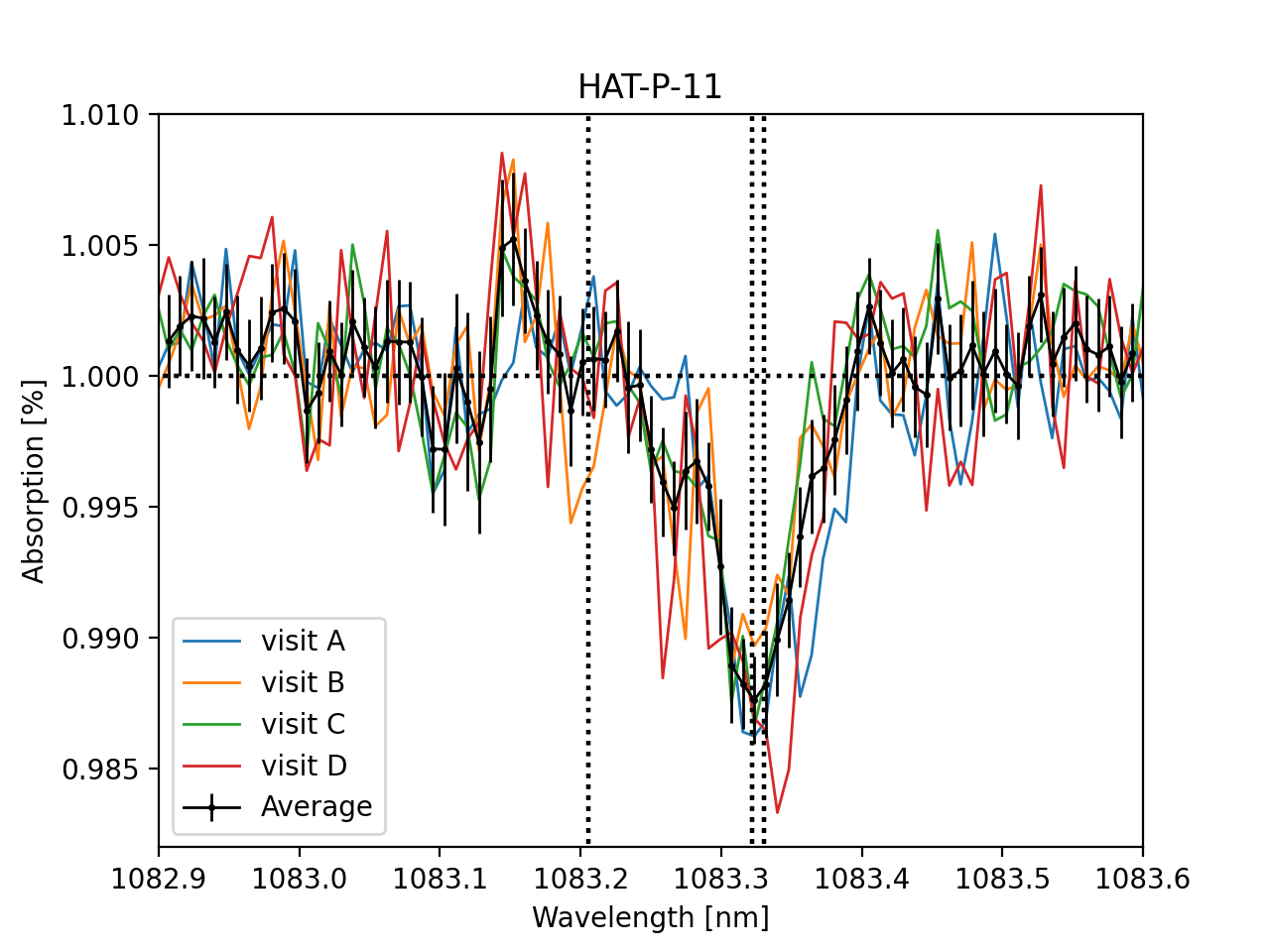"}
       \caption{Reduced spectra observed with \mbox{SPIRou} and corresponding to visits A (\change{30 Jun 2021}), B (\change{13 Aug 2021}), C (\change{18 Aug 2021}), D (\change{12 Jun 2022}), and averaged spectrum of the four transits of HAT-P-11\,b.}
       \label{hatp11_variability}
    \end{figure}

    \begin{table}[]
       \caption[]{Variability of the helium absorption depth over the four transits of HAT-P-11~b.}
       \label{hatp11_variability_tab}
       \begin{tabular}{ccc}
        \hline \hline
        Visit            & Date [yyyy-mm-dd]        & Depth [\%]             \\
        \hline
        A                & 2021-06-30               & $1.3\pm0.4$            \\ \smallskip
        B                & 2021-08-13               & $0.9\pm0.4$            \\ \smallskip
        C                & 2021-08-18               & $1.2\pm0.4$            \\ \smallskip
        D                & 2022-06-12               & $1.4\pm0.6$            \\
        \hline
        \mc{2}{l}{ Average }                          & $1.2\pm0.2$            \\
        \hline
       \end{tabular}
    \end{table}

    We performed our analysis on four transits of HAT-P-11\,b observed by \mbox{SPIRou}: two transits observed on \change{30 Jun 2021} and \change{12 Jun 2022} (hereafter visits A and D), and the \change{13 Aug 2021} and \change{18 Aug 2021} transits studied by \citet{Allart_2023} (hereafter visits B and C). All observations cover full transits with baselines before and after transit (Fig.~\ref{observation_condition}). We report an excess of absorption at the metastable helium triplet position in each transit time series (Fig.~\ref{2D_grid_reduced_data}), occurring only during the planet's transit (Fig.~\ref{TLC_grid}). We plot the spectrum for each visit in Fig.~\ref{hatp11_variability} and report the corresponding measured excess absorption in Table~\ref{hatp11_variability_tab}. We do not report significant variations of the helium signature at the 95\,\% level of confidence between the four observations.

    We report a \change{$2.3\pm1.6$\,\kms} redshift for visit A, \change{$2.5\pm1.6$\,\kms} and \change{$2.1\pm1.8$\,\kms} blue shifts for visits B and C respectively, and no significant Doppler shift for D. Our individual measurements are in agreement with \citet{Allart2018} and \citet{Allart_2023}, except for the redshift of visit A signature. The latter could be due to stellar contamination caused by variability in the stellar lines, although due to its eccentricity, the planet's radial orbital velocity, during the transit, is such that it prevents overlapping of the stellar strongest absorption lines (He and Si) with the planetary He signature during the transit. Variability of the escaping atmosphere itself could also explain the observed redshift, and further investigations would require 3D modeling of the escaping atmosphere and its interaction with stellar wind which is beyond the scope of this study.

    Combining all four transits, we obtain a 14.3-$\sigma$ detection of HAT-P-11\,b metastable helium signature with an excess of absorption of $1.2\pm0.2$\,\% and an EW of $8.8\pm0.4$ m\AA~(with 95\% confidence intervals). Our best fit corresponds to a mass-loss rate of  $(9.4^{+5.9}_{-3.5})$\,\tengs{9} and a temperature of $3700^{+500}_{-400}$ K for a H/He ratio of 98/2 (model M1), and is displayed along with the observed spectrum in  Fig.~\ref{1d_reduced_data}. When using a solar H/He ratio (model M4), we obtained a best-fit corresponding to a mass-loss rate of $8.5^{+4.2}_{-3.3}$ \,\tengs{9} and a temperature of $5000\pm600$ K. The corresponding metastable helium profiles are shown in the lower panel of  Fig.~\ref{weight_density_grid}. Finally, we obtained a mass-loss rate of $0.3^{+0.3}_{-0.2}$\tengs{11} and a temperature of $6300^{+1500}_{-1200}$ K when using a solar H/He ratio and limiting the altitude extension to the Roche lobe (model M5 from Section~\ref{models_list}), hence closer to the (\mdot, T) values derived by \citet{Allart_2023} who used a similar model parametrization.

    \subsubsection{HD\,189733\,b}

    HD\,189733\,b is one of the most well-studied exoplanets due to its large transit depth and host star brightness. This hot Jupiter orbits a relatively active K dwarf in 2.2\,d \citep{Bouchy2005}. 
    % \remove{and has become a well-known benchmark for molecule detection, with species detected in its lower atmosphere such as H$_2$O and CO \citep[\change{e.g.,}][]{Birkby2013, Cullough2014, Brogi2016, Brogi2018, Brogi2019, Alonso2019_hd189733, Cabot2019, Boucher2021, DosSantos2023}.}
     HD\,189733\,b's exosphere has been characterized through neutral and ionized species detection such as C, O, and Na \citep[\change{e.g.,}][]{Wyttenbach2015, Khalafinejad2017, DosSantos2023}, H through H$\alpha$ \citep{Jensen2012, Cauley2015, Cauley2016, Cauley2017} and Ly-$\alpha$ lines \citep[\change{e.g.,}][]{Lecavelier2010, Bourrier2013_mar, Bourrier2020}, and He \citep{Salz2018, Guilluy2020, Lampon2021_mar, Zhang2022, Lampon2023, Allart_2023}. \citet{Cauley2017} found strong variations in the strength of the H$\alpha$ transmission for seven transits that spanned almost a decade using HARPS and Keck data. These variations may be linked to stellar activity and were also reported by \citet{Lecavelier2010} and \citet{Bourrier2013_mar,Bourrier2020} using HST measurements of Ly-$\alpha$ absorption. \citet{Lecavelier2010} estimated a mass-loss rate of $10^{9}-10^{11}$ g s$^{-1}$, while observations by \citet{Bourrier2013_mar} suggest a cometary tail-like shape for the escaping neutral H. \citet{DosSantos2023} detected Ly-$\alpha$ at absorption levels in agreement with previous studies and measured escaping C II. They report atmospheric escape with a $\sim$1.1\tengs{11} mass-loss rate assuming T $\sim$12\,100\,K and a solar abundance.

    \citet{Salz2018} analyzed three transits observed with CARMENES and reported an excess of absorption in the He triplet lines of $0.88\pm0.04$\,\% with a $-3.5\pm4$\,\kms blueshift and an EW of $12.7\pm0.4$ m\AA. Reanalysis of these data using a H/He ratio of 99.2/0.8 by \citet{Lampon2021_mar}, followed by a reanalysis by \citet{Lampon2023} using a revised XUV flux of the host star, led to mass-loss rate and temperature estimations of $1.4\pm0.5$\,\tengs{11} and $12700\pm900$ K, respectively. Observations of five transits by GIANO allowed \citet{Guilluy2020} to measure a He absorption of $0.75\pm0.03$\,\%, with significant variations between these observations spanning a 1.5-year timescale. \citet{Zhang2022} measured a He absorption of $0.420\pm0.013$\,\% using Keck/NIRSPEC. \citet{Allart_2023} analyzed six transits observed with \mbox{SPIRou} from 2018 to 2021 and measured an excess of absorption varying from 0.29 to 0.75\,\%. They report an average absorption of $0.69\pm0.04$\,\% and a blueshift of $-4.2\pm0.8$ km/s, corresponding to \mdot = $0.94^{+0.82}_{-0.60}$\,\tengs{11} and T in the range $14500 - 18700$ K using a solar H/He value.

    \begin{figure}[h]
      \centering
      \includegraphics[width=\hsize]{"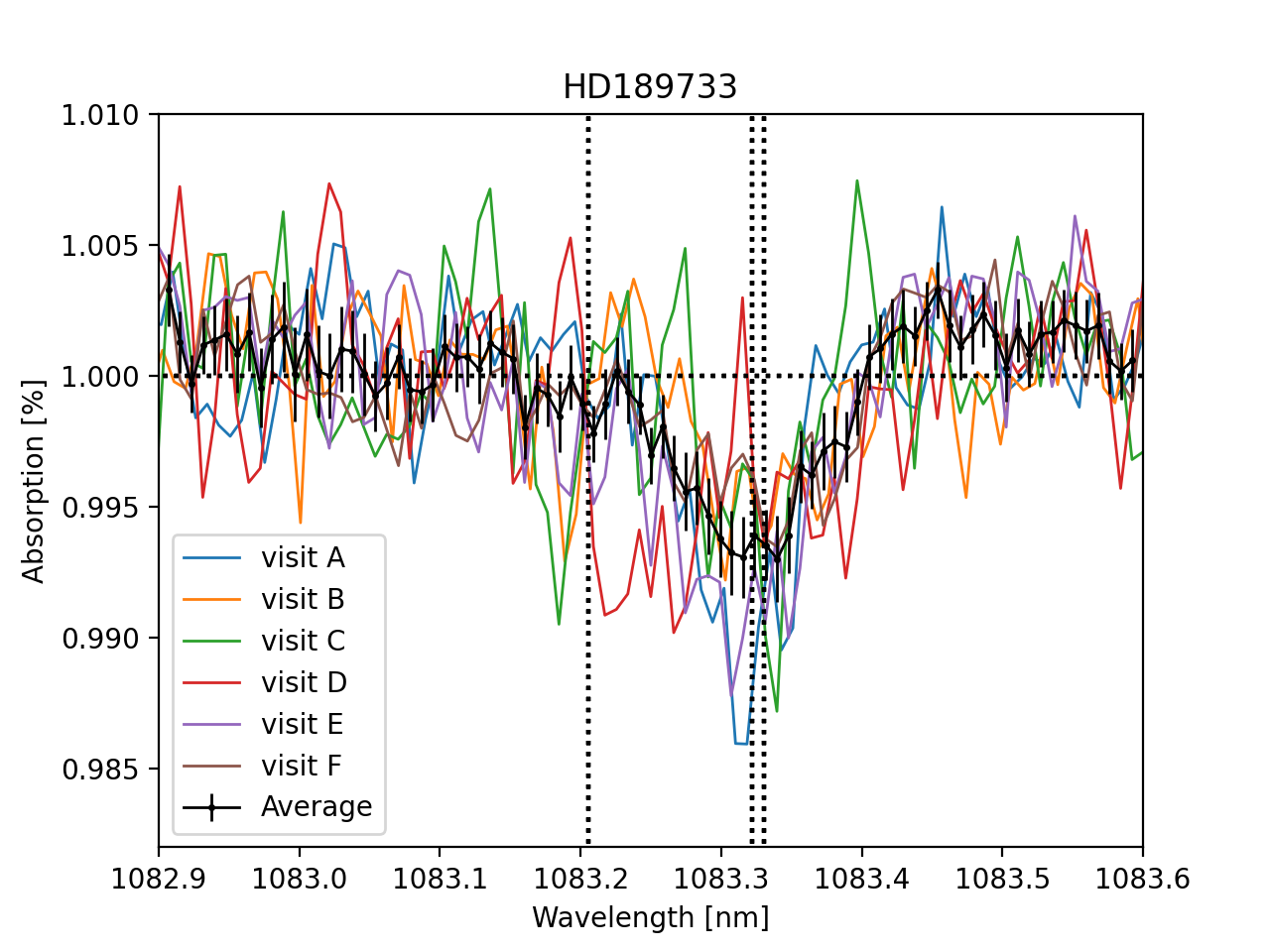"}
       \caption{Reduced spectra observed with \mbox{SPIRou} and corresponding to visits A (\change{22 Sep 2018}), B (\change{15 Jun 2019}), C (\change{3 Jul 2020}),\,D (\change{5 Jul 2020}), E (\change{25 Jul 2020}), and F (\change{24 Aug 2021}) for HD\,189733\,b.}
       \label{hd189_variability}
    \end{figure}

    \begin{table}[]
       \caption[]{Variability of the helium absorption depth over the six transits of HD~189733~b.}
       \label{hd189_variability_tab}
       \begin{tabular}{ccc}
        \hline \hline
        Visit            & Date [yyyy-mm-dd]        & Depth [\%]             \\
        \hline
        A                & 2018-09-22               & $1.2\pm0.4$            \\ \smallskip
        B                & 2019-06-15               & $0.6\pm0.4$            \\ \smallskip
        C                & 2020-07-03               & $0.8\pm0.6$            \\ \smallskip
        D                & 2020-07-05               & $0.6\pm0.4$            \\ \smallskip
        E                & 2020-07-25               & $1.0\pm0.4$            \\ \smallskip
        F                & 2021-08-24               & $0.5\pm0.2$            \\
        \hline
        \mc{2}{l}{ Average }                          & $0.7\pm0.2$            \\
        \hline
       \end{tabular}
    \end{table}

    We reanalyzed the six transits observed with \mbox{SPIRou} on \change{22 Sep 2018}, \change{15 Jun 2019}, \change{3 Jul 2020}, \change{5 Jul 2020}, \change{25 Jul 2020}, and \change{24 Aug 2021} (hereafter visits A to F). Visits B and E cover the full transit with baselines before and after transit. Visits A and F lacked observations before the end of ingress and after the start of the egress, respectively. Visits C and\,D only covered the transit partially (Fig.~\ref{observation_condition}). We report an excess of absorption in the He triplet lines in visits A, B, E, and F and a slight absorption during the end of transit for visit C and the start of transit for visit D, visible in both the time series (Fig.~\ref{2D_grid_reduced_data}) and the transit light curves (Fig.~\ref{TLC_grid}). Looking for variability in the helium absorption depth, we plot the spectrum for each visit in Fig.~\ref{hd189_variability} and report the corresponding measured excess absorption in Table~\ref{hd189_variability_tab}, which are in agreement with \citet{Allart_2023} measurements. We followed the method described in Section~\ref{average_transits_together} to average the six observations together, and we measured in our final spectrum an excess of absorption of $0.7\pm0.2$\,\%, an EW of $7.1\pm0.4$ m\AA~and a blueshift of \change{$-2.5\pm1.4$\,\kms}.

    We derived a mass-loss rate of \mdot = $2.5^{+1.6}_{-1.1}$\,\tengs{11} and a temperature of $18 000^{+4000}_{-3000}$ K (with 95\% confidence intervals) for a H/He ratio of 99/1 (model M1). Using a solar H/He ratio (model M4), we obtained a mass-loss rate of $5.1\pm{+3.1}$\,\tengs{10} and a temperature of $21 000^{+4000}_{-5000}$~K. Our \mdot values are consistent with previous studies, but we find significantly larger temperatures. These differences are explained by our model parametrization and are discussed in Section~\ref{discussion}. Our best fit is presented in Fig.~\ref{1d_reduced_data}, and the corresponding metastable helium profiles are shown in the lower panel of Fig.~\ref{weight_density_grid}. These profiles are representative of a dense atmosphere, with a maximum absorption below $\sim$2 R$_P$ and $\sim$80\,\% of the total He absorption signature contribution occurring inside the Roche lobe.

    Variations in the helium absorption depth have been reported between transits observed with the same instrument for both CARMENES, GIANO, Keck/NIRSPEC, and \mbox{SPIRou} observations. These observations span a $\sim$5 year timescale in total. The observed variability in measured excess absorption is reported by studies using the same data processed with different reduction processes (notably in terms of Earth's atmosphere correction). This variability could also be explained by variations in the stellar XUV flux, as was reported by \citet{Guilluy2020} who measured planetary helium and estimated stellar activity simultaneously using H$\alpha$. Stellar spots could also contribute up to 75\,\% of the total signature, although this effect is unlikely to explain the entire signature \citep{Salz2016, Guilluy2020, Allart_2023}. Variability in the planet's atmospheric outflow, stellar flares, or shear instability could also contribute to the observed variations \citep{Wang2021, Zhang2022}.

    \subsubsection{HD\,209458\,b}

    HD\,209458\,b is a hot Jupiter orbiting in 3.5\,d around a G-type star. This planet is both the first one for which repeated transits across the stellar disk have been observed \citep{Henry2000}, and the first one for which an evaporating atmosphere has been detected \citep{Vidal-Madjar2003}. Atmospheric escape has been detected using measurements of H (through Ly-$\alpha$ absorption), and C, O, and Mg signatures in the exosphere \citep{Vidal-Madjar2003, Vidal2004, Vidal2013}. These measures implied a lower limit of 0.4\,\tengs{11} for the mass-loss rate \citep{Vidal-Madjar2003} and a temperature of about 7000 K \citep{Lampon2020}. \citet{Moutou2003} and \citet{Nortmann2018} searched for metastable He signature and obtained upper limits for the absorption depth of 0.5\,\% and 0.84\,\%, respectively. \citet{Alonso2019} reported the first detection of He in HD\,209458\,b exosphere using CARMENES, with an absorption of $0.91\pm0.10$\,\% and a $1.8\pm1.3$\,\kms blueshift. They derived a mass-loss rate in the range $10^{8}-10^{11}$ g s$^{-1}$. These data were later reanalyzed by \citet{Linssen2022} using the CLOUDY code, who found a mass-loss rate in the range $(0.37 - 1.2)$\,\tengs{10} and T = $8200^{+1200}_{-1100}$ K.

    Several modeling efforts were performed to investigate the properties of the planet's escaping atmosphere, with a mass-loss rates in the range $10^{9} - 10^{11}$ g s$^{-1}$ \citep[\change{e.g.,}][]{Penz2008, Garcia2007, Bourrier2013_sep, Salz2016, Koll2018} which corresponds to temperatures in the range 6300 - 12500 K according to \citet{Lampon2020}. The latter authors combined He and Ly-$\alpha$ observations to constrain the H/He ratio and found a value of 98/2, higher than the solar 90/10 value used in previous studies. This low abundance of helium could be the result of the planet having formed with a small amount of helium in its primordial atmosphere, or be the result of helium mass fractionation as has been modeled in a self-consistent way by \citet{Xing2023}. \citet{Lampon2020} reanalyzed the data used by \citet{Alonso2019} and derived a mass-loss rate of ${0.4-1.0}$\,\tengs{11} and a temperature in the range $7000 - 8000$ K, with an EW of $5.3\pm0.5$\,m\AA.

    We analyzed a single transit of HD\,209458\,b observed with \mbox{SPIRou} on \change{18 Jun 2019}. Observations cover the full transit, with observations before transit but only two measures after transit (Fig.~\ref{observation_condition}). We detect an excess of absorption in the He line in both HD\,209458\,b time series (Fig.~\ref{2D_grid_reduced_data}) and transit light curve (Fig.~\ref{TLC_grid}). We measure an excess of absorption of $0.42\pm0.16$\,\%, an EW of $3.8\pm0.6$\,m\AA, and a blueshift of \change{$2\pm2$\,\kms}, corresponding to a mass-loss rate of $1.1^{+1.2}_{-0.8}$\,\tengs{11} and a temperature in the range $7000-13000$ K (with 95\% confidence intervals) for a H/He ratio of 98/2 (model M1 from Section~\ref{models_list}). Using a solar H/He ratio (model M4), we obtained a mass-loss rate of $4.6^{+1.7}_{-2.4}$\,\tengs{10} and a temperature in the range $10000-13500$ K. Our values are in good agreement with previous modeling results and measurements. Yet, given the presence of $\sim$0.4\,\% residuals in our final spectrum, we flag this detection as "tentative" and cautiously set an upper limit of 0.5\,\% in excess of absorption and 4 m\AA~in EW, leading to upper limits of 2.4\,\tengs{11} in mass-loss rate and 13\,000 K in temperature for a H/He ratio of 98/2, and a mass-loss rate not higher than 1.0\,\tengs{11} for a solar H/He ratio (for which we were not able to set an upper limit on the temperature).

    \subsubsection{K2-25\,b    }
        K2-25\,b is a Neptune-sized short-period (3.5 d, \citealp{Kokori2022_a}) exoplanet orbiting a M-dwarf star in the $727\pm75$\,Myr Hyades cluster \citep{Mann2016}. The young age of this planet, along with similar planetary and host star properties between this system and GJ\,436\,b, suggest a possible strong atmospheric escape in K2-25\,b, assuming this planet still holds a primary atmosphere. Two transits of K2-25\,b were observed using HST/STIS and analyzed by \citet{Rockcliffe2021} to search for exospheric signature using the Ly-$\alpha$ hydrogen line. Using the energy-limited regime approximation, they estimated the maximum mass-loss rate of K2-25\,b to be $10.6^{+15.2}_{-6.13}$\,\tengs{10}. They report no detection of escaping hydrogen, with possible explanations being a nonhydrogen-dominated atmosphere, stellar-wind interaction lowering the detectability of Ly-$\alpha$ absorption by confining the exosphere within a smaller radius \citep{Vidotto_2020}, or the absence of a primary atmosphere around K2-25\,b.

        A search for the metastable He signature in K2-25\,b would have been helpful to constrain the potential presence of an escaping, maybe not H-dominated, atmosphere around K2-25\,b. However, as reported by \citet{Allart_2023}, the single transit of K2-25\,b observed by \mbox{SPIRou} on \change{6 Oct 2019} exhibited very low \change{S/N} for each exposure (Fig.~\ref{observation_condition}) and was thus discarded from this analysis.

        \subsubsection{TOI-1807\,b }
        TOI-1807\,b is a young ($\sim$200 Myr) $1.26\pm0.04$\,R$_{\oplus}$ and $2.6\pm0.5$ M$_{\oplus}$ planet orbiting close a K3 star with a period of 0.55\,d \citep{Nardiello2022}. This recently discovered target \citep{Hedges2021} has yet only been studied once for atmospheric escape by \citet{Gaidos2023}, who analyzed a single transit of TOI-1807\,b using the InfraRed Doppler spectrograph on Subaru telescope. They report nondetection of He I, with an EW lower than 4 m\AA~and a corresponding mass-loss rate lower than $\sim$2\,\tengs{8} for a solar H/He value. As outlined by \citet{Gaidos2023}, the nondetection of escaping He in TOI-1807\,b is consistent with a measurement of mass compatible with a rocky planet and prediction from their hydrodynamical model that any H-He dominated primordial atmosphere around TOI-1807\,b would have already been lost.
    
        We analyzed two transits of TOI-1807\,b observed by \mbox{SPIRou} on \change{15 Apr 2022} and \change{9 Jun 2022}. The first visit did not cover the full transit event, missing its second half, and the second visit covered the full transit with proper baseline (Fig.~\ref{observation_condition}). The time series (Fig.~\ref{2D_grid_reduced_data}) and transit light curves (Fig.~\ref{TLC_grid}) of both visits are featureless, and we do not report any excess of absorption in the metastable He triplet lines in our final spectrum (Fig.~\ref{1d_reduced_data}). We thus report a nondetection of metastable He in TOI-1807\,b, with 3-$\sigma$ upper limits of 0.3\,\% in depth and 2\,m\AA~in EW, corresponding to a mass-loss rate lower than 3.8\,\tengs{12} when using a H/He value of 98/2 (model M1) and 1.4\,\tengs{12} for a solar H/He ratio (model M4). We thus improve the constraint on the metastable He absorption line's depth and EW, but not on the mass-loss rate for which the derived limit is more model-dependent. Our nondetection supports the idea of TOI-1807\,b being a rocky body whose potential primordial atmosphere would have been evaporated by its host star's strong XUV irradiation.

    \subsubsection{WASP-127\,b }
    WASP-127\,b is a 0.16 $M_{Jup}$ hot-Jupiter orbiting a G5-type star in ~4.2\,d \citep{Lam2017,Seidel2020}. Given its large radius (1.31 $R_{Jup}$), this target is thought to be inflated and this puffiness makes WASP-127\,b an ideal case for transit spectroscopy characterization. 
    % \remove{\citet{Allart2020} observed two transits of WASP-127\,b with ESPRESSO and reported a retrograde misaligned orbit along with the presence of sodium in the planet's atmosphere. \citet{Spake2021} used HST and Spitzer observations and revealed Na, H$_2$O and CO$_2$ features. \citet{Boucher2023} observed WASP-127\,b with \mbox{SPIRou} and reported the presence of H$_2$O, a tentative detection of OH, and set an upper limit for CO after combining \mbox{SPIRou}, HST and Spitzer \citep{Spake2021} data.} 
    \citet{DosSantos2020} used the Phoenix spectrograph to search for helium in the upper atmosphere of WASP-127\,b and set a 90\,\% confidence upper limit for excess absorption at 0.87\,\%. \citet{Allart_2023} analyzed three transits observed with \mbox{SPIRou} and also reported no detection for the helium triplet absorption with a 3$-\sigma$ upper limit at 0.48\,\% in excess of absorption.

    We reanalyzed the three transits observed with \mbox{SPIRou} on \change{11 Mar 2020}, \change{22 Mar 2021}, and \change{3 May 2021} (hereafter visits A, B, and C). Due to the long duration of WASP-127\,b transit, no visit covers the full transit event. Visit A almost covers the full transit, with missing observations after the start of the egress, while visits B and C are missing the end and the beginning of the transit respectively (Fig.~\ref{observation_condition}), and without enough out-of-transit observations to constitute a good baseline. We report no excess of absorption in all three visits, either from the time series (Fig.~\ref{2D_grid_reduced_data}), the transit light curves (Fig.~\ref{TLC_grid}), or in the final spectrum (Fig.~\ref{1d_reduced_data}). We thus report 3-$\sigma$ upper limits of 0.9\,\% in excess of absorption and 3.8\,m\AA~in EW, corresponding to a mass-loss rate smaller than 1.4\,\tengs{12} using a H/He ratio of 98/2 (model M1) and smaller than 0.47\,\tengs{12} using a solar H/He ratio (model M4). This nondetection is surprising, given the inflated nature of WASP-127\,b. As proposed by \citet{DosSantos2020}, the absence of metastable helium absorption in the atmosphere of this inflated hot-jupiter could be due to the low amount of ionizing irradiation from the host star, which could be related to the old age of this system (11.40 $\pm$ 1.80 Gyr as estimated by \citealp{Lam2017}).

    \subsubsection{WASP-69\,b  }
    WASP-69\,b is a 0.26 $M_{Jup}$ and 1.05 $R_{Jup}$ hot Jupiter orbiting an active K-type star in 3.87\,d \citep{Anderson2014}. Its large-scale height makes it a good target for atmospheric escape studies. \citet{Nortmann2018} reported the first detection of the He metastable triplet in WASP-69\,b after combining two transits observed with CARMENES. They report an excess absorption of $3.59\pm0.19$\,\% and a $3.58\pm0.23$\,\kms net blueshift, with 1-$\sigma$ error bars. The excess of absorption was reported to last for $22\pm3$ min after the transit ends and interpreted as potential evidence of a comet-like tail. \citet{Lampon2023} reanalyzed \citet{Nortmann2018} data and estimated a mass-loss rate of $(0.9\pm0.5)$\,\tengs{11} and a temperature of $5250\pm750$ K considering a H/He ratio higher than the solar value ($\geq$98/2), and estimated the escape to be in the recombination limited regime.

    \citet{Vissapragada2020} observed a transit of WASP-69\,b with Palomar/WIRC and detected an excess absorption of $0.498\pm0.045$\,\% with no excess of absorption after transit. A reanalysis by \citet{Vissapragada2022} reported a $0.512^{+0.049}_{-0.048}$\,\% excess depth, corresponding to a mass-loss rate of $6.3^{+9.6}_{-4.9}$\,\tengs{10}. 3D simulations by \citet{Wang2021} predicted a mass-loss rate of $\sim$9.5\,\tengs{10} with no comet-like tail. \citet{Tyler2024} analyzed a transit observed with Keck/NIRSPEC and reported an average relative helium absorption level of $2.7\pm0.4$\% and a net blueshift of $-5.9\pm1.0$ \kms, with a post-transit absorption for at least 1.28 hr blueshifted by -23 \kms suggesting an outflow strongly sculpted by ram pressure from the stellar wind. \citet{Allart_2023} analyzed a transit observed by \mbox{SPIRou} and reported a $2.35\pm0.46$\,\% excess depth and a blueshift of $5.4\pm1.2$\,\kms, with no signature after the end of the transit. This corresponds to a mass-loss rate of $0.40^{+0.58}_{-0.25}$\,\tengs{11}, and a temperature of $6900\pm1600$ K.

    We reanalyzed the \change{13 Oct 2019} transit of WASP-69\,b observed with \mbox{SPIRou} \citep{Allart_2023}. The observation covers the full transit with baselines before and after (Fig.~\ref{observation_condition}). We report a strong excess of absorption in the helium triplet lines in both the time series (Fig.~\ref{2D_grid_reduced_data}) and the transit light curve (Fig.~\ref{TLC_grid}), but we do not detect absorption after the planetary solid disk's transit as reported by \citet{Nortmann2018} and \citet{Tyler2024}. We measured in our final spectrum an excess of absorption of $3.0\pm0.8$\,\%, an EW of $25\pm2$ m\AA~and a blueshift of \change{$1.8\pm1.6$\,\kms}. Our excess of absorption agrees with the signature reported in previous studies, but we did not observe a blueshift as strong as what was derived by \citet{Nortmann2018} and \citet{Allart_2023}. Using a H/He ratio of 98/2 (model M1) as in \citet{Lampon2023}, we derived a mass-loss rate of \mdot = $1.9^{+1.8}_{-1.2}$\,\tengs{11} and a temperature of $7000^{+1600}_{-1800}$ K. Using a solar H/He ratio (model M4), we obtained a mass-loss rate of $0.8^{+0.9}_{-0.5}$\,\tengs{11} and a temperature of $8000^{+2300}_{-2100}$ K (with 95\% confidence intervals), hence higher than what was reported in previous studies using similar H/He values. For comparison with \citet{Allart_2023}, we also tested limiting the altitude extension to the Roche lobe and using a solar H/He ratio (model M5). We obtained a mass-loss rate of $5.2^{+3.6}_{-4.3}$\tengs{11} g.s$^{-1}$ and a temperature of $11000^{+1500}_{-4000}$ K. These values are higher than the ones obtained by \citet{Allart_2023}, which could be due to the different \mbox{APERO} version used for the data (\citealp{Allart_2023} used v0.7.179, while we used v0.7.288), differences in the data reduction process (\citealp{Allart_2023} performed a manual correction for the OH telluric lines near 1083.3 nm), and a different stellar XUV flux used for the modeling. Our best fit for the H/He ratio of 98/2 (model M1) is represented along with data in Fig.~\ref{1d_reduced_data}, and the corresponding metastable helium profile is shown in the lower panel of Fig.~\ref{weight_density_grid}. This profile is representative of an extended atmosphere, with about $\sim$50\,\% of the total absorption occurring inside the Roche lobe.

    \subsubsection{WASP-76\,b  }
    WASP-76\,b is an ultrahot Jupiter orbiting an F-type star in 1.8\,d \citep{West2016}. 
    % \remove{WASP-76\,b's extreme equilibrium temperatures (> 2000 K) motivated extensive studies of its atmospheric composition, with the detection of several species among which Na I \citep{Zak2019, Seidel2019, Seidel2021, Deibert2021b, Kesseli2022, Azevedo2022, Deibert2023, Pelletier2023}, Ca II \citep{Tabernero2021, Casasayas2021, Deibert2021b, Kesseli2022, Azevedo2022, Deibert2023, Gandhi2023, Pelletier2023}, Mg I \citep{Tabernero2021, Kesseli2022, Azevedo2022, Gandhi2023}, and Fe I \citep{Ehrenreich2020, Tabernero2021, Kesseli2021, Gandhi2022, Kesseli2022, Azevedo2022, Deibert2023, Gandhi2023}. The latter is of particular interest as the detection was reported to be asymmetric and interpreted as being due to condensation of iron on the nightside \citep{Ehrenreich2020, Kesseli2021, Kesseli2022}.} 
    \citet{Casasayas2021} reported possible evidence of He I in the two transits of WASP-76\,b obtained with CARMENES. Yet, given the low \change{S/N} of the tentatively detected signal, the authors reported a 3-$\sigma$ upper limit of 0.9\,\% in excess depth. \citet{Lampon2023} reanalyzed these two transits and obtained an EW of $12.4\pm1.7$ m\AA, corresponding to a mass-loss rate of $23.5\pm21.5$\,\tengs{11} and a temperature of $11500\pm5500$ K which would place the escape in the recombination limited regime.

    We analyzed two transits of WASP-76\,b observed with \mbox{SPIRou} on \change{31 Oct 2020} and \change{28 Oct 2021}. Both observations cover the full transit, with observations before and after transit (Fig.~\ref{observation_condition}). We do not detect excess absorptions in the time series (Fig.~\ref{2D_grid_reduced_data}), but a slight excess of absorption in the He line is present during visit B (Fig.~\ref{TLC_grid}). Combining the two transits, we measure an excess of absorption of $0.35\pm0.18$\,\% and an EW of $2.1\pm0.6$m\AA, in agreement with \citet{Casasayas2021}. For a H/He ratio of 98/2, this signature would correspond to \mdot and T in the range 1\,\tengs{9} $-$ 1.5\,\tengs{11}  and $4000-11500$ K with a best-fit at \mdot $\sim$0.6\,\tengs{11} and T\,$\sim$\,8000 K (Fig.~\ref{chi2_map_grid}). For a solar H/He ratio, the signature corresponds to a mass-loss rate in the range 2\,\tengs{8} $-$ 0.8\,\tengs{11} and a temperature in the range $5400 - 12500$ K (with 95\% confidence intervals). Yet, given the presence of residuals with similar amplitude in the vicinity of this signature, we flag this detection as "tentative" and cautiously set an upper limit of 0.4\,\% in excess of absorption and 2.2\,m\AA~in EW, leading to a mass-loss rate lower than 3.2\,\tengs{11} for a H/He ratio of 98/2 (model M1) and 0.9\,\tengs{11} for a solar H/He ratio (model M4). Our tentative detections and upper limits are lower than what was estimated by \citet{Lampon2023}. This could be explained by stellar variability and variations in the He escape during the $\sim$1 yr timescale between CARMENES observations and the two transits observed with \mbox{SPIRou}. More observations of WASP-76\,b are thus required to understand this discrepancy.

    \subsubsection{WASP-80\,b  }
    WASP-80\,b is a 0.54 M$_{Jup}$ and $\sim$1 R$_{Jup}$ warm Jupiter orbiting a relatively active \citep{Salz2016,King2018} K-type star in 3\,d \citep{Triaud2013,Triaud2015}. 1D hydrodynamic simulations of WASP-80\,b's upper atmosphere, conducted by \citet{Salz2016}, predicted this target to be one of the most promising targets for atmospheric escape detection due to its large radius and Roche lobe extension compared to the stellar radius. Yet, a previous study by \citet{Fossati2022} led to no detection of the He signature using four transits observed with GIANO in 2019 and 2020. They set an upper limit of 0.7\,\% in excess depth. This nondetection can be explained by a H/He ratio higher than the solar value, by a strong stellar wind, or by a combination of the two, while a planetary magnetic field is unlikely to be the cause of the nondetection \citep{Fossati2023}. \citet{Allart_2023} analyzed a transit of WASP-80\,b observed with \mbox{SPIRou} in 2019 and reported no signature, with an upper limit set at 1.24\,\% corresponding to a mass-loss rate lower than 0.14\,\tengs{11}.

    We reanalyzed the \change{7 Oct 2019} transit of WASP-80\,b observed with \mbox{SPIRou} using our reduction process and models. The observation covers the full transit event, with baselines before and after the transit (Fig.~\ref{observation_condition}). We do not detect excess absorption in the He triplet lines, either in the time series (Fig.~\ref{2D_grid_reduced_data}) or in the transit light curve (Fig.~\ref{TLC_grid}). The final spectrum (Fig.~\ref{1d_reduced_data}) is featureless and we thus set a $3-\sigma$ upper limit of 2.4\,\% in excess depth and 21\,m\AA~in EW, corresponding to a mass-loss rate lower than 4.2\,\tengs{11} for a H/He ratio of 98/2 (model M1) and 1\,\tengs{11} for a solar H/He ratio (model M4). Our nondetection, in agreement with previous studies, is surprising given \citet{Salz2016} predictions of this target being one of the most promising for atmospheric escape detection. Our results do not improve the constraints previously reported for this target.

    \begin{table*}[]
      \begin{threeparttable}
       \caption[]{Summary of the results obtained from the $\chi^2$ minimization}

       \smallskip
       \label{table_results}
       \begin{tabular}{l|lcc|cc|cr}
        \hline \hline
        \multirow{2}{*}{Target   }           & \multirow{2}{*}{Depth [\%]}             & \multirow{2}{*}{EW [m\AA]}              & \multirow{2}{*}{\change{S/N}    }          & \mdot [$10^{11}$g/s]            & \mdot [$10^{11}$g/s]              & T [K]                              & T [K]                    \\
                                            &                                          &                                         &                                   & ($\geq$98/2)                    & (90/10)                           & ($\geq$98/2)                       & (90/10)                  \\
        \hline
        55\,Cnc\,e                    & $<0.1$                         & $<1$                           & ---                      & $< 2.2 $                & $< 0.8 $                & ---                       & ---                      \\[0.1cm] % use this instead of a \smallskip to prevent breaking vertical lines. The value in [] is the space between lines.
        AU\,Mic\,b                    & ---                            & ---                            & ---                      & ---                     & ---                     & ---                       & ---                      \\[0.1cm]
        GJ\,1214\,b                   & $<3.7$                         & $<16$                          & ---                      & $< 30 $                 & $< 10$                  & ---                       & ---                      \\[0.1cm]
        GJ\,3470\,b                   & $<0.9$                         & $<11.2$                          & 5.6                      & $< 260 $                & ---                     & $< 7000                $  & ---                      \\[0.1cm]
        GJ\,436\,b                    & $<0.3$                         & $<2.6$                         & ---                      & ---                     & ---                     & ---                       & ---                      \\[0.1cm]
        GJ\,486\,b                    & $<0.2$                         & $<1.4$                         & ---                      & $< 8.3 $                & $< 0.9 $                & ---                       & ---                      \\[0.1cm]
        HAT-P-11\,b                  & $1.2\pm0.2$                    & $8.8\pm0.4$                    & 14.3                     & $0.09^{+0.06}_{-0.04}$  & $0.08^{+0.05}_{-0.04}$  & $3700^{+500}_{-400}    $  & $5000\pm600            $ \\[0.1cm]
        HD\,189733\,b                 & $0.7\pm0.2$                    & $7.1\pm0.4$                    & 13.3                     & $2.5^{+1.6}_{-1.1}$     & $0.5\pm{0.3}$           & $18000^{+4000}_{-3000} $  & $21000^{+4000}_{-5000} $ \\[0.1cm]
        HD\,209458\,b                 & $<0.5$                         & $<4$                           & 5.8                      & $<2.4  $                & $< 1.0 $                & $< 13000               $  & ---                      \\[0.1cm]
        K2-25\,b                     & ---                            & ---                            & ---                      & ---                     & ---                     & ---                       & ---                      \\[0.1cm]
        TOI-1807\,b                  & $<0.3$                         & $<2$                           & ---                      & $<38 $                  & $< 14 $                 & ---                       & ---                      \\[0.1cm]
        WASP-127\,b                  & $<0.9$                         & $<3.8$                           & ---                      & $<14 $                  & $< 4.7     $              & ---                       & ---                      \\[0.1cm]
        WASP-69\,b                   & $3.0\pm0.8$                    & $25\pm2$                       & 8.2                      & $1.9^{+1.8}_{-1.2}$     & $0.8^{+0.9}_{-0.5}$     & $7000^{+1600}_{-1800}  $  & $8000^{+2300}_{-2100}  $ \\[0.1cm]
        WASP-76\,b                   & $<0.4$                         & $<2.2$                         & 4.0                      & $< 3.2 $                & $< 0.9 $                & $< 12500               $  & $< 17000               $ \\[0.1cm]
        WASP-80\,b                   & $<2.4$                         & $<21$                          & ---                      & $< 4.2  $               & $< 1 $                  & ---                       & ---                      \\
        \hline
       \end{tabular}

       \begin{tablenotes}[flushleft]
         \footnotesize{\item{\textbf{Notes.} Uncertainties for detections and upper limits for tentative and nondetections correspond to a 95\,\%, and a 3-$\sigma$ confidence interval respectively. For nondetections, we assumed that a signature should have a FWHM covering at least 10 \mbox{SPIRou} pixels. mass-loss rate (\mdot) and temperature of the escape (T) were estimated with the reduced $\chi^2$ maps (Fig~\ref{chi2_map_grid}), using the model parameters in Table~\ref{table_model_parameter_targets} and computed with our wrapper of p-winds (see Section~\ref{Modeling}). This corresponds to our model M1 (see Section~\ref{models_list}), which provided equivalent results to our model M2. Error and upper limits estimations are detailed in Appendix~\ref{error_estimation_depth_EW} for the depth and EW, and in Appendix~\ref{error_estimation_mlr_T} for \mdot and T. Constraints on \mdot and T are reported for $\geq$98/2 (model M1) and 90/10 values (model M4) for the H/He ratio for comparison. The corresponding ratio is indicated in parenthesis below the relevant parameter. We were not able to derive upper limits for K2-25\,b due to its extremely low \change{S/N} (see Fig.~\ref{observation_condition}), and for AU\,Mic\,b due to the overlapping of a strong stellar residual with the expected planetary He lines positions.}}
       \end{tablenotes}

     \end{threeparttable}
    \end{table*}

    %______________________________________________________________
    \section{Discussion}
    \label{discussion}

    \begin{table*}[]
      \caption{Mass-loss rate (in $10^{11}$ g s$^{-1}$ units) comparison between the five models described in Section~\ref{models_list}.}
      \begin{tabular}{llllll}
      \hline
      \hline
      Model              & M1                        & M2                      & M3                      & M4                        & M5                   \\
      \hline
      HAT-P-11\,b        &  $0.09^{+0.06}_{-0.04}$   & $0.09^{+0.06}_{-0.04}$  & $0.4^{+0.5}_{-0.3}$     &  $0.08^{+0.05}_{-0.04}$   & $0.3^{+0.3}_{-0.2}$  \\[0.1cm]
      HD\,189733\,b      &  $2.5^{+1.6}_{-1.1}$      & $2.6^{+1.5}_{-1.1}$     & $2.5^{+1.8}_{-1.3}$     &  $0.5\pm{0.3}$            & $0.8^{+0.2}_{-0.4}$  \\[0.1cm]
      WASP-69\,b         &  $1.9^{+1.8}_{-1.2}$      & $2.0^{+1.8}_{-1.2}$     & $2.0^{+1.8}_{-1.2}$     &  $0.8^{+0.9}_{-0.5}$      & $5.2^{+3.6}_{-4.3}$  \\[0.1cm]
      \hline
      \end{tabular}
      \label{mlr_model_comparison}
    \end{table*}

    \begin{table*}[]
      \caption{Temperature (in K) comparison between the five models described in Section~\ref{models_list}.}
      \begin{tabular}{llllll}
      \hline
      \hline
      Model              & M1                        & M2                          & M3                       & M4                         & M5                          \\
      \hline
      HAT-P-11\,b        &  $3700^{+500}_{-400}    $ & $3800^{+500}_{-400}    $    & $5000^{+1300}_{-1000}$   &  $5000\pm600            $  & $6300^{+1500}_{-1200}    $  \\[0.1cm]
      HD\,189733\,b      &  $18000^{+4000}_{-3000} $ & $18800^{+5000}_{-3000} $    & $18000^{+5000}_{-4000}$  &  $21000^{+4000}_{-5000} $  & $23000^{+1400}_{-4000} $    \\[0.1cm]
      WASP-69\,b         &  $7000^{+1600}_{-1800}  $ & $7000^{+1500}_{-1700}    $  & $7000^{+1600}_{-1800}$   &  $8000^{+2300}_{-2100}  $  & $11000^{+1500}_{-4000}  $   \\[0.1cm]
      \hline
      \end{tabular}
      \label{T_model_comparison}
    \end{table*}
    % Rq: M2 & M3 devrait donner les mêmes résultats pour WASP-69\,b car la limite SW est au delà du disk stellaire, mais la modification d'autres paramètres (nb de layers et transit grid size) pour le tps de calcul donne un résultat légèrement différent mais ok à 1 sigma

    \begin{figure*}[]
      \centering
      \includegraphics[width=\hsize,trim={8cm 2cm 8cm 2cm},clip]{"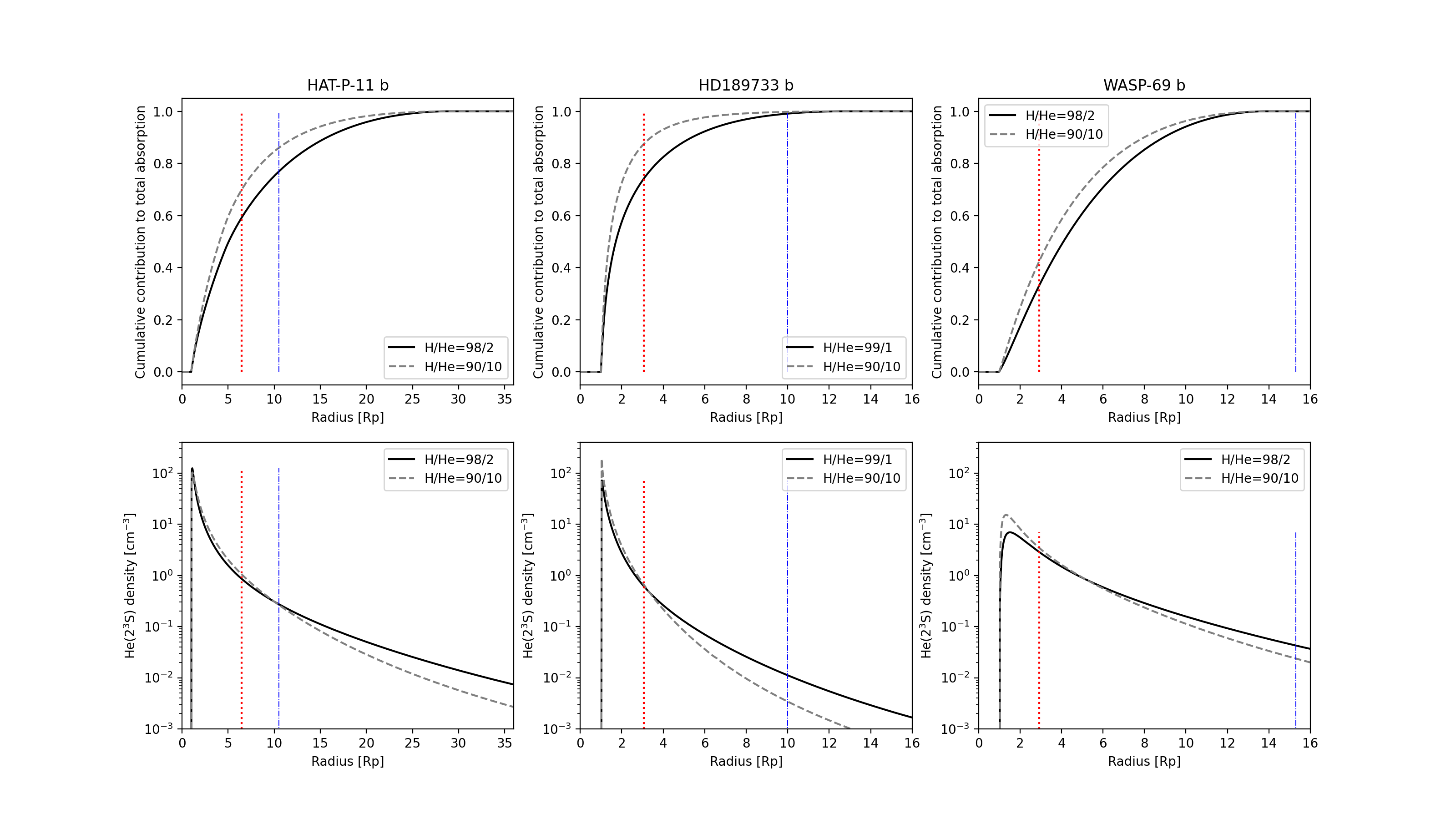"}
      \caption{\change{Structure of the escape for the three detections}. Upper panel: Cumulative contribution of each altitude grid layer to the total absorption in the best model. Lower panel: Metastable helium density profile corresponding to the best model. In both panels, the solid black curve corresponds to the best parameters found for model "M1" (H/He $\geq$ 98/2) and the gray dashed line corresponds to the best solution for model "M4" (H/He = 90/10). The dotted vertical red line indicates the Roche lobe radius and the dash-dotted vertical blue lines correspond to our calculation of the ionopause standoff distance \citep{Khodachenko2019}.}
      \label{weight_density_grid}
    \end{figure*}

    The best fits obtained using the different models presented in Section~\ref{models_list} are reported in Table~\ref{mlr_model_comparison} and Table~\ref{T_model_comparison}.
        As a sanity check, we first compared the results obtained with models M1 and M2, \change{i.e.,} between our wrapper of p-winds (which integrate eccentricity and spin-orbit angle) and our adapted \texttt{Python} version of \citet{Lampon2023}. Both models yield very similar results, which was expected as they are based on the same formulations of \citet{Oklopcic2018} and \citet{Lampon2020}. We compare the other models and detail the influence of these parameters in the next parts.

        \subsection{\mdot/T correlation}
        \label{mlr_T_correlation}
        We observe a clear correlation between the mass-loss rate \mdot and temperature T of the escape in the case of the detections shown in Fig.~\ref{chi2_map_grid}. The helium absorption signature indeed directly increases with the mass-loss rate. The influence of the temperature is twofold. On one hand, at a given altitude, a higher temperature results in a decrease of the atmosphere’s density, and then also on the He triplet density leading to a weaker absorption. On the other hand, a decrease of the atmosphere’s density allows the stellar XUV flux to penetrate deeper in the atmosphere, thus shifting the hydrogen ionization front to lower altitudes which results in a smaller area of absorption for the metastable helium triplet and thus less absorption. Both weakening absorptions compensate for the deepening of the signature when \mdot increases \citep{Lampon2020}. The influence of these two parameters on the line shape and absorption depth is displayed in Fig.~\ref{param_influence_on_line_shape}.
        \subsection{H/He ratio}
    \label{h_he_ratio}
        Comparison between M1 (H/He $\geq 98/2$) and M4 (H/He = 90/10) leads to higher temperature and lower mass-loss rate estimations when decreasing the H/He ratio. This result is expected, as decreasing the H/He ratio results in an atmosphere with a higher He abundance and thus more metastable He absorption. Fitting the same absorption signature than in the H/He$\geq$98/2 case thus requires either to i) decrease the mass-loss rate, which directly decreases the modeled absorption depth, or ii) increase the temperature, which will result in a weaker He signature as explained in Section~\ref{mlr_T_correlation}.

        Similarly, we obtained lower upper limits for \mdot and higher upper limits for T when using a solar H/He ratio instead of the $\geq$ 98/2 values (see Tab.~\ref{table_results}, and the red/gray contours comparison in Fig.~\ref{chi2_map_grid}). These results are in agreement with what \citet{Lampon2023} reported, doing a similar test on CARMENES data. Using a solar value for the H/He ratio yielded closer results to previous solar ratio-based studies: using model M4 (H/He = 90/10) we obtained mass-loss rate values closer to the values derived for example by \citet{Allart_2023} for the same targets (Table~\ref{mlr_model_comparison}).
    % using model M1 (H/He $\geq$ 98/2) we obtained mass-loss rate values of $0.09^{+0.06}_{-0.04}$\,\tengs{11}, $2.5^{+1.6}_{-1.1}$\,\tengs{11}, and $1.9^{+1.8}_{-1.2}$\,\tengs{11} for HAT-P-11\,b, HD\,189733\,b, and WASP-69\,b respectively, while with M4 (H/He = 90/10) we obtained values of $0.08^{+0.05}_{-0.04}$\,\tengs{11}, $0.5\pm{0.3}$\,\tengs{11}, and $0.8^{+0.9}_{-0.5}$\,\tengs{11} respectively, hence closer to the values derived for example by \citet{Allart_2023} for the same targets: $(0.67)^{+0.27}_{-0.24}$\,\tengs{11} for HAT-P-11\,b, $0.94^{+0.82}_{-0.60}$\,\tengs{11} for HD\,189733\,b, and $0.40^{+0.58}_{-0.25}$\,\tengs{11} for WASP-69\,b}. 
    
    However, recent works on constraining the H/He ratio from observations and predictions from 3D MHD models have highlighted that higher H/He ratios are also plausible and could explain the absence of detection in some of the targets studied here (\change{e.g.,} WASP-80\,b, GJ\,436\,b) for which atmospheric escape was expected \citep{Fossati2022,Rumenskikh2023}. So far, constraints on the H/He ratio have been reported in 10 exoplanets: HD\,209458\,b \citep{Lampon2020, Khodachenko2021_hd209, Xing2023}, HD\,189733\,b \citep{Lampon2021b_apr, Rumenskikh2022}, GJ\,3470\,b \citep{Shaikhislamov2021, Lampon2021b_apr, Lampon2023}, HAT-P-32 b \citep{Lampon2023}, GJ\,1214\,b \citep{Lampon2023}, GJ\,436\,b \citep{Rumenskikh2023}, WASP-52 b \citep{Yan2022}, HAT-P-11\,b \citep{DosSantos2022}, WASP-80\,b \citep{Fossati2022} and WASP-107 b \citep{Khodachenko2021}. All of these studies reported a trend toward high H/He values, with WASP-107 b being the only exception with a solar H/He ratio.

    The high H/He ratio estimated in these upper atmospheres remains to be understood. \citet{Xing2023}, using a multi-fluid model accounting for mass fractionation, shows that the high H/He ratio in the upper atmosphere of HD 209458 b is naturally explained by this process, even if the lower atmosphere has a solar H/He ratio. Similar results are also found with evolutionary models that predict an enrichment of He in the lower atmosphere by mass fractionation in hydrodynamic escape (\change{e.g.,} \citealp{Hu2015,Gu2023}). Other processes associated with planet formation, the formation of an H-He immiscibility layer in the deep atmosphere (\change{e.g.,} \citealp{Salpeter1973}), or diffusive separation above the homopause could also play an important role in enhancing the H/He ratio in the upper atmosphere of these planets.

    \subsection{Stellar wind influence}
      \subsubsection{Roche lobe limit}
      Comparing results from models M4 and M5, we tested the influence of stopping the integration of the simulated He absorption signature altitude layers at the Roche lobe, as was done by \citet{Allart_2023}. When limiting the model integration to the Roche lobe (model M5), we dramatically increased the retrieved mass-loss rate and temperature for each of the three detections (Table~\ref{mlr_model_comparison} and \ref{T_model_comparison}), with a factor of 2.5 for the mass-loss rate of HAT-P-11\,b and a factor of 15 for WASP-69\,b.

      To understand this effect, we estimated how much He I triplet absorption was being excluded beyond the Roche lobe. We did so by plotting the cumulative contribution of each altitude layer to the total absorption and compared it to the Roche lobe radius, using both a solar and a super-solar H/He ratio (Fig.~\ref{weight_density_grid}). We estimate that altitude layers beyond the Roche lobe can contribute up to $\sim$10\,\%, $\sim$25\,\% and $\sim$50\,\% to the total absorption detected in HD\,189733\,b, HAT-P-11\,b, and WASP-69\,b, respectively, which explains the difference in the retrieved mass-loss rate and temperature.

      %Our default model M1 yielded \mdot = $0.09^{+0.06}_{-0.04}$\,\tengs{11} and T = $3700^{+500}_{-400}    $ K for HAT-P-11\,b, \mdot = $2.5^{+1.6}_{-1.1}$\,\tengs{11} and T = $18000^{+4000}_{-3000} $ K for HD\,189733\,b, and \mdot = $1.9^{+1.8}_{-1.2}$\,\tengs{11} and T = $7000^{+1600}_{-1800}  $ K for WASP-69\,b. With M5, we obtained \mdot = $0.3^{+0.3}_{-0.2}$\,\tengs{11} and T = $6300^{+1500}_{-1200}    $ K for HAT-P-11\,b, \mdot = $0.8^{+0.2}_{-0.4}$\,\tengs{11} and T = $23000^{+1400}_{-4000} $ K for HD\,189733\,b, and \mdot = $5.2^{+3.6}_{-4.3}$\,\tengs{11} and T = $11000^{+1500}_{-4000}  $ K for WASP-69\,b. 
      Although model M5 aims at reproducing the same assumptions as \citet{Allart_2023} (H/He ratio fixed to a solar value and limiting the model's vertical extension to the Roche lobe), we obtained results closer to their analysis only in the case of HAT-P-11\,b (both in mass-loss rate and temperature) and for the mass-loss rate of HD\,189733\,b (Table~\ref{mlr_model_comparison} and \ref{T_model_comparison}). The remaining differences for the temperature of HD\,189733\,b, and the mass-loss rate and temperature of WASP-69\,b, can be explained by the different stellar XUV fluxes used for the modeling: \citet{Allart_2023} computed the stellar SED following equations from \citet{Linsky2014}, while we used stellar SED from Sanz-Forcada (priv. comm.), \citet{Bourrier2020}, and the \mbox{MUSCLES} database for HAT-P-11\,b, HD\,189733\,b, and WASP-69\,b respectively. We also note that, in the case of HD\,189733\,b, \citet{Allart_2023} let the vertical extension of the model as a free parameter and obtain a top radius at r = $1.41^{+0.20}_{-0.03}$\,Rp, while in model M5 our top radius was set to HD\,189733\,b's Roche lobe at 3.0\,Rp.

      \subsubsection{Ionopause limit}
      In order to approximately take into account the stellar wind effects, \citet{Lampon2023} proposed to stop the integration at the so-called "ionopause standoff distance", where the pressure of the escaping planetary wind equilibrates with the stellar wind pressure \citep{Khodachenko2019}. This approach takes into account the compression of the escaping planetary wind by the stellar wind in a simplified 1D framework, thus not letting the escaping atmosphere cover the entire stellar disk. We computed this ionopause distance using a similar method as in \citet{Lampon2023}. We first used the stellar wind parameters (density profile, temperature, and velocity) characterizing the fast solar wind \citep{Johnstone2015}, and rescaled them to the stellar mass-loss rate of the target's host star which was computed by \citet{Lampon2023} using Eq. 7 in \citet{Johnstone2015}. We then injected the rescaled stellar wind density in Eq. 8 from \citet{Khodachenko2019} to obtain the ionopause distance:

      \begin{center}
        $r_{ionopause} / R_{P} \sim \sqrt{ \frac{M'_{pw} (V_{pw} + kT_{pw}/(m_{p}V_{pw}))}{4\pi R_{p}^2 m_{p} V_{orb}^{2} n_{sw}} }$
      \end{center}

      \noindent where $r_{ionopause}$ is the ionopause altitude, $M'_{pw}$ is the mass-loss rate of the planetary wind (equivalent to \mdot), $V_{pw}$ is the velocity of the escape at the ionopause, $k$ the Boltzmann constant, $T_{pw}$ the temperature of the escaping planetary wind, $m_{p}$ the proton mass, $V_{orb}$ the instantaneous orbital velocity of the planet, and $n_{sw}$ the density of the stellar wind. We computed $V_{pw}$ iteratively, by estimating the velocity at $r_{ionopause}$ from the velocity profile then updating the value of $r_{ionopause}$ until convergence.

      In the same way as we did for the Roche lobe limit, we excluded altitude layers beyond the ionopause to get model M3 and compared it to model M1 (Fig.~\ref{weight_density_grid}). Stopping the integration at the ionopause results in increased temperature and mass-loss rate estimations for HAT-P-11\,b and HD\,189733\,b. We obtained similar results for WASP-69\,b when limiting the model integration to the ionopause (M3) and when integrating the model over the full stellar disk, which was expected as our estimation of WASP-69\,b's ionopause distance lies beyond the stellar disk surface (Fig.~\ref{weight_density_grid}). Although this method proposes a very simplistic estimation of the stellar wind's impact on the escape, it offers an interesting compromise between excluding absorption beyond the Roche lobe and totally ignoring the effect of the stellar wind, without adding computation time to the 1D retrieval process.

    \subsection{Impact of stellar spots}
    Inhomogeneities on the stellar surface at the He I triplet wavelengths can create pseudo-signals that mimic the planetary He signature during the transit \citep[\change{e.g.,}][]{Salz2018, Guilluy2020}. The impact of this effect has been estimated by \citet{Allart_2023} for several targets of the present study, using filling factors based on \citet{Andretta2017}. They report maximum pseudo-absorption contribution from stellar spots to be below the $\sim$0.05\,\% level for AU\,Mic\,b, GJ\,1214\,b, and HAT-P-11\,b due to their low $R_P$/$R_\star$ ratio, and the high eccentricity of HAT-P-11\,b allowing a clear separation between the planetary and stellar signatures even at mid-transit. \citet{Allart_2023} estimated maximum pseudo-signature contributions of $\sim$0.21\,\% for GJ\,3470\,b, $\sim$0.53\,\% for WASP-69\,b and $\sim$0.75\,\% for HD\,189733\,b. The presence of stellar spots could thus not entirely reproduce the He signatures seen in HAT-P-11\,b and WASP-69\,b, and the tentative detection in GJ\,3470\,b, but could significantly affect the HD\,189733\,b detection.

  %______________________________________________________________
  \section{Summary}
    \label{summary}
    In this work, we reduced and analyzed data from 32 transits of 15 different short-period exoplanets observed with \mbox{SPIRou} at CFHT. Our target list spans from mini-Neptunes and potential super-Earths to inflated and ultrahot Jupiters. We present our reduction pipeline\footnote{\change{Will be made available on \url{https://github.com/admasson/HR-SpARTA}}}, involving preprocessing with \mbox{APERO} and correction of the telluric absorption lines, the OH emission, center-to-limb variations, and the Rossiter-McLaughlin effect. Although its influence can be corrected to some extent, the overlap between the telluric OH emission lines and the He triplet has to be taken into account when planning helium surveys. The dates on which observations are made should be chosen carefully in order to optimize the BERV to avoid this contamination.

    We used 1D spherically symmetric Parker-wind hydrodynamic models coupled with a NLTE radiative transfer code to derive the escaping outflow parameters. For comparison purposes, we performed this study with two different models: our wrapper of the p-winds code \citep{DosSantos2019}, and our \texttt{Python} implementation of the model from \citet{Lampon2023}, both with improved transit geometry taking into account the spin--orbit angle and eccentricity\footnote{\change{The radiative transfer module will be made available at \url{https://github.com/admasson/art_he}}}. We retrieved 95\,\% confidence level estimations or set 3$\sigma$ upper limits for the mass-loss rate and the high atmosphere temperature in 13 of the studied targets, for both solar (90/10) and supersolar ($\geq$98/2) values for the H/He ratio.

    We report detections of escaping metastable He triplet in HAT-P-11\,b, HD\,189733\,b, and WASP-69\,b. For HAT-P-11\,b, we report a 14.3$\sigma$ detection with $1.2\pm0.2$\,\% excess depth, corresponding to \mdot = $(9.4^{+5.9}_{-3.5})$\,\tengs{9} and T = $3700^{+500}_{-400}$ K for a H/He ratio of 98/2. For HD\,189733\,b, we report a 13.3$\sigma$ detection with $0.7\pm0.2$\,\% excess depth, \mdot = $2.5^{+1.6}_{-1.1}$\,\tengs{11} and T = $18000^{+4000}_{-3000}$ K for a H/He ratio of 99/1. For WASP-69\,b, we report a 8.2$\sigma$ detection with $3.0\pm0.8$\,\% excess depth, \mdot = $1.9^{+1.8}_{-1.2}$\,\tengs{11} and T = $7000^{+1600}_{-1800}$ K for a H/He ratio of 98/2. Following the analysis by \citet{Allart_2023} of the influence of  stellar spots on the He signature, we confirm that our detections cannot entirely be reproduced by pseudo-signatures. However, for HD\,189733\,b, this effect could contribute to the detected signatures in a non-negligible way (up to $\sim$0.75\%) and may partly explain the observed variability.

    We report tentative detections of escaping He in HD\,209458\,b, GJ\,3470\,b, and WASP-76\,b, along with an asymmetric absorption in the He triplet lines during AU\,Mic\,b's transit, which we attribute to stellar residuals.
    Nevertheless, due to the presence of residuals exceeding the expected noise level in their final spectra, we set upper limits for HD\,209458\,b, GJ\,3470\,b, and WASP-76\,b. We also report nondetections in the remaining targets, with first-time constraints on GJ\,486\,b and improved constraints for TOI-1807. Our nondetections for 55\,Cnc\,e and TOI-1807\,b support the hypothesis of rocky bodies without H/He-dominated atmosphere \citep{Zhang2021, Gaidos2023}.
    Our nondetection in GJ\,436\,b, also reported by \citet{Nortmann2018}, does not agree with the predictions of \citet{Oklopcic2018}. Our absence of a signature in WASP-80\,b is compatible with previous observations \citep{Fossati2022, Allart_2023} and does not agree with the predictions from \citet{Salz2016}.

    We explored the influence of limiting the vertical extension of 1D atmospheric escape models.
    We estimate that the absorbing layers beyond the Roche lobe make a significant contribution to the total absorption. Excluding these layers therefore has a clear impact on the retrieved parameters. Another approach that we tested was to stop the vertical integration at the ionopause, thus taking into account the compression effect of the stellar wind in a simplified 1D toy model. This method can be seen as a compromise between limiting 1D models to the Roche lobe and ignoring the influence of the stellar wind, by approximately accounting for possible effects of the stellar wind. Each of these three approaches can be relevant, depending on the expected influence of the stellar wind on the atmospheric escape of a specific target.

    We also explored the influence of the H/He ratio value on the retrieved properties of the escape. Due to the degeneracy between the H/He ratio, the mass-loss rate, and the temperature, fixing the H/He ratio to a solar or a supersolar value has an important impact on the retrieved parameters. Among the ten exoplanets for which H/He constraints have been obtained so far, nine of them show a trend towards high H/He values. These results, combined with the impact of the H/He ratio on the retrieved parameters, highlight the need to constrain the H/He ratio in He escape studies. As proposed by \citet{Lampon2023}, this could be achieved by simultaneously probing the atmospheric escape through helium (He I) and hydrogen (Ly-$\alpha$ or H$\alpha$) lines. This would not only allow us to constrain the H/He ratio using the retrieved H profile but also to monitor the stellar activity during transit as in \citet{Guilluy2020}. This second point would allow a better understanding of the impact of stellar activity on escaping atmospheres and of its role in the observed variability of the He triplet in some targets. He escape studies would thus benefit from larger samples and monitoring of stellar activity and temporal variability \citep{Farret2023}.

  %______________________________________________________________

  \begin{acknowledgements}
      We thank the anonymous referee for thoughtful reading and suggestions that improved the quality of the manuscript. We thank M. López-Puertas and M. Lampón for providing the original IDL code from which our \texttt{Python} radiative transfer code has been developed. We thank W. Dethier for providing us with the code for modeling the metastable He triplet lines shape from a stellar chromosphere.
      This work received funding from the Programme National de Planétologie (PNP) of the Institut National des Sciences de l'Univers (INSU) of the Centre National de la Recherche Scientifique (CNRS), co-funded by the Centre National d'\'Etudes Spatiales (CNES).
      This work received funding from the "Action Pluriannuelle Incitative (API) Exoplanètes" from the Paris Observatory.
      This work is based on observations obtained at the Canada-France-Hawaii Telescope (CFHT) which is operated from the summit of Maunakea by the National Research Council of Canada, the Institut National des Sciences de l’Univers of the Centre National de la Recherche Scientifique of France, and the University of Hawaii. The observations at the Canada-France-Hawaii Telescope were performed with care and respect from the summit of Maunakea which is a significant cultural and historic site. \mbox{SPIRou} data are accessible through the Canadian Astronomy Data Centre: \url{https://www.cadc-ccda.hia-iha.nrc-cnrc.gc.ca/en/}. We acknowledge funding from the French ANR under contract number ANR18CE310019 (SPlaSH).
      The IAA team acknowledges financial support from the Agencia Estatal de Investigación, MCIN/AEI/10.13039/501100011033, through grants PID2022-141216NB-I00 and CEX2021-001131-S.
      X.D. and A.C. acknowledge funding by the French National Research Agency in the framework of the Investissements d’Avenir program (ANR-15-IDEX-02), through the funding of the “Origin of Life” project of the Université Grenoble Alpes.
      W.D. acknowledges funding from the European Research Council (ERC) under the European Union’s Horizon 2020 research and innovation program (grant agreement No 742095; SPIDI: Star-Planets-Inner Disk-Interactions; http://www.spidi-eu.org)
      B.K. acknowledges funding from the European Research Council under the European Union’s Horizon 2020 research and innovation
      programme (grant agreement No 865624, GPRV)
      E.M. acknowledges funding from FAPEMIG under project number APQ-02493-22 and research productivity grant number 309829/2022-4 awarded by the CNPq, Brazil.
      V.B. acknowledges funding from the NCCR PlanetS supported by the Swiss National Science Foundation under grants 51NF40$\_$182901 and 51NF40$\_$205606, and from the European Research Council (ERC) under the European Union's Horizon 2020 research and innovation programme (project \mbox{Spice Dune}, grant agreement No 947634).
      R. A. is a Trottier Postdoctoral Fellow and acknowledges support from the Trottier Family Foundation. This work was supported in part through a grant from the Fonds de Recherche du Quebec - Nature et Technologies (FRQNT). This work was funded by the Institut Trottier de Recherche sur les Exoplanetes (iREx).
      O. V. acknowledges funding from the ANR project ‘EXACT’ (ANR-21-CE49-0008-01).
      This work has made use of the VALD database, operated at Uppsala University, the Institute of Astronomy RAS in Moscow, and the University of Vienna.
      This work has made use of the MUSCLES Treasury Survey High-Level Science Products; doi:10.17909/T9DG6F.
      This work has made use of data from the European Space Agency (ESA) mission \mbox{Gaia} (\url{https://www.cosmos.esa.int/gaia}), processed by the \mbox{Gaia} Data Processing and Analysis Consortium (DPAC, \url{https://www.cosmos.esa.int/web/gaia/dpac/consortium}). Funding for the DPAC has been provided by national institutions, in particular, the institutions participating in the \mbox{Gaia} Multilateral Agreement.
      This research has made use of the NASA Exoplanet Archive, which is operated by the California Institute of Technology, under contract with the National Aeronautics and Space Administration under the Exoplanet Exploration Program.
      This research has made use of data obtained from or tools provided by the portal exoplanet.eu of The Extrasolar Planets Encyclopaedia.
      This work has made use of the following \texttt{Python} packages: Numpy \citep{harris2020array}; SciPy \citep{2020SciPy-NMeth}; Astropy \citep{astropy:2013, astropy:2018, astropy:2022}; PyAstronomy \citep{pya}; Jupyter \citep{jupyter}; Matplotlib \citep{Hunter:2007}; p-winds \citep{DosSantos2022}. For the purpose of Open Access, a CC-BY 4.0 public copyright license (\url{https://creativecommons.org/licenses/by/4.0/}) has been applied by the authors to the present document and will be applied to all subsequent versions up to the Author Accepted Manuscript arising from this submission.

  \end{acknowledgements}

  \bibliographystyle{aa}
  \bibliography{references}

  \begin{appendix}

    \onecolumn
    \section{Observation conditions}
      \begin{figure}[!htb]
        \centering
        \includegraphics[width=0.9\hsize]{"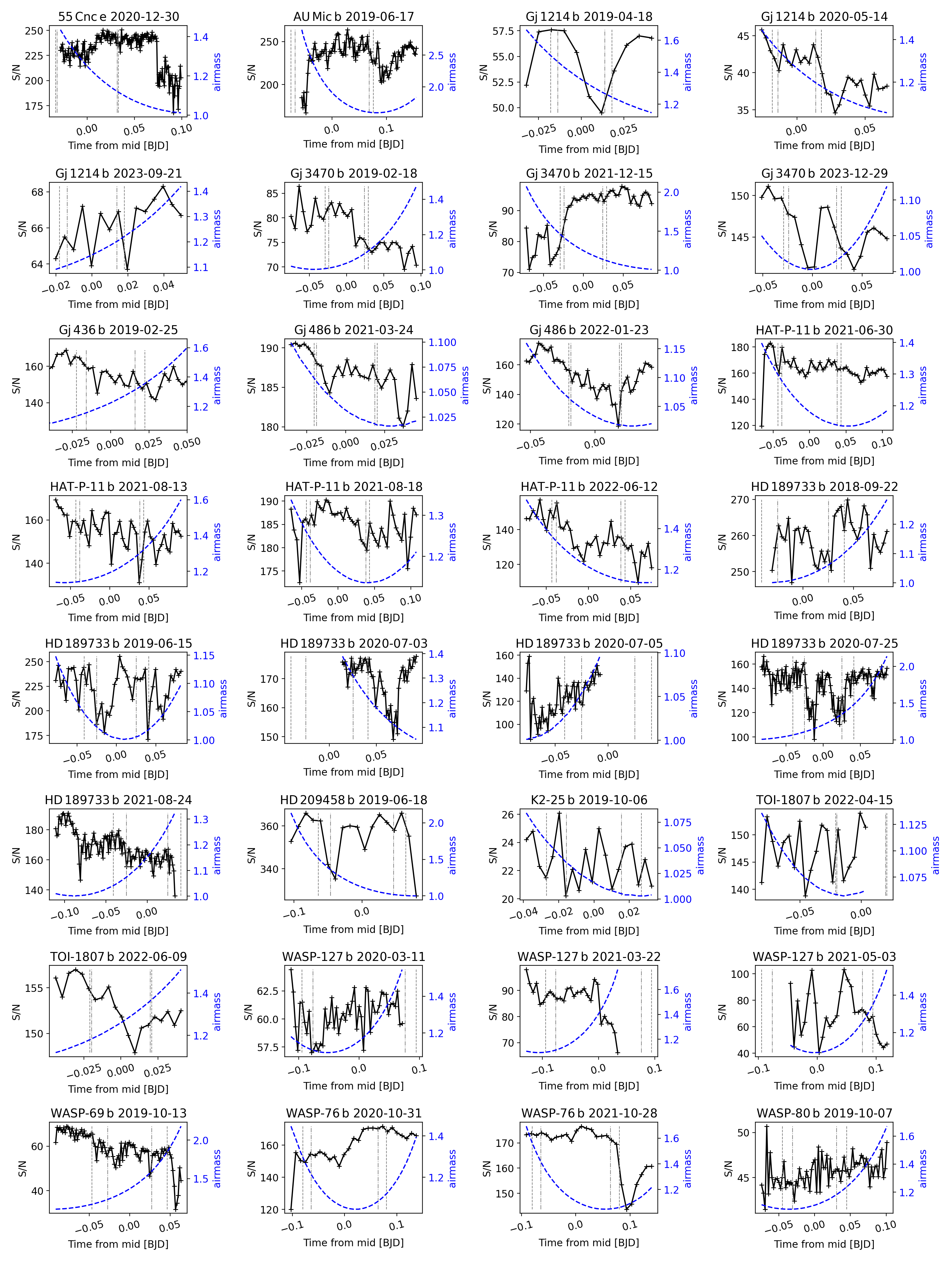"}
        \caption{Observation conditions for each studied transit: signal-to-noise ratio (\change{S/N}) evolution with respect to time from mid-transit is shown in the solid black curve. The \change{S/N} provided by \mbox{SPIRou} is measured in the H band (1.65 $\mu$m). The blue dashed curve shows the airmass evolution along time from mid-transit. The vertical gray lines indicate the four contact points: the dashed ones correspond to T1 (start of transit) and T4 (end of transit), and the dash-dotted ones to T2 (end of ingress) and T3 (beginning of egress).}
        \label{observation_condition}
      \end{figure}

    \newpage

    \onecolumn
    \section{Telluric correction for each transit}
      \label{telluric_correction}
      \begin{figure}[!htb]
        \centering
        \includegraphics[width=\hsize,trim={0cm 20cm 0cm 0cm}]{"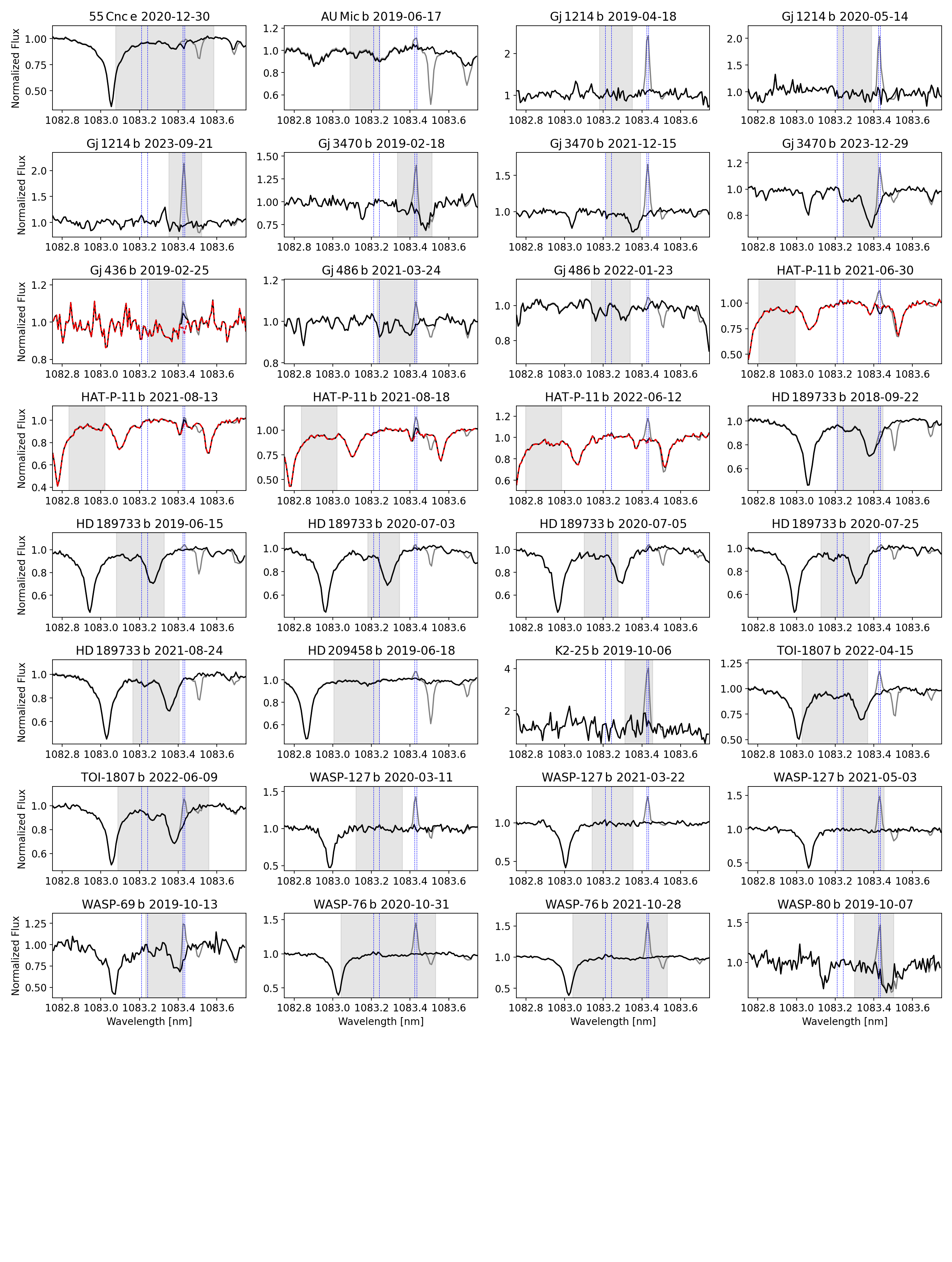"}
        \caption{Data from the E2DS "e" files before (solid gray) and after correction of the telluric lines by \mbox{APERO} ("t" files, solid black), using for each transit an arbitrary observation in the Earth rest frame as an example. The dotted blue vertical lines indicate the position of the OH emission lines, while the gray areas indicate the range of theoretical positions spanned by the planetary helium triplet lines in the Earth rest frame during the transit. For the transits of HAT-P-11\,b and GJ\,436\,b we used our custom correction (dotted red line) of the OH emission lines as residuals of these lines were still visible in the "t" files even after \mbox{APERO} correction.}
        \label{telluric_correction_grid}
      \end{figure}
    \newpage

    \section{Reduced data for each transit}
      \label{reduced_data}
      \begin{figure}[!htb]
        \centering
        \includegraphics[width=0.89\hsize]{"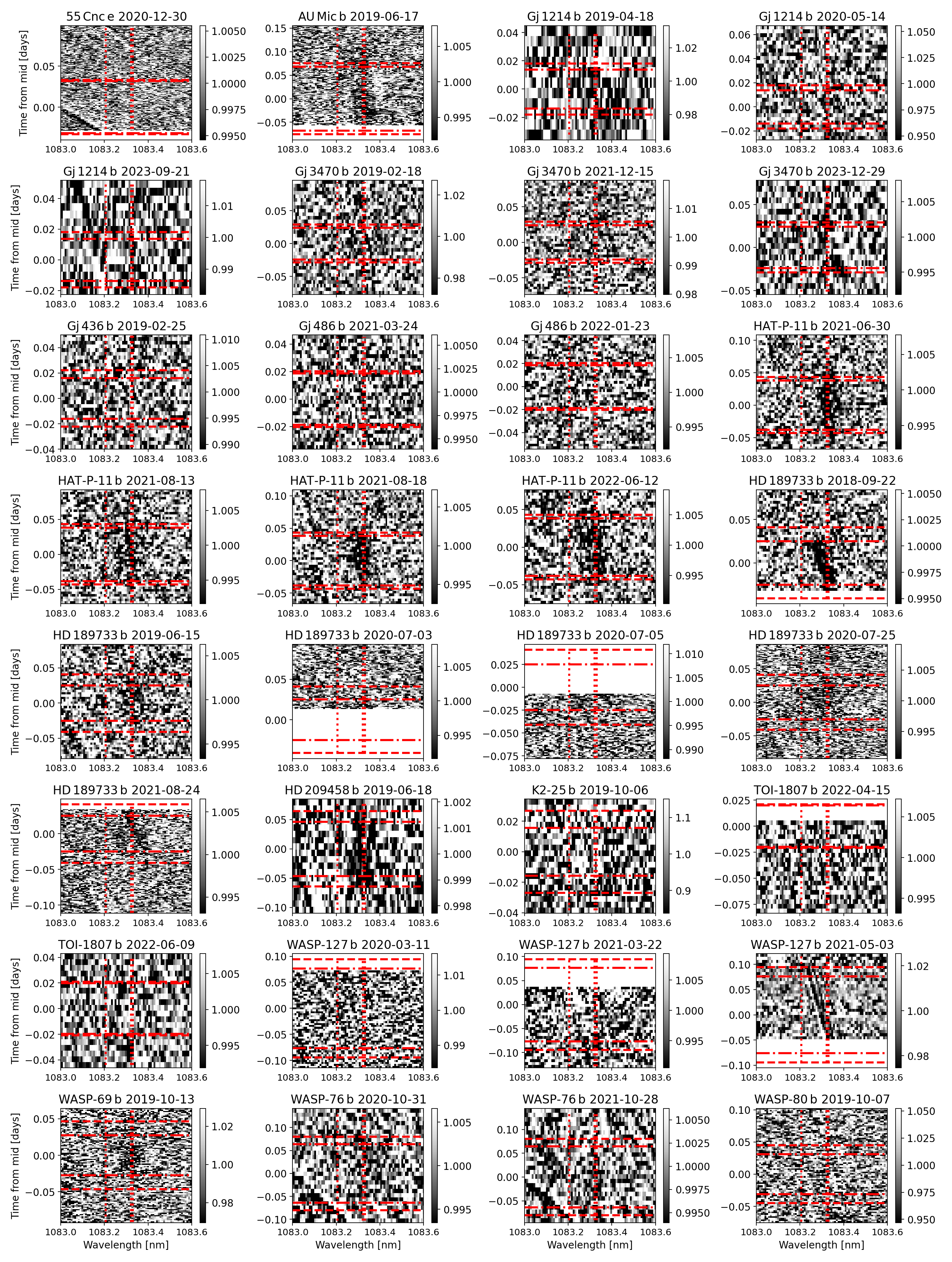"}
         \caption{Time series of the data after reduction: axes are wavelengths in nm and time from mid-transit in days, with the grayscale mapping the normalized intensity. The gray scale span is set to $\pm$\,3 times the standard deviation of the data. The four contact points are indicated in horizontal red lines, the dashed ones corresponding to T1 and T4, and the dash-dotted ones to T2 and T3. Vertical dotted red lines indicate the theoretical position of the He triplet. An excess of absorption, attributed to planetary He, is visible during the transits of HAT-P-11\,b, HD\,189733\,b, and WASP-69\,b.}
         \label{2D_grid_reduced_data}
      \end{figure}

    \newpage

    \section{Transit Light Curves}
      \begin{figure}[!htb]
        \centering
        \includegraphics[width=0.9\hsize]{"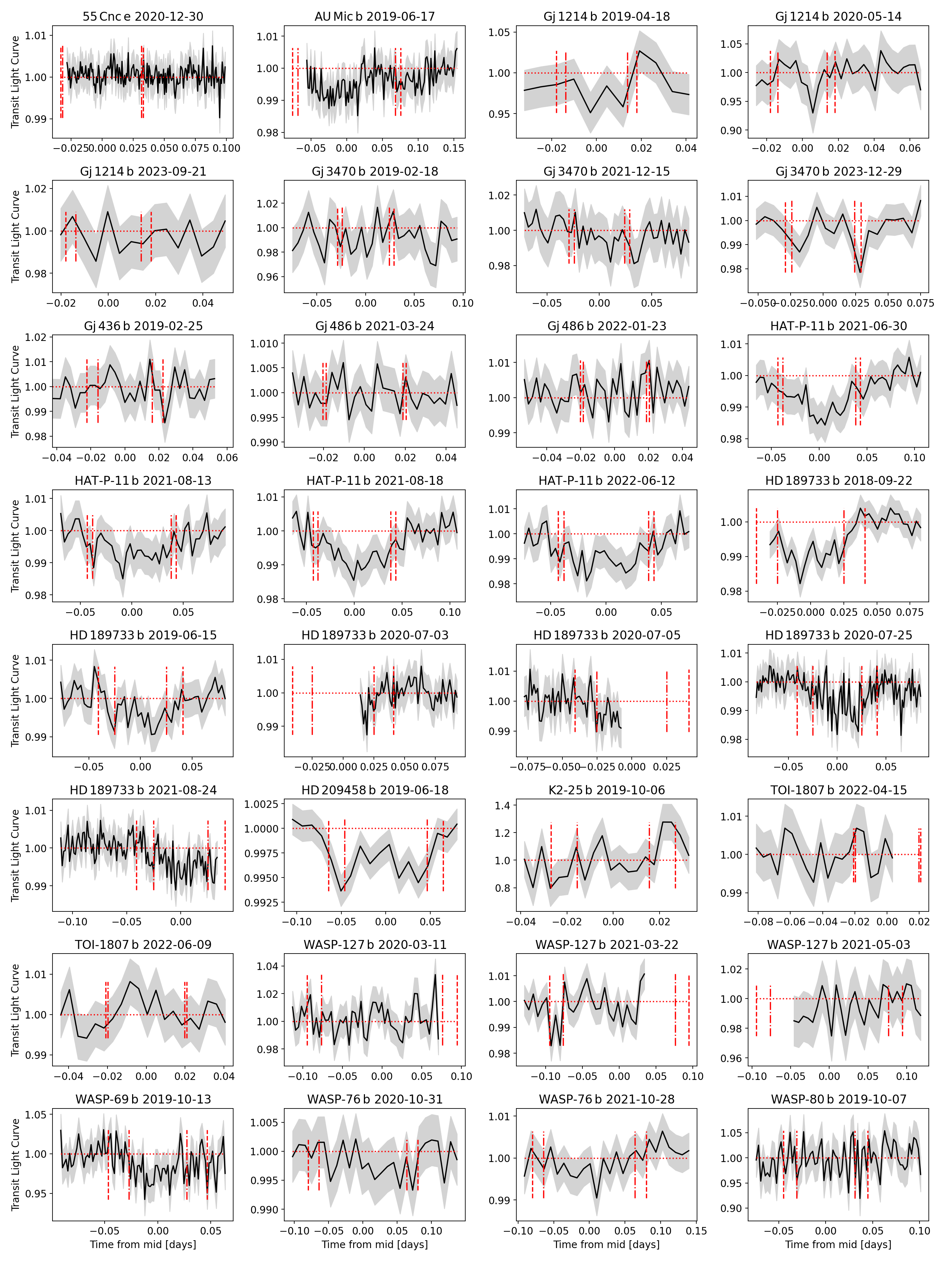"}
        \caption{Light curve of each transit, computed by integrating the intensity of each spectrum in a spectral bin centered at the He triplet position for each observation. The horizontal axis shows the time from mid-transit in days, with the vertical dashed red lines showing contact points T1 (start of transit) and T4 (end of transit), and the dash-dotted red ones showing the contact points T2 (end of ingress) and T3 (beginning of egress). Error bars are shown in light gray and correspond to a 95\% confidence interval.}
        \label{TLC_grid}
      \end{figure}

    \newpage

    \section{Influence of parameters on the Helium line shape}
      \begin{figure}[!htb]
        \centering
        \includegraphics[width=0.75\hsize]{"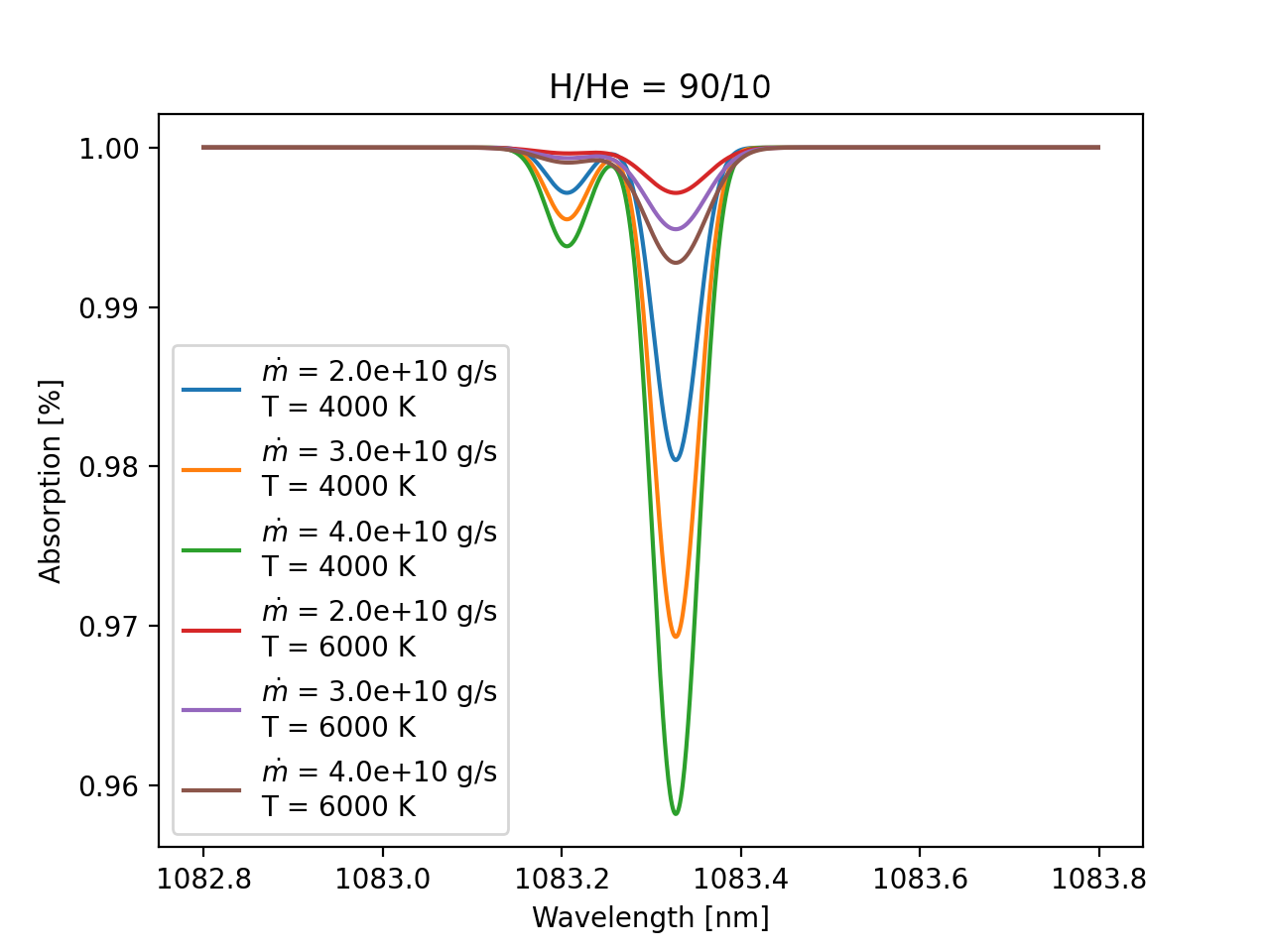"}
        \includegraphics[width=0.75\hsize]{"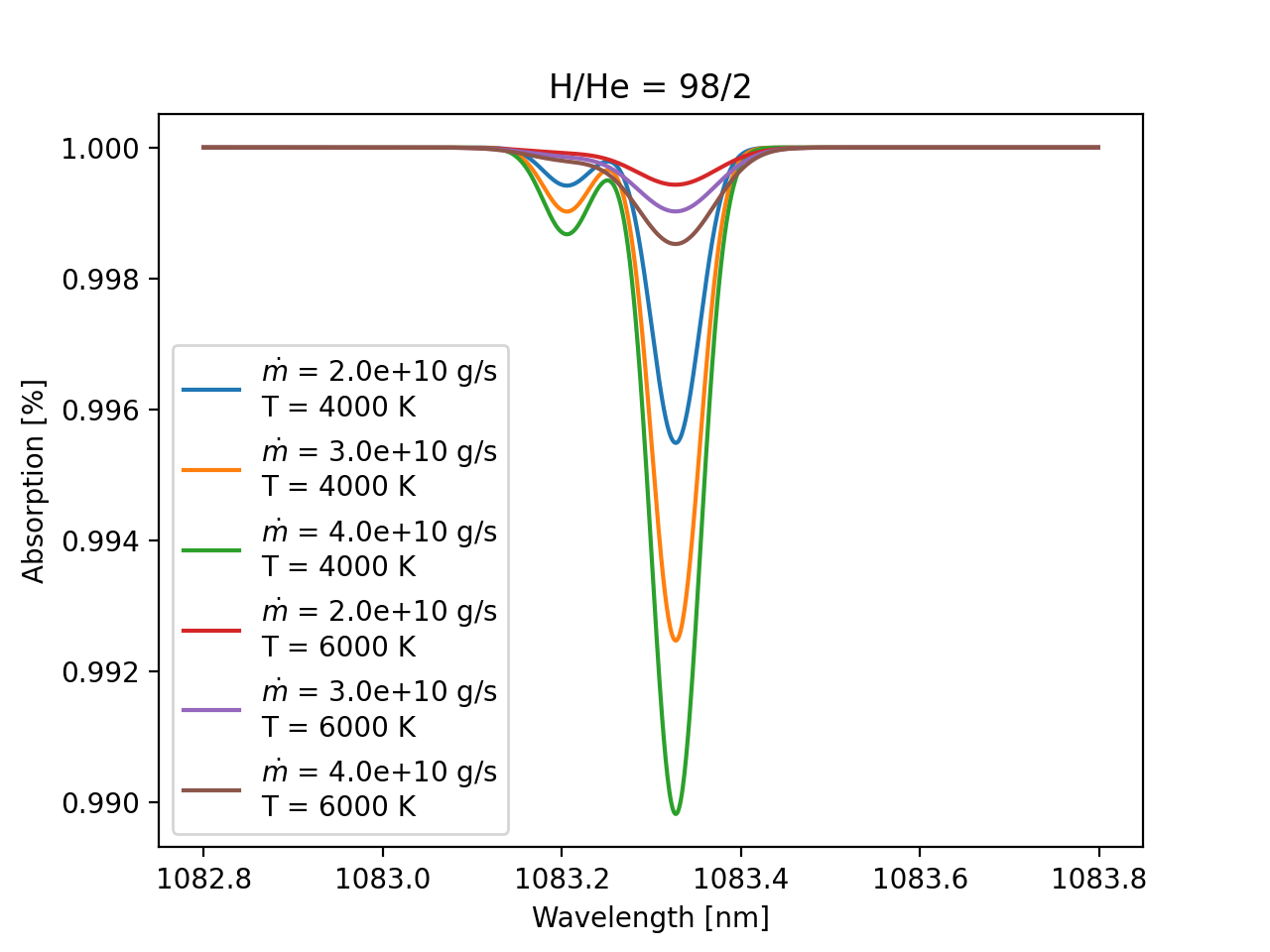"}
        \caption{Influence of the mass-loss rate (\mdot), temperature (T), and H/He ratio on the predicted depth and shape of the He triplet lines in the case of HAT-P-11\,b. The upper panel corresponds to a solar (90/10) H/He ratio, while the lower panel corresponds to a H/He ratio of 98/2. The influence of \mdot and T, as detailed in Section~\ref{mlr_T_correlation}, is clearly visible: for a given H/He ratio, increasing \mdot results in a deeper line while increasing the temperature decreases the depth and increases the broadening of the lines. For a fixed \mdot and T, using a H/He ratio of 98/2 (lower panel) instead of 90/10 (upper panel) results in a decrease of the line depth (note the change of scale on the y-axis).}
        \label{param_influence_on_line_shape}
      \end{figure}

    \newpage

    \twocolumn
    \section{Error estimation}
      \subsection{Error bars of the reduced spectrum}
      \label{error_bar_estimation}
      The final spectrum of a given transit is computed by averaging all in-transit observations. Our analysis focuses on a small region containing $\sim$100 spectral bins, centered around the He triplet lines. This region falls at the center of \mbox{SPIRou}'s echelle order \#71, and thus corresponds to a region where the instrumental blaze function is i) maximum, thus maximizing the \change{S/N} of the individual spectral bins and ii) can be approximated as a constant, given the size of this $\sim$100 pixels region compared with the 4088 spectral bins that make a full \mbox{SPIRou} order.

      We thus assume that the statistics of the noise are the same for all $\sim$100 pixels contained in this region of interest. Under this assumption, each observation of a single spectral bin will contain a realization of the noise. Regarding the averaged spectrum of a single transit, the error on each spectral bin can thus be estimated using its standard deviation along the time axis during the transit. This standard deviation is computed, for each spectral bin "$\lambda_i$", following:

      \begin{center}
        $\sigma_{\lambda_{i}}^2 = \frac{N_{obs}}{N_{obs}-1} \frac{\sum_{k}^{N_{obs}}w_{k}(I(\lambda_i,k)-\bar{I}(\lambda_i))^{2}}{\sum_{k}^{N_{obs}}w_{k}}$
      \end{center}

      \noindent with $\sigma_{\lambda_i}$ the temporal standard deviation of the spectral bin $\lambda_i$, $N_{obs}$ is the number of in-transit observations, $I(\lambda_i,k)$ is the intensity measured at the spectral bin $\lambda_i$ during the $k^{th}$ in-transit observation, and $\bar{I}(\lambda_i)$ is the mean in-transit intensity value of this spectral bin:

      \begin{center}
        $\bar{I}(\lambda_i) = \frac{\sum_{k}^{N}w_{k} I(\lambda_i,k)}{\sum_{k}w_{k}}$
      \end{center}

      \noindent $w_k$ is a weight corresponding to the squared \change{S/N} value of the $k^{th}$ observation (measured by the instrument in a dedicated spectral bin, see Section~\ref{observation}), ensuring an optimal combination of observations with different \change{S/N} values.
      We then estimate the error on the average value $\bar{I}(\lambda_i)$ of each spectral bin by dividing its temporal standard deviation $\sigma_{\lambda_i}$ by the square root of the number of observations $N_{obs}$ (see, for example, eq 4.23 in \citealp{Bevington} for a more complete justification):

      \begin{center}
        $\sigma_{\bar{I}(\lambda_i)} = \frac{\sigma_{\lambda_{i}}}{\sqrt{N_{obs}}}$
      \end{center}

      This $\sigma_{\bar{I}(\lambda_i)}$ represents the 1-$\sigma$ error bar associated with the spectral bin $\lambda_{i}$ in the averaged spectrum $\bar{I}(\lambda_i)$, for a given transit. When multiple transits were available for a given target, we combined these transits by averaging them, weighting each transit's spectrum using its error bars:

      \begin{center}
        $\bar{I}_{comb} = \frac{ \sum_{j}^{N_t}  \sigma_{\bar{I}^{j}}^{-2}  \bar{I}^{j} }{ \sum_{j}^{N_t} \sigma_{\bar{I}^{j}}^{-2}}$
      \end{center}

      \noindent where $\bar{I}_{comb}$ is the final spectrum of a given target, computed by combining a number $N_{t}$ of averaged spectra of individual transits of the same target. $\bar{I}^{j}$ is the averaged spectrum of the $j^{th}$ transit, and $\sigma_{\bar{I}^{j}}$ contains the associated error bars for each spectral bins in the $\bar{I}^{j}$ spectrum. We note that this equation applies for each spectral bin $\lambda_{i}$ individually, we just dropped the notation showing the spectral dependence for readability purposes. Lastly, the error bar $\sigma_{\bar{I}_{comb}}(\lambda_{i})$ on each spectral bin $\lambda_{i}$ in the final spectrum $\bar{I}_{comb}$ is computed by combining the error bars of the corresponding spectral bin from the different $\bar{I}^{j}$ spectra:

      \begin{center}
        $\sigma_{\bar{I}_{comb}} = \frac{ 1 }{ \sqrt{ \sum_{j}^{N_t} \sigma_{\bar{I}^{j}}^{-2}} }$
      \end{center}

      \subsection{Upper limits and uncertainties on the signature depth and equivalent width}
      \label{error_estimation_depth_EW}
        The above formulas were used to compute the 95\,\% confidence interval error bars displayed in Fig.~\ref{1d_reduced_data}. We however note the presence of residuals at a level higher than the expected noise level (Fig.~\ref{1d_reduced_data}). To properly account for these features in our uncertainties and upper limits estimations of the signature depth and equivalent width (EW), we used the following procedure: we first fitted a Gaussian model with free amplitude, full width at half maximum (FWHM), and position around the expected He line rest location. The best fit was obtained using the non-linear least squares method implemented in SciPy \citep{2020SciPy-NMeth}.

        This method also provides an estimation of the Hessian matrix of the fitting model with respect to the free parameters that naturally account for the higher standard deviation of the data along the spectral axis induced by the residuals. For the detections of metastable He, we inverted the Hessian matrix to estimate the covariance matrix of the free parameters. The 1-$\sigma$ uncertainty on each parameter (depth, FWHM, and position) was computed as the square root of the diagonal elements in the covariance matrix, from which we derived our estimations of the signature's depth, EW, and Doppler shifts with their associated 95\,\% confidence interval for each detection reported in Tables~\ref{table_model_parameter_targets} and \ref{table_results}.

        For tentative and nondetections, the 3-$\sigma$ upper limits on the signature's depth and EW were obtained by iteratively computing Gaussian models with increasing depth and fixed FWHM and position. The FWHM and position were fixed to the best-fit values in case of tentative detection, while for nondetection we fixed the Gaussian center to the theoretical mean position of the metastable He triplet strongest lines, and set the FWHM to 10 SPIRou pixels (\change{i.e.,} 0.08\,nm). The Gaussian depth was then iteratively increased, computing at each iteration the $\Delta\chi^2$ between the iterating model and the best fit in case of tentative detection or a null model for nondetections. We stopped the iterations when reaching a $\Delta\chi^2$ of 9, which corresponds to a 3-$\sigma$ confidence interval for a $\chi^2$ statistics with one degree of freedom. The depth and EW of this $\Delta\chi^2 = 9$ model were set as the 3-$\sigma$ upper limits for the depth and EW of the tentative and nondetections reported in Table~\ref{table_results}.

        We note that the $\Delta\chi^2$ computation was performed in a spectral bin of $\pm0.1$\,nm around the planetary He lines position. As described in Section~\ref{master_spectrum_reduction}, the master spectrum was computed differently within and outside this spectral bin, by averaging the observations from the entire transit outside the bin and only the out-of-transit observations within the bin. Although this process ensured optimal correction of the stellar lines outside the He signature position, it also introduced statistical differences between data inside and outside the spectral bin. Limiting the $\Delta\chi^2$ computation to wavelengths within this spectral bin hence ensured that our error estimations were based on statistically consistent data.

      \subsection{Upper limits and uncertainties on the mass-loss rate and temperature}
        \label{error_estimation_mlr_T}
        Uncertainties at a 95\,\% confidence interval and 3-$\sigma$ upper limits on the mass-loss rate and temperature of the escaping atmospheres were obtained from the reduced $\chi^2$ ($\chi^{2}_{red}$) maps and the confidence interval contours displayed in Fig.~\ref{chi2_map_grid}. Estimation of the $\chi^{2}_{red}$ requires an estimation of the error, as appears in the $\chi^{2}_{red}$ definition:

        \begin{center}
        $\chi^{2}_{red} = \frac{1}{N-n} \sum^{N}_{i}\frac{(d(\lambda_{i}) - m(\lambda_{i}))^2}{\sigma(\lambda_{i})^2}$
        \end{center}

        \noindent where $m(\lambda_i)$, $d(\lambda_{i})$ and $\sigma(\lambda_i)$ respectively represent the model, the data, and the associated error at a wavelength $\lambda_i$. N is the number of points considered, and n is the number of fitted parameters ($n=2$ here). As explained above, our data contain residuals at a higher level than the $\sigma_{\bar{I}_{comb}}(\lambda_i)$ error bars estimated from the temporal standard deviation of the data during transit.
        To account for these residuals in our $\chi^{2}_{red}$ estimation, we proceeded in two steps: first, we computed $\chi^{2}_{red}$ maps for each target using the temporal error bars $\sigma_{\bar{I}_{comb}}(\lambda_i)$ (\change{i.e.,} estimated from the temporal standard deviation) as the $\sigma(\lambda_i)$ error in the $\chi^{2}_{red}$ formula:

        \begin{center}
        $\chi^{2}_{red} = \frac{1}{N-n} \sum^{N}_{i}\frac{(d(\lambda_{i}) - m(\lambda_{i}))^2}{\sigma_{\bar{I}_{comb}}(\lambda_i)^2}$
        \end{center}

        Doing so, we underestimated the actual error in our data and obtained minimum $\chi^{2}_{red}$ values higher than 1, but we obtained a first estimation of the best fit for each detection and tentative detection. In a second step, we reestimated this error by computing the standard deviation of the data along the spectral axis to better take into account potential systematic signals, due to the star for instance. For detections and tentative detections, this standard deviation was computed on the residuals obtained after subtracting the best model found in the previous $\chi^{2}_{red}$ minimization to the data. For nondetection, this standard deviation was directly computed on the data, which was equivalent to considering a null model as the best model. Computing the standard deviation along the spectral axis allowed us to include the influence of the stellar residuals in our error estimation. This error (hereafter $\sigma_{spec}$) does not depend on wavelength and was used as the error estimation in our final $\chi^{2}_{red}$ maps:

      \begin{center}
        $\chi^{2}_{red} = \frac{1}{N-n} \sum^{N}_{i}\frac{(d(\lambda_{i}) - m(\lambda_{i}))^2}{\sigma_{spec}^2}$
      \end{center}

      Similarly to the depth and EW upper limits estimations in the previous section, the $\chi^{2}_{red}$ computations were performed in a spectral bin of $\pm0.1$\,nm around the planetary He lines position to ensure statistical consistency between the data used for the computations. The uncertainties and upper limits on the mass-loss rate and temperature were then estimated by finding the extremal (\mdot, T) values yielding models with a $\Delta\chi^2$ corresponding to a 95\,\% confidence interval for detections and a 3-$\sigma$ confidence interval for tentative and nondetections (the confidence intervals are displayed in Fig.~\ref{chi2_map_grid}). As in Section~\ref{error_estimation_depth_EW}, the $\Delta\chi^2$ was computed as the difference between each model's $\chi^2$ and the best model for detections and tentative detections, and a null model for nondetections.
  \end{appendix}

\end{document}